\documentclass[12pt]{article}

\usepackage{amsmath}
\usepackage{graphicx,psfrag,epsf,epstopdf}
\usepackage{enumerate}
\usepackage{natbib}
\usepackage{url} 

\usepackage{amssymb}
\usepackage{amsthm}
\usepackage{color}
\usepackage{caption}
\usepackage{float}
\usepackage{booktabs}
\usepackage{bbm}
\usepackage{multicol}
\usepackage{adjustbox}
\usepackage{multirow}
\usepackage{rotating}
\usepackage{pdflscape}
\usepackage[hypertexnames=false]{hyperref}
\hypersetup{colorlinks=true,citecolor=blue}

\def\bmath0{{\boldmath 0}}
\def\bm1{{\boldsymbol 1}}

\def\bmu{{\boldsymbol u}}

\def\bmx{{\boldsymbol x}}
\def\bmy{{\boldsymbol y}}
\def\bmB{{\boldsymbol B}}

\def\bmI{{\boldsymbol I}}

\def\bmp{{\boldsymbol p}}

\def\bmz{{\boldsymbol z}}
\def\bmA{{\boldsymbol A}}

\def\bmX{{\boldsymbol X}}
\def\bmY{{\boldsymbol Y}}

   \def\0{\boldsymbol 0}

\DeclareMathOperator*{\argmin}{arg\,min}
\newcommand{\RomanNumeralCaps}[1]{\MakeUppercase{\romannumeral #1}}
\newcommand{\ol}{\overline}
\usepackage{algorithm}
\usepackage[noend]{algpseudocode}

\algnewcommand{\IIf}[1]{\State\algorithmicif\ #1\ \algorithmicthen}
\algnewcommand{\EndIIf}{\unskip\ \algorithmicend\ \algorithmicif}

\newcommand{\ip}[2]{\left\langle#1,#2\right\rangle}

\begin{document}

\title{\LARGE\bf Abnormal component analysis}
\author{    Romain Valla \\    {\small LTCI, Télécom Paris, Institut Polytechnique de Paris}\\
    Pavlo Mozharovskyi \\    {\small LTCI, Télécom Paris, Institut Polytechnique de Paris}\\
    Florence d'Alché-Buc \\    {\small LTCI, Télécom Paris, Institut Polytechnique de Paris}}
    
\date{June 05, 2025}
\maketitle

\vspace{5ex}

\begin{abstract}
At the crossroads of machine learning and data analysis, anomaly detection aims at identifying observations that exhibit abnormal behaviour. Be it measurement errors, disease development, severe weather, production quality default(s) (items) or failed equipment, financial frauds or crisis events, their on-time identification and isolation constitute an important task in almost any area of industry and science. While a substantial body of literature is devoted to detection of anomalies, little attention is payed to their explanation. This is the case mostly due to intrinsically non-supervised nature of the task and non-robustness of the exploratory methods like principal component analysis (PCA).

We introduce a new statistical tool dedicated for exploratory analysis of abnormal observations using data depth as a score. Abnormal component analysis (shortly ACA) is a method that searches a low-dimensional data representation that best visualises and explains anomalies. This low-dimensional representation not only allows to distinguish groups of anomalies better than the methods of the state of the art, but as well provides a---linear in variables and thus easily interpretable---explanation for anomalies.
In a comparative simulation and real-data study, ACA also proves advantageous for anomaly analysis with respect to methods present in the literature.
\end{abstract}

\indent\\


\newpage

\section{Introduction}
\label{sec:intro}

Anomaly detection is a branch of machine learning which aims at finding unusual patterns in the data and allows to identify observations that deviate significantly from normal behavior; see, \textit{e.g.},~\cite{ChandolaBK09,AhmedMI16,RousseeuwH18,ThudumuBJS20} (and references therein) for surveys on existing anomaly detection methods. Anomalies can be represented by abnormal body cells or deviating health parameters, failed equipment or default items, network intrusions or financial frauds, and need to be identified for undertaking further action. Detecting anomalies can help to start timely treatment or handling, improve product's quality, and ensure operational safety. To develop a reaction policy, a deeper insight into anomalies' nature is required, which further demands to explain the reasons for abnormality. A number of works underline importance of explainability in statistics and machine learning, \textit{e.g.}, \citep{DOSHI17, MURDOCH19, BARREDOARRIETA20}, including the recent survey by~\cite{LiZVL23}. This task of explainability, undergoing active development with several proposed solutions in the supervised setting (\textit{e.g.}, variable importance for random forest~\citep{CARLETTI23} or concept-based explanation for neural networks~\citep{KimWGCWVS18, PAREKH21}), is particularly challenging in the unsupervised setting not only due to the absence of the feedback, but also because of potentially infinite variety of abnormalities.

In the current article, we focus on the multivariate setting, where observations possess $d$ (metric) quantitative properties. More precisely, we consider a (training) data set $\bmX=\{\bmx_1,...,\bmx_n\}\subset\mathbb{R}^d$ that consists of $n$ observations in a $d$-dimensional Euclidean space $\mathbb{R}^d$. This data set $\bmX$ may or may not contain anomalies, with this information being unknown (on the training stage) in the unsupervised framework considered here. Explanation of an anomaly $\bmx\in\mathbb{R}^d$ in this case can be done, \textit{e.g.}, by importance ranking of $\bmx$ constituting variables (or their combinations, to account for non-linearity). Possibly based on this information, even more important is insightful data visualization, which allows to identify anomalies and (simultaneously) the features that are causing them.

Providing a meaningful and easily interpretable visualization cannot be overestimated in practice, and in reality is of highest importance for solving a number of practical tasks. Several methods serving this purpose have are widely used and implemented in numerous software packages employed in various areas of industry and science. These---being shortly over-viewed right below---fail to underline anomalies, mostly for two reasons: either (a) they lack robustness necessary to ``notice'' anomalies or (b) they are simply not aiming at highlighting them.

\subsection{Existing methods for meaningful visualization of anomalies}\label{ssec:allmethods}

A number of methods at hand, though not intrinsically designed for anomaly detection framework, can be useful to provide meaningful visualization. In particular, dimension reduction techniques are  effective in providing representation spaces that can, in certain cases, highlight anomalies. These can be enhanced by explanation capacity (if available), see, \textit{e.g.},~\cite{ANOWAR21} for a recent survey.


\paragraph{Linear methods}
Linear methods provide explainable data visualization by searching for a new basis in $\mathbb{R}^d$ with components being linear combinations of input variables:
\emph{Principal component analysis} (PCA) computes (up to) $d$ components---mutually orthogonal---such that in projection on each of them variance is maximized~\citep{Pearson01,Hotelling33}. In this way, first principal component corresponds to the direction in projection on which data variance is maximized. Second principal component then maximizes the variance of the data in linear subspace of $\mathbb{R}^d$ orthogonal to the first component. Third principal component maximizes variance in the linear subspace of $\mathbb{R}^d$ orthogonal to the first two components; this process continues until either the required number of components is found or the entire variance is explained. Plotting pairwise components provides insightful visualization, together with other visualizations employed for clustering or revealing hidden structure in the data. Due to it's simplicity of understanding and speed of execution, through decades PCA remains one of the most used data visualization and explanation tools for practitioners.
\emph{Robust principal component analysis} (robPCA) has been designed to compensate for presence of anomalies in the data, because anomalies' values---amplified being squared---distract found by traditional PCA variance-maximizing directions~\citep{HUBERT05}.
While the classic principal component analysis methods describe well Gaussian (elliptical) data, \emph{independent component analysis} (ICA) allows to departure from this limitation by searching for non-Gaussian statistically-independent features~\citep{COMON92}.

\paragraph{Non-linear methods}
Non-linear methods, different to those exploiting first-order stochastic dependency (and thus categorized as linear), are based on non-linear geometric transform, often performed via applying a kernel function to between-point distances. Thus, \emph{kernel principal component analysis} (kPCA) can be seen as an extension of traditional PCA using the ``kernel trick''~\citep{SCHOLKOPF97} to handle data in the (infinite-dimensional) reproducing kernel Hilbert space (RKHS) based on a properly chosen kernel function \citep{SCHOLKOPF02}, in which in order the principal components are searched. \emph{Multi-dimensional scaling} (MDS) makes use of kernel-transformed pair-wise dissimilarities (often expressed as distances) of centered data to construct a lower-dimensional representation by means of the eigenvalues decomposition of the kernel matrix~\citep{COX08}. To construct an insightful visualization, \emph{$t$-distributed stochastic neighbor embedding} ($t$-SNE) first defines a similarity (using Euclidean distance or alternative measure) distribution on the space of objects, and then maps it to another low-dimensional distribution by minimizing the asymmetric Kullback-Leibler divergence between the two~\citep{VANDERMAATEN08}.

\paragraph{Further methods}
Further methods have been developed that can naturally serve for insightful visualization, which logically do not fall under any of the two mention above categories.
\emph{Non-negative matrix factorization} (NMF) decomposes the data matrix into a product of two tentatively smaller (and thus naturally lower-rank) matrices under the non-negativity constraint, in order to minimize, \textit{e.g.}, the Frobenius norm or the Kullback-Leibler divergence of the product~\citep{LEE99}.
\emph{Locally linear embedding} (LLE) is another non-linear dimension-reduction method which proceeds in two stages~\citep{ROWEIS00}: first each point is reconstructed as a weighted sum of its neighbors, and second a lower-dimensional space is constructed (based on eigenvalue decomposition) searching for the reconstruction using the weight from the first stage. The local linearity is then governed by the predefined number of neighbors and the distance used. \emph{Laplacian eigenmaps} (LE) approximate data in a lower-dimensional manifold using the neighborhood-based graph with eigenfunctions of the Laplace–Beltrami operator forming the embedding dimensions~\citep{BELKIN03}. \emph{Autoencoder}~\citep{SAKURADA14} consists of artificial neuronal encoder and decoder connected by an (information compressing) bottleneck. The latent (neuronal) signals of this bottleneck can be then used to visualize the data, as well as reconstruction error allows to detect anomalies.

\paragraph{Explainability of anomalies} 
Explainability of anomalies constitutes an open question and an active field of research with very little explicit available solutions in the unsupervised setting (different to the supervised one, see, \textit{e.g.},~\citep{GornitzKRB13} for a survey). One of them is \emph{depth-based isolation forest feature importance}~\citep{CARLETTI23}, a variable-importance method for isolation forest~\citep{LIU08} that ensures both global (\textit{i.e.}, on the level of the trained procedure) and local (\textit{i.e.}, for the particular (new) observation in question) explainability by providing quantitative information on how much each variable influenced the abnormality decision. In the same group can be put \emph{cell-wise outlier detection}~\citep{ROUSSEEUW18}, which identifies the cells (\textit{i.e.}, observation's variables, avoiding labeling the entire observation as an outlier) for outlying observations which contaminate the data. Generally speaking, though explainability of anomalies can be seen as an unresolved issue, insightful (linear) visualization methods provide variable-wise information about anomalies, if those can be identified. With explainability of anomalies constituting an important contemporary challenge, it seems that---in view of the potentially rich nature of anomalies---their side-effect identification (and interpretation) is unlikely, though not excluded. That is, special methods---focused on the search of anomalies---are required not only to find them but also to interpret.

\subsection{The proposed approach}

In the current article, we propose a versatile method for anomaly visualization and interpretation, targeting directions relevant for anomalies. The so-called \emph{abnormal component analysis} ACA sequentially constructs an ortho-normal basis that best unveils the anomalies (additionally splitting them according to geometric grouping) to the human's eye and at the same time allows to perform their automatized interpretation. The proposed method largely exploits the concept statistical data depth function, and in particular depth notions satisfying the \emph{weak projection property} introduced by~\cite{DYCKERHOFF04} and extensively studied later in the computational context by~\cite{NAGY20} and~\cite{DYCKERHOFF21}. More precisely, a direction is being searched which allows for identification of the most outlying (cluster of) anomalies, while in the subsequent steps such a direction is searched in the linear orthogonal complement of the previous directions. Except for the intrinsic (and indispensable) robustness and depth-inherited affine invariance, ACA possesses attractive computational complexity of $\mathcal{O}(pkdn^2)$ for the entire data set of size $n$ in dimension $d$, with $p$ being the number of the searched components and $k$ being the number of necessary directions, with its choice discussed in Section~\ref{ssec:parameters}.

\subsection{Outline of the article}

The rest of the article is organized as follows. After a short reminder on data depth, Section~\ref{sec:meth} introduces the ACA method, suggests an algorithm for its computation, and discusses the choice of relevant parameters. Section~\ref{sec:comparevisu} is focused on the visual comparison of ACA as a dimension-reduction tool with existing methodologies, on simulated data sets possessing different properties. Section~\ref{sec:compareinter} provides insights on explainable anomaly detection with data depth employed following the ACA-based philosophy, in a simulated setting (where the correct direction is known) in a comparison with PCA, robPCA, ICA. Section~\ref{sec:realdata}---in an application to real data sets---provides insightful visualization as well as explanations to them, unknown to the preceding literature. Section~\ref{sec:conc} concludes, and enumerates the contents of the Supplementary Materials.
 
\section{Method}\label{sec:meth}

In this section, we briefly remind the notion of data depth (Section~\ref{ssec:depth}), and after this we introduce the method of abnormal component analysis (Section~\ref{ssec:aca}) followed by the algorithm (Section~\ref{ssec:algorithm}) and a discussion on the choice of its parameters (Section~\ref{ssec:parameters}). Denote $\bmX=\{\bmx_1,...,\bmx_n\}$ a data set of $n$ points in $\mathbb{R}^d$ (we use set operator in a slight abuse of notation, since ties are possible but do not distort the proposed methodology), and let $\bmz\in\mathbb{R}^d$ be an arbitrary point of the space.

\subsection{Background on data depth}\label{ssec:depth}

In the multivariate setting, \textit{i.e.}, for data which elements are points in the $d$-variate Euclidean space $\mathbb{R}^d$, statistical data depth function is a mapping
\begin{equation*}
    D\,:\,\mathbb{R}^d \times \mathbb{R}^{n \times d} \rightarrow [0,\,1],\, (\bmz,\bmX) \mapsto D(\bmz|\bmX)\,,
\end{equation*}
which satisfies the properties of~\citep[see also][ for a slightly different (but equivalent) set of requirements]{ZuoS00a}: \emph{affine invariance}, \emph{monotonicity on rays}, \emph{vanishing at infinity}, \emph{upper-semicontinuity}.


While the definition is general, a number of particular depth notions have been developed throughout the recent decades, with these notions differing in statistical as well as computational  properties and suitable for various applications. In what follows we will focus on the projection depth and its asymmetric version, since these are very robust \citep[following][asymptotic breakdown point of projection depth attains highest possible value of $1/2$]{Zuo03} and everywhere positive, thus allowing to identify and distinguish anomalies, also beyond the convex hull of the data.

\emph{Projection depth}~\citep{ZuoS00a} is defined in the following way:
\begin{equation*}
    D^{pd}(\bmz|\bmX) = \min_{\bmu\in\mathbb{S}^{d-1}} 1 \Big/ \Bigl( \frac{|\bmz^\top\bmu - \text{med}(\bmX\bmu)|}{\text{MAD}(\bmX\bmu)} + 1 \Bigr) \,,
\end{equation*}
with $\bmX\bmu$ being a shortcut for $\{\bmx_1^\top\bmu,...,\bmx_n^\top\bmu\}$ where $\text{med}$ and $\text{MAD}$ denote (univariate) \emph{median} and \emph{median absolute deviation from the median}, respectively, and $\mathbb{S}^{d-1}$ stands for the unit hyper-sphere in $\mathbb{R}^d$.

With projection depth retaining certain degree of symmetry (of its depth regions), \emph{asymmetric projection depth}~\citep{DYCKERHOFF04} has been designed to reflect the non-symmetric behaviour of the data:
\begin{equation*}
    D^{apd}(\bmz|\bmX) = \min_{\bmu\in\mathbb{S}^{d-1}} 1 \Big/ \Bigl( \frac{(\bmz^\top\bmu - \text{med}(\bmX\bmu))_+}{\text{MAD}^+(\bmX\bmu)} + 1 \Bigr) \,,
\end{equation*}
with $(a)_+ = \max\{a,0\}$ being the positive part of $a$ and $\text{MAD}^+$ denoting the \emph{median of the positive deviations from the median}.\\

As mentioned above, both projection and asymmetric projection depths belong to the class of depths satisfying the (weak) projection property, which includes depths for which it holds:
\begin{equation}\label{equ:prjprop}
    D(\bmx|\bmX) = \inf_{\bmu\in\mathbb{S}^{d-1}} D^{1} (\bmz^\top\bmu|\bmX\bmu)
\end{equation}
with $D^{1}$ standing for univariate depth. More precisely, in what follows, we shall make much use of the optimal direction $\bmu^*\in\mathbb{S}^{d-1}$. Furthermore, it is noteworthy that, following~\eqref{equ:prjprop}, such depths can be (well) approximated (from above) by means of multiple computations of solely univariate depths; \cite{DYCKERHOFF21} develop time-efficient algorithms for approximate computation of depths satisfying the projection property. In addition, any notion of data depth that satisfies the projection property, as well as possible another directional anomaly score defined in a similar manner can be used for ACA.

\subsection{Abnormal component analysis}\label{ssec:aca}

In this subsection we introduce the novel method---\emph{abnormal component analysis}, or shortly ACA. This method searches for orthogonal components in a subspace of $\mathbb{R}^d$ to provide a meaningful basis-representation that highlights and explains anomalies in the data. Different to existing visualization and explanation methods which optimize a predefined criterion (normally based on majority of the data) for obtaining a meaningful basis, here the goal is to focus on underlining anomalies, and thus these should be the object of optimization. We tackle this question by identifying anomalies based on minimal depth value and use the minimizing direction(s) to construct an (orthogonal) basis in $\mathbb{R}^d$.

\begin{figure}[!t]
\centering
\begin{adjustbox}{width=1\textwidth}
    \begin{tabular}{ccc}
        Original data in $\mathbb{R}^3$ & $1^{st}$ AC, space of search for $2^{nd}$ AC & ACA subspace with 2 ACs\\
        \includegraphics[width=0.33\textwidth, trim=3cm 3cm 3cm 3cm, clip]{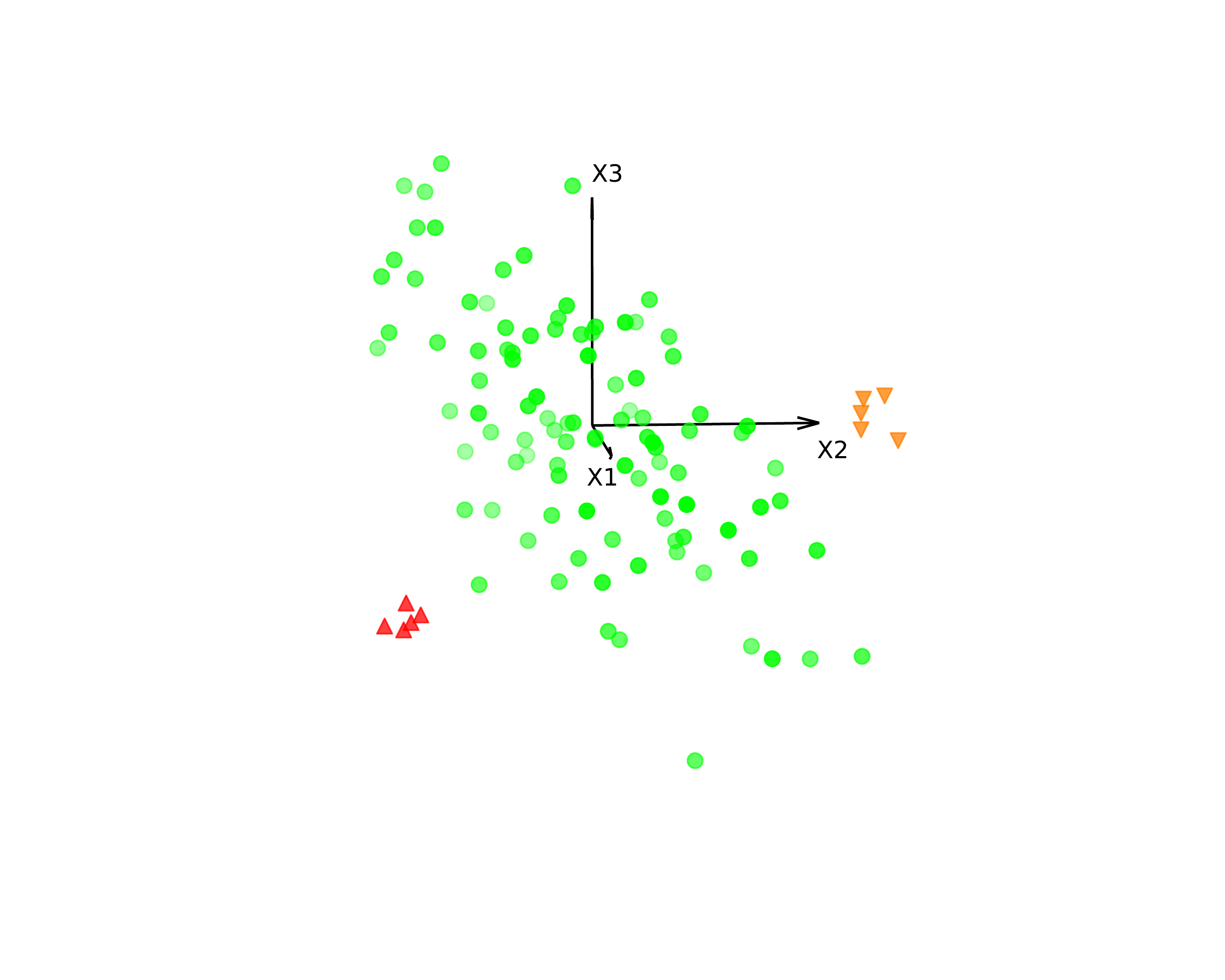} &
        \includegraphics[width=0.33\textwidth, trim=1cm 3cm 1cm 0.1cm, clip]{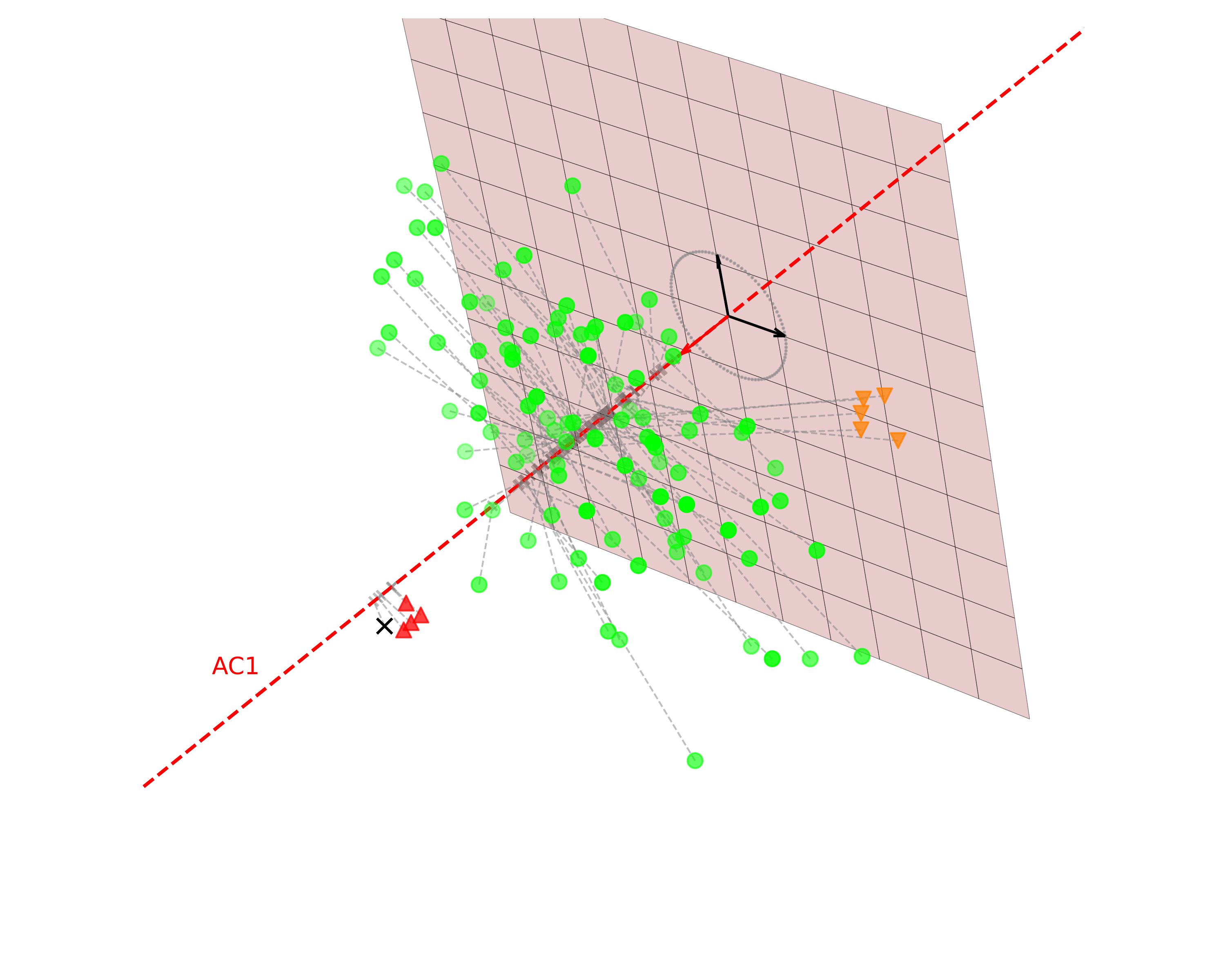} &
        \includegraphics[width=0.33\textwidth]{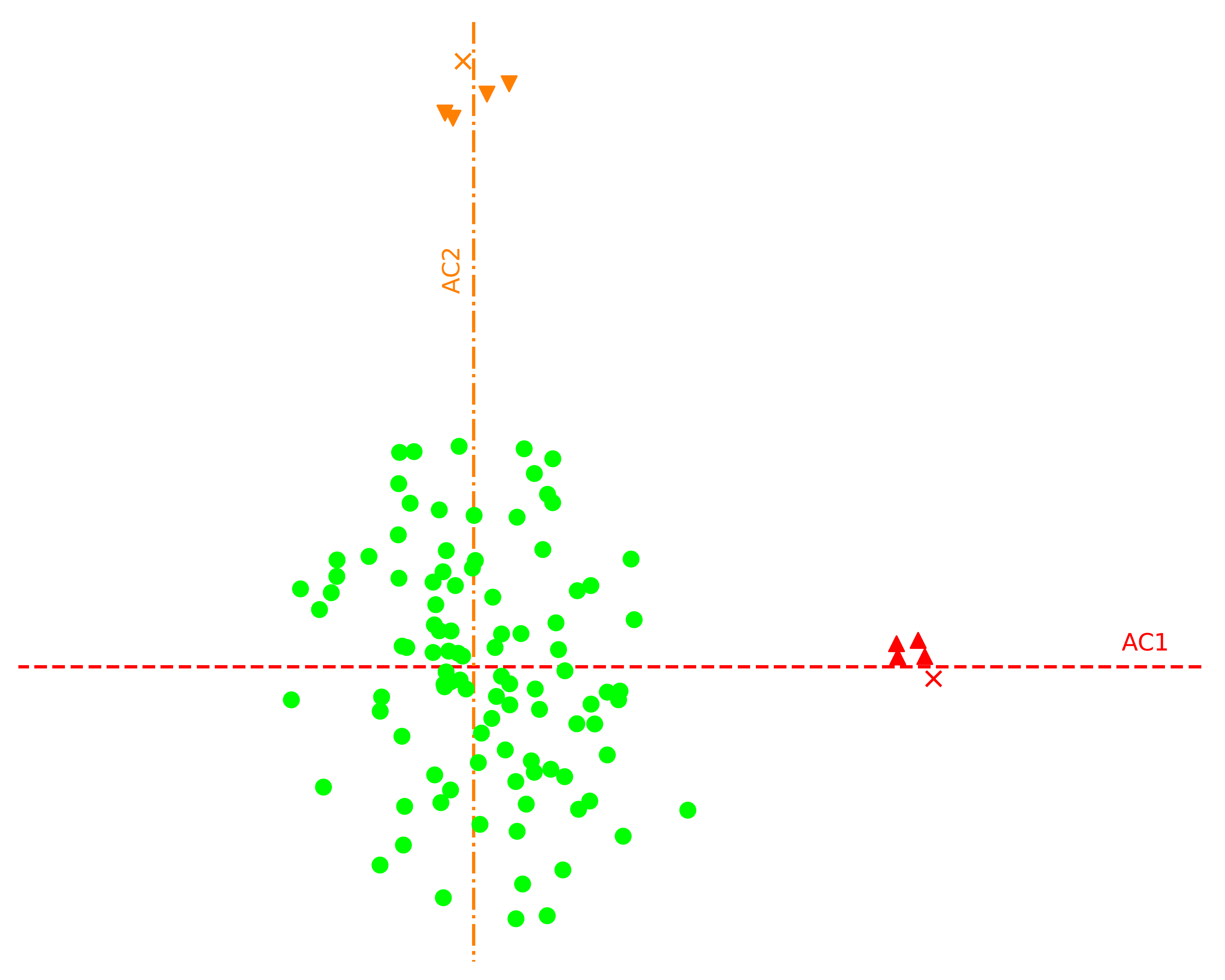}\\
    \end{tabular}
    \end{adjustbox}
\caption{An example of the ACA procedure on a data set containing $n=100$ points in $\mathbb{R}^3$ with $10$ anomalies in two groups of $5$.}\label{fig:acamain}
\end{figure}

We start with an intuitive explanation of the ACA method. To facilitate the exposition, let us consider as an example a data set $\bmX$ containing $n=100$ points in $\mathbb{R}^3$ with $10$ anomalies, in two groups of $5$ anomalies each (red triangles and orange reverse triangles); see Figure~\ref{fig:acamain}, top left. It is important to mention, that for data in higher dimensions no visualization is possible. On the first step, a point $\in\bmX$ is searched with minimal depth (among the $n$ points), and it's direction being the argument of~\eqref{equ:prjprop} $\bmu_1$ is taken as the first abnormal component (AC). This direction shall clearly identify the most significant group of anomalies (red triangles); see Figure~\ref{fig:acamain}, top right. On the second step, again a point $\in\bmX$ with smallest depth is searched, while the search space (for $\bmu$ in~\eqref{equ:prjprop}) is now limited to the orthogonal complement of $\bmu_1$ (red plane in Figure~\ref{fig:acamain}, top right; see also Figure~\ref{fig:acamain}, bottom left for this bivariate linear space). The minimizing direction $\bmu_2$, which in order identifies the second group of anomalies (orange reverse triangles) is taken as the second component. Figure~\ref{fig:acamain}, bottom right, depicts the constructed bivariate space on the basis of $\bmu_1$ and $\bmu_2$, which clearly distinguishes the two groups of anomalies. Although we stop here for our example in $\mathbb{R}^3$, the process continues, each time searching in the orthogonal complement of all abnormal components found on the earlier steps.

In the following subsection, we shall formally state the algorithm for ACA in pseudo-code, accompanied with a brief step-wise explanation.

\subsection{The algorithm}\label{ssec:algorithm}

We start by introducing the following depth computation problem (valid for an arbitrary univariate depth notion $D^{1}$), which is very similar to~\eqref{equ:prjprop}:
\begin{equation}\label{equ:acaroutine}
    D_{\bmB}(\bmz|\bmX) = \min_{\bmu\in\mathbb{S}^{\bmB}} D^{1} (\bmz^\top\bmu|\bmX\bmu)
\end{equation}
where $\mathbb{S}^{\bmB}$ stands for the unit hyper-sphere in the space spanned by columns of the basis matrix~$\bmB$. Note, that though real dimension of $\bmu$ here is limited by the number of columns in $\bmB$, it is a vector (of length $1$) in the original space $\mathbb{R}^d$. To be used in what follows, denote \texttt{projDepth} an algorithmic routine which computes~\eqref{equ:acaroutine} and returns both the depth values and its minimizing direction, taking as parameters:\\
(\RomanNumeralCaps{1}) point $\bmz\in\mathbb{R}^d$ for which the depth is computed, (\RomanNumeralCaps{2}) the data set $\bmX\subset\mathbb{R}^d$, (\RomanNumeralCaps{3}) the search-basis matrix $\bmB$, (\RomanNumeralCaps{4}) the chosen \texttt{\textit{depth}} notion, (\RomanNumeralCaps{5}) the number of $k$ directions used to approximate the depth value (see the next Section~\ref{ssec:parameters} to get insights about the choice of $k$), (\RomanNumeralCaps{6}) further parameters (\texttt{\textit{params}}) necessary for the optimization procedure~\citep[more details are contained in][]{DYCKERHOFF21}.\\
A number of useful algorithms for $\texttt{projDepth}$ can be found in~\cite{DYCKERHOFF21}; we thus simply refer the reader to this article for the computational questions.

\begin{algorithm}[t]
\caption{Abnormal component analysis}\label{alg:ACA}
\begin{algorithmic}[1]
\Require{$\bmX = \bmx_1, \cdots ,\bmx_n\in\mathbb{R}^d$, $p$, \texttt{depth}, $k$, \texttt{params}.}
\Ensure{$\boldsymbol{A}$.}
\Statex \textbf{Initialisation:} $\boldsymbol{A} = [\,]$, $\boldsymbol{B} = \boldsymbol{I}_{d}$.
\For{$i=1$ to $p$}\Comment{{\bf Find $p$ abnormal components}}
    \State{\bf Set $\bmu^* = \boldsymbol{0}_d$, $D^*=1$}
    \For{$j=1$ to $n$}\Comment{{\bf Minimize over data points}}
        \State $(D,\bmu)=$ \texttt{projDepth}$(\bmx_j,\,\bmX,\,\boldsymbol{B},\,\texttt{depth},\, k,\,\texttt{params})$
        \If{$D < D^*$}
            \State $m=\text{med}(\bmx_1^\top\bmu,...,\bmx_n^\top\bmu)$
            \IIf{$\bmx_i^\top\bmu < m$} $\bmu = -\bmu$
            \State $D^* = D$, $\bmu_i = \bmu$
        \EndIf
        \State $\boldsymbol{A} = [\boldsymbol{A}, \bmu_i]$
        \State $\boldsymbol{B} = \boldsymbol{A}^{\bot}$
    \EndFor
\EndFor
\end{algorithmic}
\end{algorithm}

Algorithm~\ref{alg:ACA} starts with the empty set of abnormal components and the full-space basis $[\boldsymbol{\mathrm{e}}_1,...,\boldsymbol{\mathrm{e}}_n]$ encoded by matrix $\bmB=\bmI_d$, with $\bmI_d$ being the $d \times d$ identity matrix. On each step of Algorithm~\ref{alg:ACA}, $i$th abnormal component is (found and) added to the set of components (matrix $\bmA$)
until the pre-specified number of abnormal components $p$ has been reached. Further, on each step, the size of the basis matrix $\bmB$ is reduced by one column so that the basis remains orthogonal to all the found (until $i$th step) components (saved in matrix $\bmA$).


Algorithmic routine \texttt{projDepth} simply goes through all points $\in\bmX$ and selects the one delivering minimal depth:
\begin{equation*}
    \text{min}D_{\bmB}(\bmX) = \min_{\bmx\in\bmX} \min_{\bmu\in\mathbb{S}^{\bmB}} D^{1} (\bmx^\top\bmu|\bmX\bmu)\,
\end{equation*}
The search is performed in the linear subspace of $\mathbb{R}^d$ defined by matrix $\bmB$, and the minimal depth is returned in addition to the depth-minimizing direction with anomalies on its positive side. In order to do this, the depth notion, the number of directions used for depth approximation, as well as further algorithmic parameters shall be chosen; we discuss these right below in the following subsection.

Algorithm~\ref{alg:ACA} possesses complexity $\mathcal{O}(pkdn^2)$ which can be decomposed as follows: with number of searched components $p$, number of depth-approximating directions $k$, dimension $d$ obviously entering linearly in the complexity, $n^2$ is explained by the fact that, for each component, all $n$ points should be revisited while each time all $n$ points should be projected on each direction (inside the optimisation routine \texttt{projDepth}). While approximation accuracy is clearly dependent on $k$, one can suppose it's polynomial dependence on $d$. In this article, we employed the spherical modification of the Nelder-Mead algorithm delivering best results as studied by~\cite{DYCKERHOFF21}. Regarding the accuracy, which (though not exact) can still be sufficient for components' search: (a) the work by~\cite{NAGY20} sheds the light on algorithmic convergence of the simplest approximation techniques and (b) chosen values of $k$ in practice delivered highly satisfactory results in all experiments conducted for this article. Right below, we discuss further the choice of parameters when performing ACA.

\subsection{Choice of parameters}\label{ssec:parameters}

When applying ACA, several parameters need to be set: (\RomanNumeralCaps{1}) number of abnormal components to search, (\RomanNumeralCaps{2}) the notion of data depth, (\RomanNumeralCaps{3}) number of directions used to approximate point's depth, (\RomanNumeralCaps{4}) optimization parameters. We discuss these choices in detail right below.

\paragraph{Number of abnormal components} As in any dimension reduction (or visualization) method, dimension of the obtained space $p\in\{1,...,d\}$ is guided either by prior knowledge about the data (generating process) or by computational resources (needed for further analysis). Though the choice of $p$ is entirely heuristic, it can be guided by \textit{a priori} information about (expected) anomalies in the data, \textit{e.g.}, possible dimension of anomalies' subspace or number of their groups.

\paragraph{Depth notion} The chosen notion of statistical depth function can have an important influence on the ACA's performance. Throughout this article, we stick to the notion of the projection depth, with sparse use of the asymmetric projection depth, for the reasons mentioned in Section~\ref{ssec:depth}. For a detailed discussion on the choice of depth notion in the multivariate setting we refer the reader to the recent survey by~\cite{MoslerM22}. Furthermore, theoretically, any (efficiently optimizable) univariate directional score can be used instead as well; a precaution shall be exercised regarding its eventual statistical properties though.


\paragraph{Number of directions} The number of directions $k\in\mathbb{N}_+$ has a profound influence on the precision of depth computation, and thus on the found direction(s) of abnormal component(s). \cite{NAGY20} prove that, even for the probability distribution, the number of $k$ shall grow exponentially with dimension for uniformly good depth approximation, if these directions are drawn at random. Further, \cite{DYCKERHOFF21} show that this number can be substantially reduced when using (adapted to the task) optimization algorithm, hopefully (and at least heuristically) departing from the exponential dependency on $d$, and indicate that in experiments where dimension is up to $d=20$ the (zero-order) optimization algorithm can converge fast requiring only few hundred directions. In their work, the authors propose a comprehensive comparison of number of algorithms in various settings with the lead taken by the Nelder-Mead algorithm, the sphere-adjusted coordinate descent, and the refined random search.


\paragraph{Optimization parameters} To obtain more insights about the choice of the optimization parameters, including the optimization algorithm itself, the reader is referred to the article by~\cite{DYCKERHOFF21}. Furthermore, the above described methodology on a subset of $\bmX$ can be employed here, as well as in a larger simulation setting resembling the real data at hand.



\section{Visual comparison}
\label{sec:comparevisu}

In this section, visualization capacity of ACA is explored in a comparative simulation study. After a brief discussion on present visualization tools and existing problems (Section~\ref{ssec:presentvistools}) we present the simulation settings (Section~\ref{ssec:simset}). Further, Section~\ref{ssec:vissimresults} presents the results compared to those obtained with most used (interpretable) dimension reduction methods, while the rest is preserved for the real-data study in Section~\ref{sec:realdata}.

\subsection{On existing dimension-reduction tools}\label{ssec:presentvistools}

With the task of Section~\ref{sec:comparevisu} being examination of ACA's performance in dimension-reduction compared to existing methods, we shall start with selecting from the state of the art. For interpretability reasons, methods with linear---in input variables---components will be preferred. 

With PCA being the natural candidate due to it's wide spread. For fair comparison in presence of anomalies robPCA is also employed in the same manner as traditional PCA where the mean and covariance matrix are estimated robustly, using the minimum covariance determinant \citep[MCD, see][]{RousseeuwL87,LopuhaaR91}(results are detailed in Supplementary Section~\ref{sup:ssec:simset}). To allow for `non-Gaussian' methodology, we further include ICA. Finally, we include auto-encoder, being a widely used neural-network-based tool able to learn highly non-linear components, it provides a visualization in form of (latent) variables of the `bottle-neck' layer. (The auto-encoder used has 10-5-2-5-10 layers and was trained using the stochastic gradient descent algorithm with $\mathbb{L}_2$ loss during 100 epoch with mini-batch size 10 and learning rate 0.005.)

Further methods like kPCA, $t$-SNE, MDS, LLE, or LE, though based on components non-linear in input variables, constitute powerful dimension-reduction machinery and can also provide insightful visualization. Due to this non-linearity property, but also for conciseness, we skip them in this simulation comparison, while include later in Section~\ref{sec:realdata} for analysis of real data.
Though mentioned above for completeness, we exclude NMF due to the positivity of components (which is not justified for general types of data, but only in specific applications, \textit{e.g.}, audio signals, images, text \textit{etc}).
For fair comparison in presence of anomalies robPCA is also presented in Supplementary Figure~\ref{fig:robpca2d} in the same manner as traditional PCA where the mean and covariance matrix are estimated robustly, using the minimum covariance determinant \citep[MCD, see][]{RousseeuwL87,LopuhaaR91}. However one should notice robPCA is usually applied to draw distance-distance plot rather than data visualisation in Euclidean space.

\subsection{Explored simulation settings}\label{ssec:simset}

Below, we describe five distributional settings (for normal data) used in the current visualization comparison, but also later throughout the article. Here, all of them follow the (famous and most adapted in the literature) Huber's contamination model~\cite{Huber64,Huber65}:
\begin{equation}\label{equ:HTCM}
    Y \stackrel{\mathcal{D}}= (1 - \epsilon)X + \epsilon\widetilde X\,,
\end{equation}
where the random vector $X$ (\textit{e.g.}, here, in $\mathbb{R}^d$) represents normal data while $\widetilde X$ stands for outliers (and $\stackrel{\mathcal{D}}=$ denotes equality in distribution). More precisely, in a sample of size $n$, the $\lfloor n \cdot (1 - \epsilon) \rfloor$ points of normal data are generated according to one of the following scenarios:

\begin{itemize}
    \item Setting~1 -- MVN(A09): Normal data are generated as i.i.d. copies of the multivariate-normally distributed random vector:
    $X \sim \mathcal{N}\bigl(\boldsymbol{0}_d, \boldsymbol{\Sigma}_{A09(d)}\bigr)\,,
    $
    where $\boldsymbol{\Sigma}_{A09(d)}$ consisting of $\{\sigma_{i,j}\}_{i,j=1}^d$ is the Toeplitz matrix with $\sigma_{i,j}=0.9^{|i-j|}$, \textit{i.e.}, to ensure various values of correlation between different variables~\citep{ROUSSEEUW18}.
    \item Setting~2 -- MVN(hCN): This setting copies MVN(A09) where the covariance matrix is a matrix with high condition number~\citep[$=100$, following][see Section~4 for the exact matrix-generating procedure]{AGOSTINELLI15}.
    \item Setting~3 -- ELL($t(5)$): Multivariate elliptical Student-$t(5)$ distribution:\\
    $
        X \stackrel{\mathcal{D}}= \boldsymbol{\mu} + \boldsymbol{\Lambda}UR\,,
    $
    where $U\sim\mathcal{U}(\mathbb{S}^{d-1})$ is uniformly distributed on the unit hypersphere, $R\sim St(5)$ is a Student-$t(5)$-distributed random variable,  $\boldsymbol{\Lambda}\boldsymbol{\Lambda}^\top=\boldsymbol{\Sigma}$, and $\mathbb{R}^d\ni\boldsymbol{\mu}(=\boldsymbol{0}_d)$ and $\boldsymbol{\Sigma}=\boldsymbol{\Sigma}_{A09(d)}$ are the distribution's center and scatter, respectively. ELL($t(5)$) has heavier tails than MVN(A09) (additional simulations available in Supplementary Section~\ref{sup:ssec:simset} Figure\ref{fig:sim2d:ell}).
    \item Setting~4 -- EXP: Random vector of normal data is here:
    $
        X = (X_1,...,X_d)^\top$ with mutually independent $X_i\sim\mathcal{E}(\lambda_i)\,, i=1,...,d\,,
    $
    where $X_i$s are $d$ exponentially distributed random variables with parameters $\lambda_i=\frac{1}{\beta_i}\text{ and }\beta_i\sim\mathcal{U}([0.1,1])$. EXP is asymmetric and possesses high degrees of skewness w.r.t. different variables.
    \item Setting~5 -- MV-Sk: Bivariate normal distribution skewed along the first variable \cite{AZZALINI99} (with $\alpha=10$ being the skewness parameter):
    \begin{equation*}
        X = (X_1, X_2)^\top \quad \text{with} \quad X_1 \sim \mathcal{N}_{Sk}\bigl(0, 1, \alpha\bigr)\quad \text{and} \quad
        X_2 \sim \mathcal{N}\bigl(0, 1/4\bigr)\,.
    \end{equation*}
\end{itemize}

\subsection{Two-dimensional plots}\label{ssec:vissimresults}

\begin{figure}[H]
    \centering
    \begin{tabular}{cccc}
         & PCA & ICA & Autoencoder\\
        \raisebox{1.75\normalbaselineskip}[0pt][0pt]{\rotatebox[origin=c]{90}{MVN(A09)}} & \includegraphics[width=0.3\textwidth, page=1, trim=1cm 0.95cm 0 0.75cm, clip]{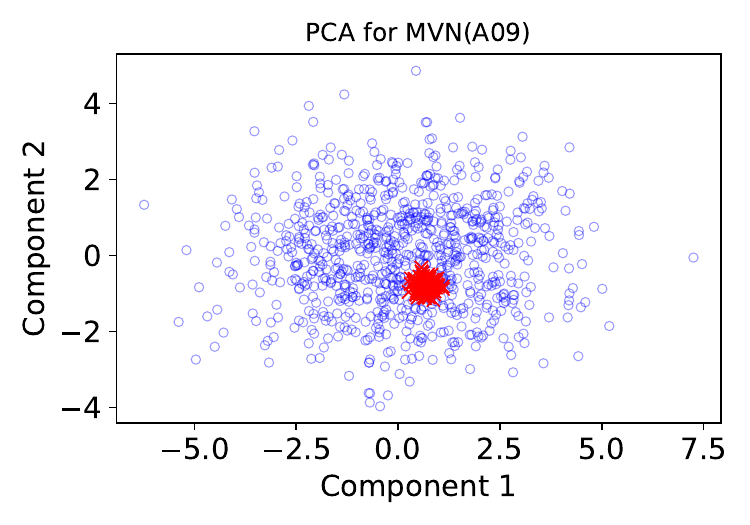} & \includegraphics[width=0.3\textwidth, page=1, trim=1cm 0.95cm 0 0.75cm, clip]{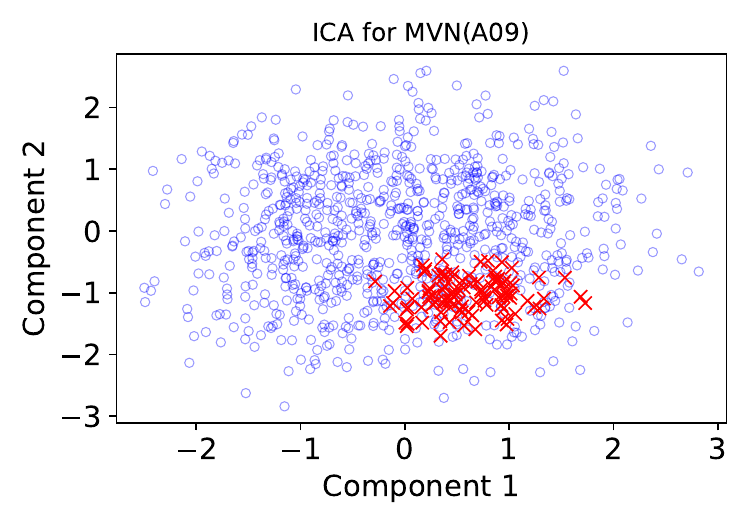} & \includegraphics[width=0.3\textwidth, page=1,trim=1cm 0.95cm 0 0.75cm, clip]{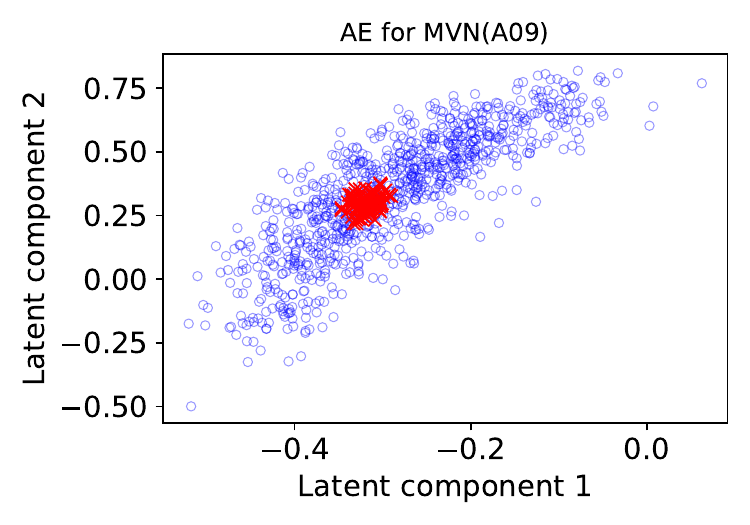}\\
        \raisebox{1.75\normalbaselineskip}[0pt][0pt]{\rotatebox[origin=c]{90}{MVN(hCN)}} & \includegraphics[width=0.3\textwidth, page=2, trim=1cm 0.95cm 0 0.75cm, clip]{figures/settings/by_model/pca_10.pdf} & \includegraphics[width=0.3\textwidth, page=2, trim=1cm 0.95cm 0 0.75cm, clip]{figures/settings/by_model/ica_10.pdf} & \includegraphics[width=0.3\textwidth, page=2, trim=1cm 0.95cm 0 0.75cm, clip]{figures/settings/by_model/ae_10.pdf}\\
        \raisebox{1.75\normalbaselineskip}[0pt][0pt]{\rotatebox[origin=c]{90}{ELL($t(5)$)}} & \includegraphics[width=0.3\textwidth, page=4, trim=1cm 0.95cm 0 0.75cm, clip]{figures/settings/by_model/pca_10.pdf} & \includegraphics[width=0.3\textwidth,  page=4, trim=1cm 0.95cm 0 0.75cm, clip]{figures/settings/by_model/ica_10.pdf} & \includegraphics[width=0.3\textwidth, page=4, trim=1cm 0.95cm 0 0.75cm, clip]{figures/settings/by_model/ae_10.pdf}\\
        \raisebox{1.75\normalbaselineskip}[0pt][0pt]{\rotatebox[origin=c]{90}{EXP}} & \includegraphics[width=0.3\textwidth, page=3, trim=1cm 0.95cm 0 0.75cm, clip]{figures/settings/by_model/pca_10.pdf} & \includegraphics[width=0.3\textwidth, page=3, trim=1cm 0.95cm 0 0.75cm, clip]{figures/settings/by_model/ica_10.pdf} & \includegraphics[width=0.3\textwidth, page=3, trim=1cm 0.95cm 0 0.75cm, clip]{figures/settings/by_model/ae_10.pdf}\\
    \end{tabular}
    \caption{Two-dimensional plots for PCA, ICA and Autoencoder applied to each setting when $\epsilon=0.1$. Axis are the 2 first components for PCA and ICA. Concerning the Autoencoder we visualise the $2d$ latent space.}
    \label{fig:sim2d:0.1}
\end{figure}

In this section we shall focus on first four settings from Section~\ref{ssec:simset} in order to compare the components-based visualization. Fixing the portion of anomalies to $\epsilon=0.05$ (cases with $\epsilon=0.01$ and $\epsilon=0.10$ are available in Supplementary Section~\ref{sup:ssec:simset} with Figure\ref{fig:sim2d:0.01} and Figure~\ref{fig:sim2d:0.1}), in each case we generate the $\lceil n \cdot \epsilon \rceil$ contaminating data from $\widetilde X \sim \mathcal{N}(\boldsymbol{0}_d, \boldsymbol{I}_d / 20)$ placed in direction of the last principal component of PCA of normal data centered at the distance of $1.25\,\times$ the largest Mahalanobis distance among normal data points (let us fix $\boldsymbol{\tilde{\mu}}$, the location of generated anomalies after the contamination process described just before). We thus obtain $\boldsymbol{Y}=\{\bmy_1,...,\bmy_n\}$ by  fixing $n=1000$ and $d=10$. It is important to note that concentrating the contaminating cluster of $\widetilde X$ on the last principal component does not influence the generality of conclusions but simplifies the presentation; we shall comment this more in detail right below when analysing the PCA output. In case more abnormal components are to be inspected at once, one can resort to a three-dimensional visualization or use pairwise two-dimensional plots.


\begin{figure}
    \centering
    \begin{tabular}{cccc}
         & PCA & ICA & Autoencoder\\
        \raisebox{1.75\normalbaselineskip}[0pt][0pt]{\rotatebox[origin=c]{90}{MVN(A09)}} & \includegraphics[width=0.3\textwidth, page=1, trim=1cm 0.95cm 0 0.75cm, clip]{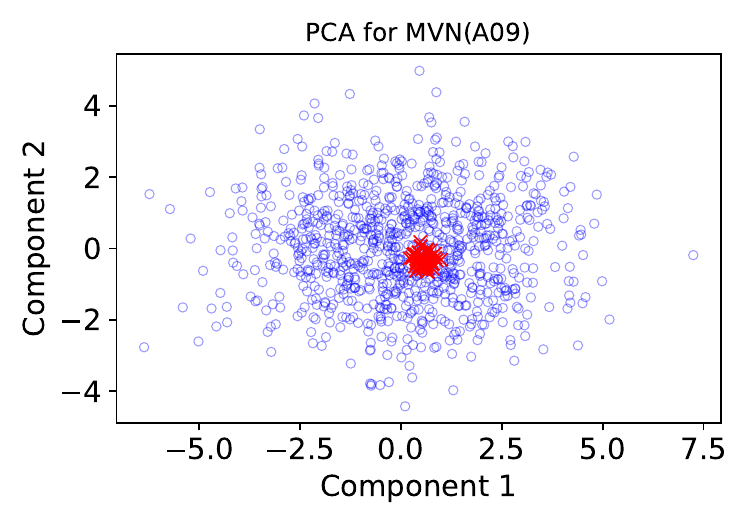} & \includegraphics[width=0.3\textwidth, page=1, trim=1cm 0.95cm 0 0.75cm, clip]{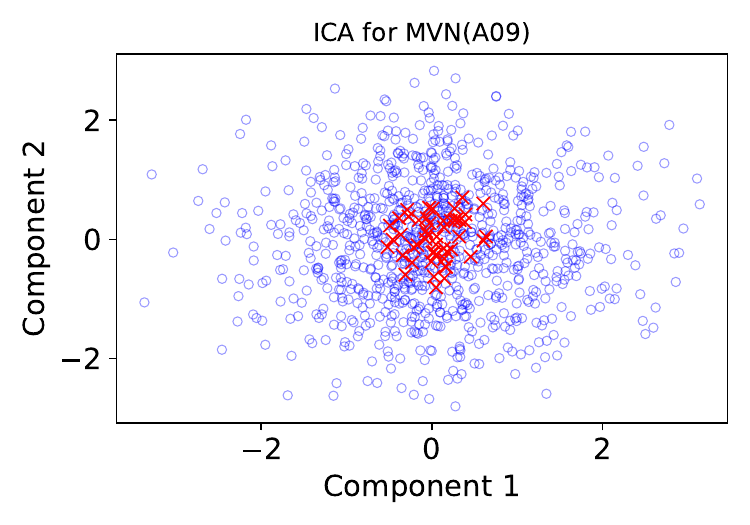} & \includegraphics[width=0.3\textwidth, page=1,trim=1cm 0.95cm 0 0.75cm, clip]{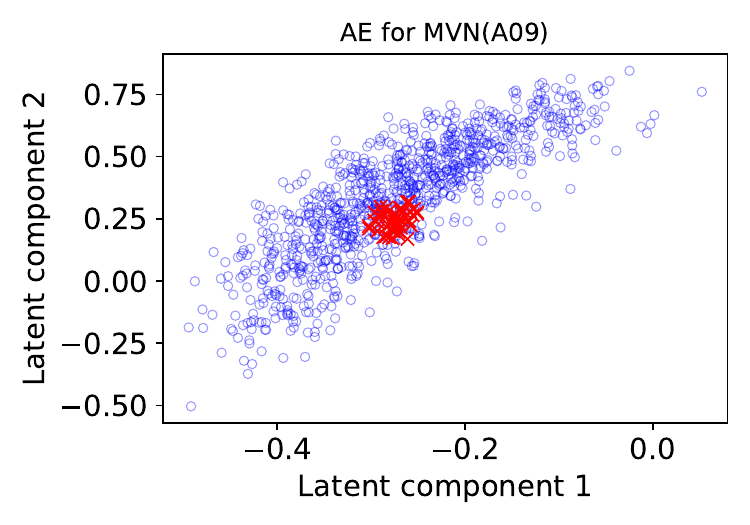}\\
        \raisebox{1.75\normalbaselineskip}[0pt][0pt]{\rotatebox[origin=c]{90}{MVN(hCN)}} & \includegraphics[width=0.3\textwidth, page=2, trim=1cm 0.95cm 0 0.75cm, clip]{figures/settings/by_model/pca.pdf} & \includegraphics[width=0.3\textwidth, page=2, trim=1cm 0.95cm 0 0.75cm, clip]{figures/settings/by_model/ica.pdf} & \includegraphics[width=0.3\textwidth, page=2, trim=1cm 0.95cm 0 0.75cm, clip]{figures/settings/by_model/ae.pdf}\\
        \raisebox{1.75\normalbaselineskip}[0pt][0pt]{\rotatebox[origin=c]{90}{ELL($t(5)$)}} & \includegraphics[width=0.3\textwidth, page=4, trim=1cm 0.95cm 0 0.75cm, clip]{figures/settings/by_model/pca.pdf} & \includegraphics[width=0.3\textwidth,  page=4, trim=1cm 0.95cm 0 0.75cm, clip]{figures/settings/by_model/ica.pdf} & \includegraphics[width=0.3\textwidth, page=4, trim=1cm 0.95cm 0 0.75cm, clip]{figures/settings/by_model/ae.pdf}\\
        \raisebox{1.75\normalbaselineskip}[0pt][0pt]{\rotatebox[origin=c]{90}{EXP}} & \includegraphics[width=0.3\textwidth, page=3, trim=1cm 0.95cm 0 0.75cm, clip]{figures/settings/by_model/pca.pdf} & \includegraphics[width=0.3\textwidth, page=3, trim=1cm 0.95cm 0 0.75cm, clip]{figures/settings/by_model/ica.pdf} & \includegraphics[width=0.3\textwidth, page=3, trim=1cm 0.95cm 0 0.75cm, clip]{figures/settings/by_model/ae.pdf}\\
    \end{tabular}
    \caption{Two-dimensional plots for PCA, ICA and Autoencoder applied to each setting with $\epsilon=0.05$. Axis are the 2 first components for PCA and ICA. Concerning the Autoencoder we visualise the $2d$ latent space.}
    \label{fig:sim2d}
\end{figure}

\paragraph{PCA} Figure~\ref{fig:sim2d} (left column) plots $\bmY$'s projection on first two components obtained by application of PCA. While it is not surprising that---with first two components---PCA is not able to `notice' anomalies located in direction of the $10$th component in $\mathbb{R}^{10}$, this example is illustrative and by no means restrictive, for the three following reasons:

\begin{itemize}
    \item It is a reasonable frequent practice, to pre-process data using dimension-reduction techniques (like PCA) and then apply statistical (\textit{e.g.}, anomaly detection) method in the space of several first and several last components. From this point of view, if the component with anomalies (independent of it's number) is not taken over into the reduced space, anomalies remain unnoticed for most visualization and analysis tools.
    \item If correlation (or a higher-order stochastic dependency) is present in the data, (small number of) anomalies cannot be noticed in any (\textit{e.g.,} $2=$) $k$-dimensional projection if they are placed on, \textit{e.g.}, the average of $k+1$ ($=3$) correlated components, being hidden by $k$-dimensional marginals.
    \item For a practitioner interested in identifying (and explaining) anomalies, it is in any case advantageous if anomalies are explicitly highlighted by first component(s).
\end{itemize}

\paragraph{ICA} Further, though anomalies-highlighting directions are (obviously) linear in variables, as expected, without an anomaly-specific criterion employed when searching components, the picture does not change for ICA; see Figure~\ref{fig:sim2d} (middle column).

\paragraph{Auto-encoder} To visualize the data representation using auto-encoder, we exploit its bottle-neck where information about noisy observations is expected to be filtered out. Thus, we use output of the two neurons of the third layer as latent variables, plotted for $\bmY$ in Figure~\ref{fig:sim2d} (right column). The intrinsic smoothness assumption on the approximable by neural network function makes them generally more vulnerable to increasing portions of anomalies (in the training sample) than traditional methods of robust statistics. Generally speaking, even when auto-encoder properly distinguishes anomalies by their reconstruction loss (which is not the case here), finding a low-dimensional anomaly-insightful representation can still be challenging, especially in realistic situations where the number of bottle-neck neurons is substantially larger than $2$.

\begin{figure}[!t]
\centering
    \begin{tabular}{cc}
        \hspace{1cm}ACA for MVN(A09) & \hspace{1cm}ACA for MVN(hCN) \\
        \includegraphics[width=0.4\textwidth, page=1, trim=0 0.25cm 0 0.75cm, clip]{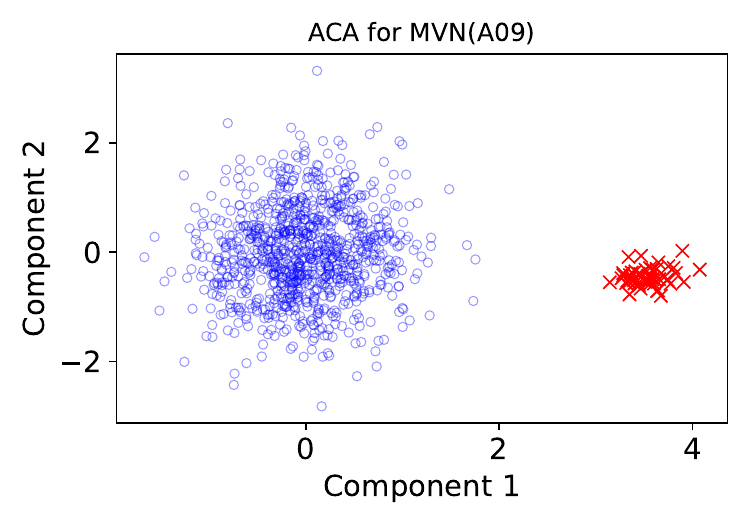} &
        \includegraphics[width=0.4\textwidth, page=2, trim=0 0.25cm 0 0.75cm, clip]{figures/settings/by_model/aca.pdf}\\
        \hspace{1cm}ACA for ELL($t(5)$) & \hspace{1cm}ACA for EXP \\
        \includegraphics[width=0.4\textwidth, page=4, trim=0 0.25cm 0 0.75cm, clip]{figures/settings/by_model/aca.pdf} &
        \includegraphics[width=0.4\textwidth, page=3, trim=0 0.25cm 0 0.75cm, clip]{figures/settings/by_model/aca.pdf}\\
    \end{tabular}
    \caption{Projection of data set $\bmY$ on first two components obtained by ACA in settings with $\epsilon=0.05$.}
\end{figure}

\paragraph{ACA} Non surprisingly, being designed for highlighting anomalies, and directly targeting them when searching for components, ACA copes with this artificial task, with contaminating $\bmY$ anomalies being located (and centered) on the first component. With $\bmY$ being rather a simple illustrative example, in what follows we switch to further aspects of ACA as well as to more challenging settings.

The above figures are two-dimensional, for illustration purposes. In case if more abnormal components are to be inspected at once, one can resort to a three-dimensional visualization or use pair-wise two-dimensional plots.

\section{Explainability}\label{sec:compareinter}


The intrinsic linearity of the ACA method positions it as a powerful tool for explainability---a highy demanded property in the domain of unsupervised anomaly detection. Here, different to supervised setting, no feedback can provide a criterion to decide about importance of a variable, with the goal being to explain how a variable contributes to the method's decision about abnormality of an observation. The framework of ACA suggests possibilities to identify most deviating variable(s) for each anomaly.
With each ACA's abnormal component (further AC) being a linear combination of the (input) variables, their contributions (perhaps properly re-scaled) highlight variables' abnormalities.

\subsection{Direction that highlights abnormality}

According to the principle of projected outlyingness \citep{STAHEL81,DONOHO82}, an AC (\textit{i.e.}, the corresponding direction) should be chosen in a way so that---in projection on it---abnormal observation(s) (cluster) is best separated from normal data. The goal of this subsection is to benchmark usefulness of the generated ACs in the elliptical setting, where a reasonable guess is easier to find theoretically. More precisely, the found direction (let us name it $\bmu^*_1$) should be such that it best separates the anomaly (or anomalies' cluster) from normal data. In the elliptical setting, a good candidate is the direction orthogonal to the---tangent to any ellipsoid---hyperplane that contains the abnormal point of interest. For comparison, let us fix this point to the earlier defined $\boldsymbol{\tilde\mu}\in\mathbb{R}^d$ with $d=10$ (see Section~\ref{ssec:vissimresults}). Then, a good candidate for the searched direction is $\bmu^*_1 = \frac{\boldsymbol{\Sigma}^{-1}\boldsymbol{\tilde\mu}}{\|\boldsymbol{\Sigma}^{-1}\boldsymbol{\tilde\mu}\|}$, with $\boldsymbol{\Sigma}$ standing for the covariance matrix of the elliptical distribution. From this point of view, effectiveness of a dimension-reduction method, which---when applied to data set $\bmY\subset\mathbb{R}^d$---returns up to $m\le d$ ordered (with decreasing importance) component vectors $\bmu_1,...,\bmu_m$, can be naturally measured by two indicators.

\begin{figure}[!t]
\centering
    \begin{tabular}{cc}
        \includegraphics[width=0.4\textwidth, trim=0 0.25cm 0.25cm 0.85cm, clip=true, page=2]{figures/settings/angles/best_angle_ind.pdf} & \includegraphics[width=0.4\textwidth, trim=0 0.25cm 0.25cm 0.85cm, clip=true, page=1]{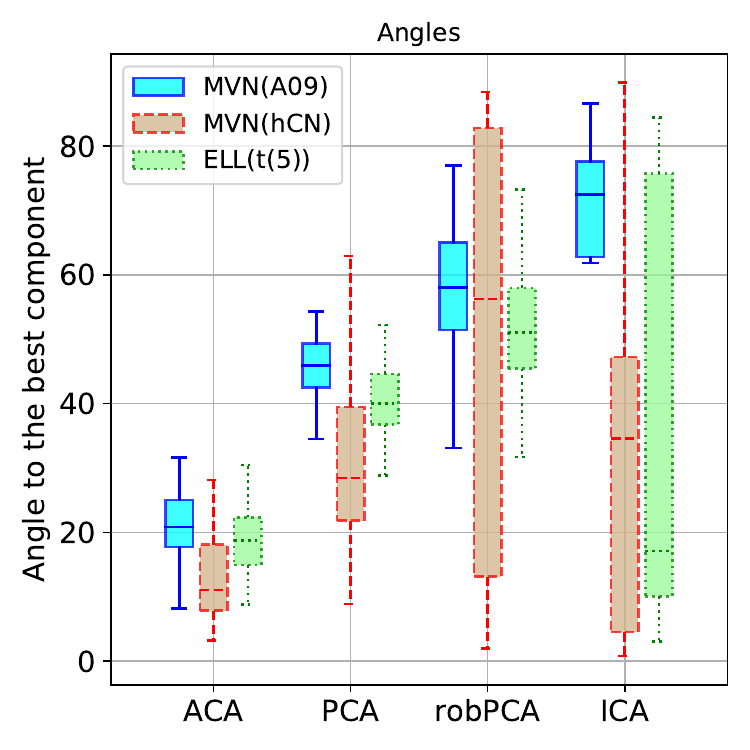} \\
    \end{tabular}
    \caption{For the component having smallest angle with $\bmu_1^*$ obtained with ACA, PCA, robPCA, and ICA, boxplots of its index ($\hat i$, left) and angle to $\bmu_1^*$ ($\hat \alpha$, right); for MVN(A09), MVH(hCN) and ELL($t(5)$), over $50$ random draws of $\bmY\subset\mathbb{R}^{10}$.}\label{fig:bestcomponent}
\end{figure}

The index of component that is most aligned with $\bmu^*_1$ (for most fair comparison),
	$
		\hat i = \argmin_{i\in\{1,...,m\}} \, \text{arc\,cos}(\bmu_i^\top \bmu^*_1)\,,
	$
 and the corresponding angle:
	$
		\hat \alpha = \text{arc\,cos}(\bmu_{\hat i}^\top \bmu^*_1)\,.
	$
    
For the three elliptical distributions from Section~\ref{ssec:simset} (MVN(A09), MVN(hCN), and ELL($t(5)$)) contaminated as described in Section~\ref{ssec:vissimresults}, we plot in Figure~\ref{fig:bestcomponent} $\hat i$ and $\hat \alpha$ for ACA, PCA, robPCA, and ICA. Note, that since the angle is measured as a positive value, the error in itself is inevitable, in particular for the empirical case. While it is expected that components with higher index numbers can be better aligned with the anomalies-explaining direction for other methods, their angles are still much higher that those of ACA (which always identifies anomalies with the first component). Furthermore, variables contributing more to this direction can be seen as responsible for the abnormality.
Supplementary Section~\ref{sup:sec:compDDC} and Section~\ref{sup:sec:compDIFFI} provides two additional comparisons related to explainability. The first comparison involves the DDC method, which utilizes robust statistical estimators to pinpoint the variables that contribute to outlying observations. The second comparison employs DIFFI, which leverages Isolation Forest technique to evaluate the contributions of each variables in anomalies.

\section{Application to real data sets}\label{sec:realdata}

\begin{table}[h]
\centering
\begin{tabular}{l|rrccc}
\textbf{Name} & $n$ & $d$ & \textbf{\# Anomalies} & \textbf{\% Anomalies} & \textbf{Category} \\ \hline
Musk       & 3062 & 166 & 97  & 3.17  & Chemistry    \\
Satellite image (2) & 5803 & 36  & 71  & 1.22  & Astronautics \\
Thyroid    & 3772 & 6   & 93  & 2.47  & Healthcare   \\
\end{tabular}
\caption{Information on real-world data sets.}\label{tab:readdatades}
\end{table}

Next, we explore the performance of both visualization and explainability provided by ACA in a comparative real-data study. For this, we address $33$ real-world data sets downloaded from~\cite{Rayana16,han2022adbench}, and present $3$ of them right below (see Table~\ref{tab:readdatades} for brief information), with further $30$ (for space reasons) being shifted to the Supplementary Materials, see Section(~\ref{sup:sec:realdata}). For each of them, we contrast ACA with $9$ dimension-reduction methods PCA, robPCA, kPCA, ICA, AE, $t$-SNE, MDS, LLE, and LE (see Section~\ref{ssec:allmethods} for a brief overview with references), by visualizing projection on first two components.

\begin{figure}[!h]
\centering
\includegraphics[width=\textwidth]{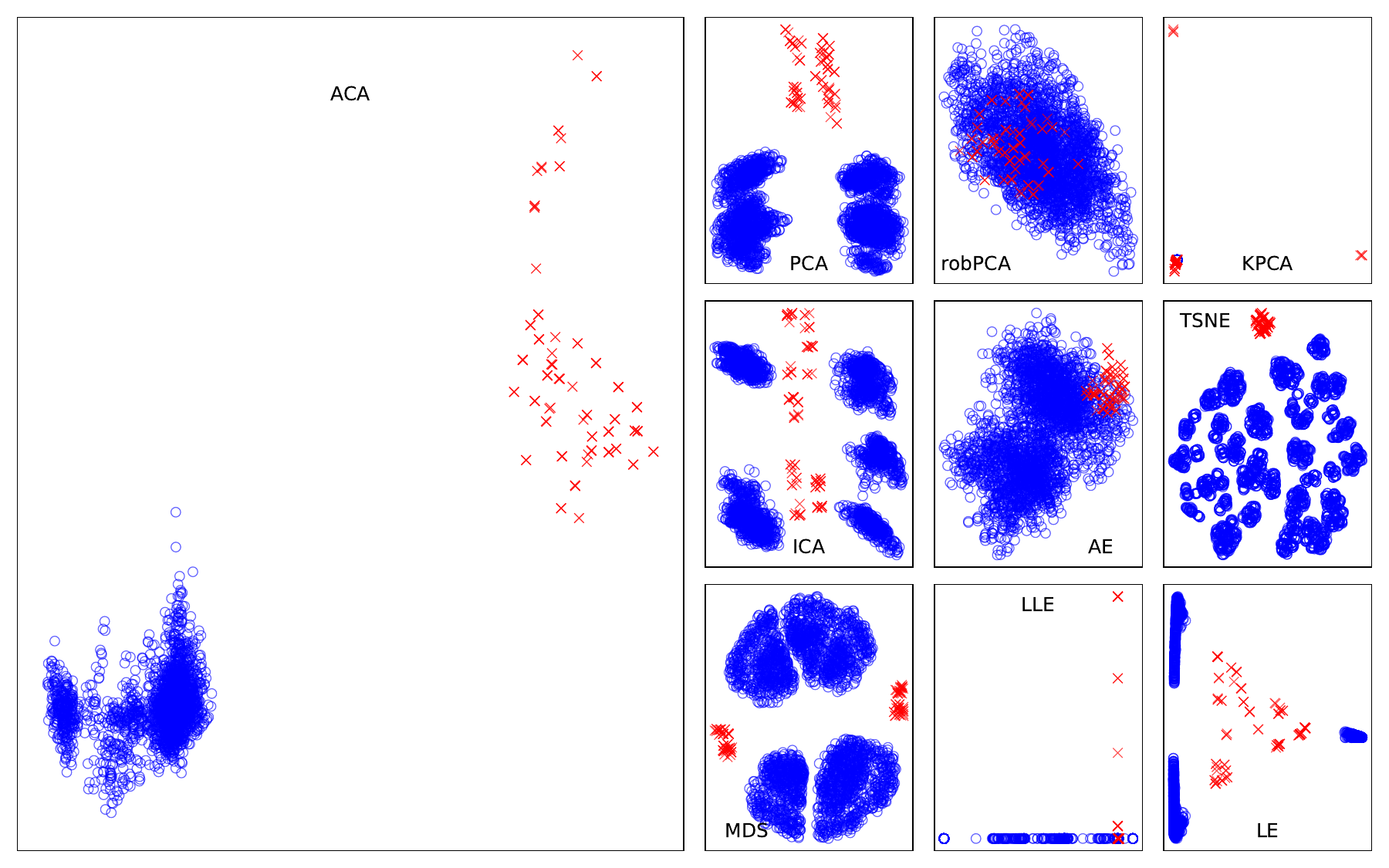}
\caption{Visualization of projection on two first components obtained by different methods for Musk data set.}\label{fig:realMusk}
\end{figure}


\paragraph{Musk data set} Musk data set contains molecules described by $166$ features extracted from low-energy conformation, with a part being marked as musk ($=$anomalies) by experts; variables are hence further indexed by integers here. Visualization of projection on first two components (Figure~\ref{fig:realMusk}) reveals that ACA precisely distinguishes anomalies by their projection on AC1 (with projection on AC2 being shifted as well). While some other methods (PCA, ICA, $t$-SNE, MDS, LE) also isolate them, detecting those further (say, in an automatic way) is less obvious. It is further interesting to consider the components' constitution (for methods with linear variables' combinations); see Table(~\ref{tab:realMuskVars}), with percentages being low due to high number of variables. Thus, only ACA identifies importance of variables ($=$conformations' features) `44' and `105' for abnormality, while solely variable `155' (in AC2) also appears in PCA.

\begin{figure}[!h]
\centering
\includegraphics[width=\textwidth]{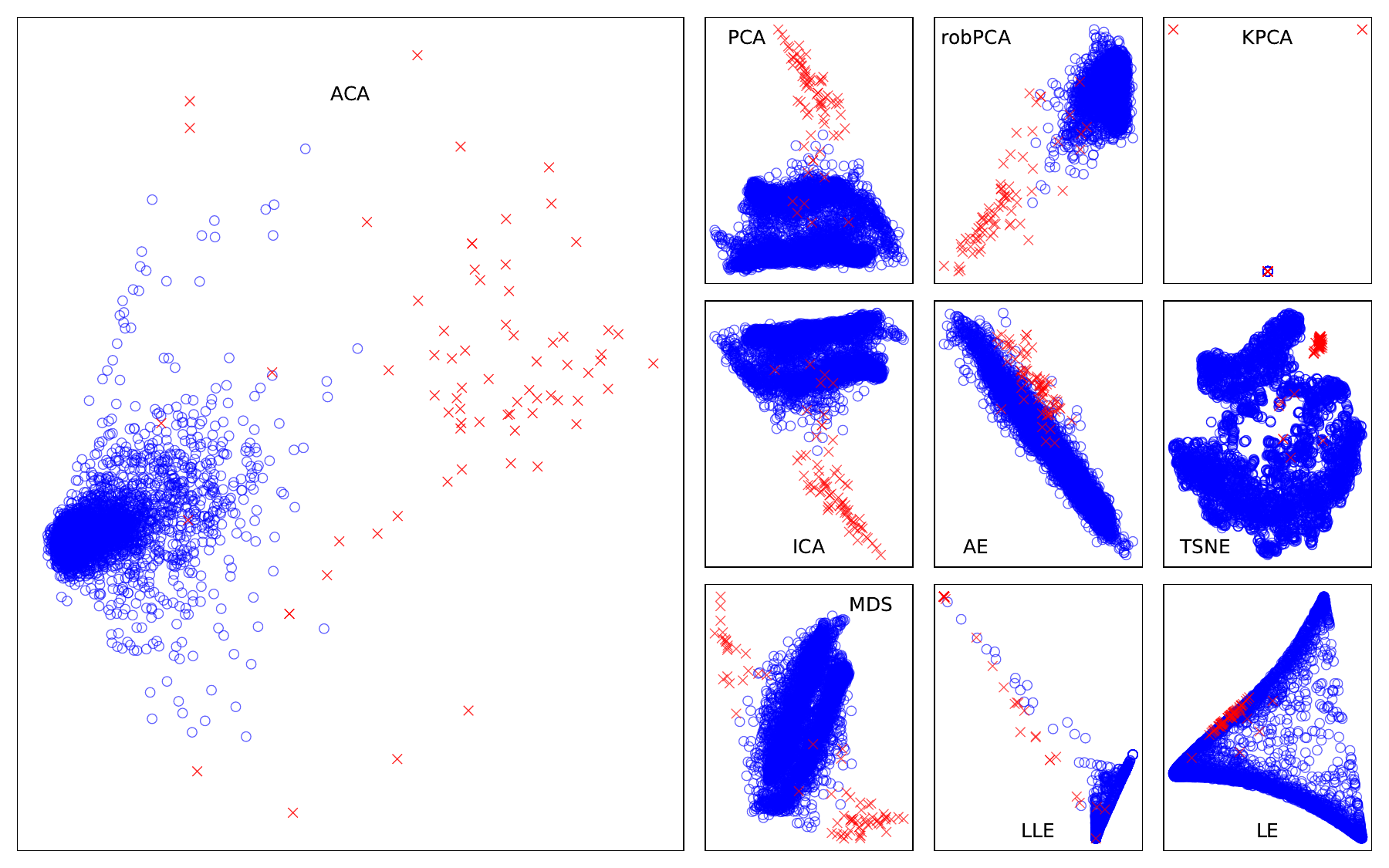}
\caption{Visualization of projection on two first components obtained by different methods for Satellite image (2) data set.}\label{fig:realSatimage}
\end{figure}


\paragraph{Satellite image (2) data set} This data set contains multi-spectral values of pixels in $3 \times 3$ neighborhoods in a satelite image, with one of the classes (Class 2) being downsampled to $71$ anomalies. (Variables are also numbered by integers here.) As one can observe from the bi-component visualization, ACA distances anomalies already on AC1, while only PCA, robPCA and ICA preserve them on a second (linear) component; see Figure~\ref{fig:realSatimage}. Out of variables `18', `2', and `26' spotted by AC1, only the first one finds itself 3rd on PC1; see Table(~\ref{tab:realSatimageVars}).

\begin{figure}[!h]
\centering
\includegraphics[width=\textwidth]{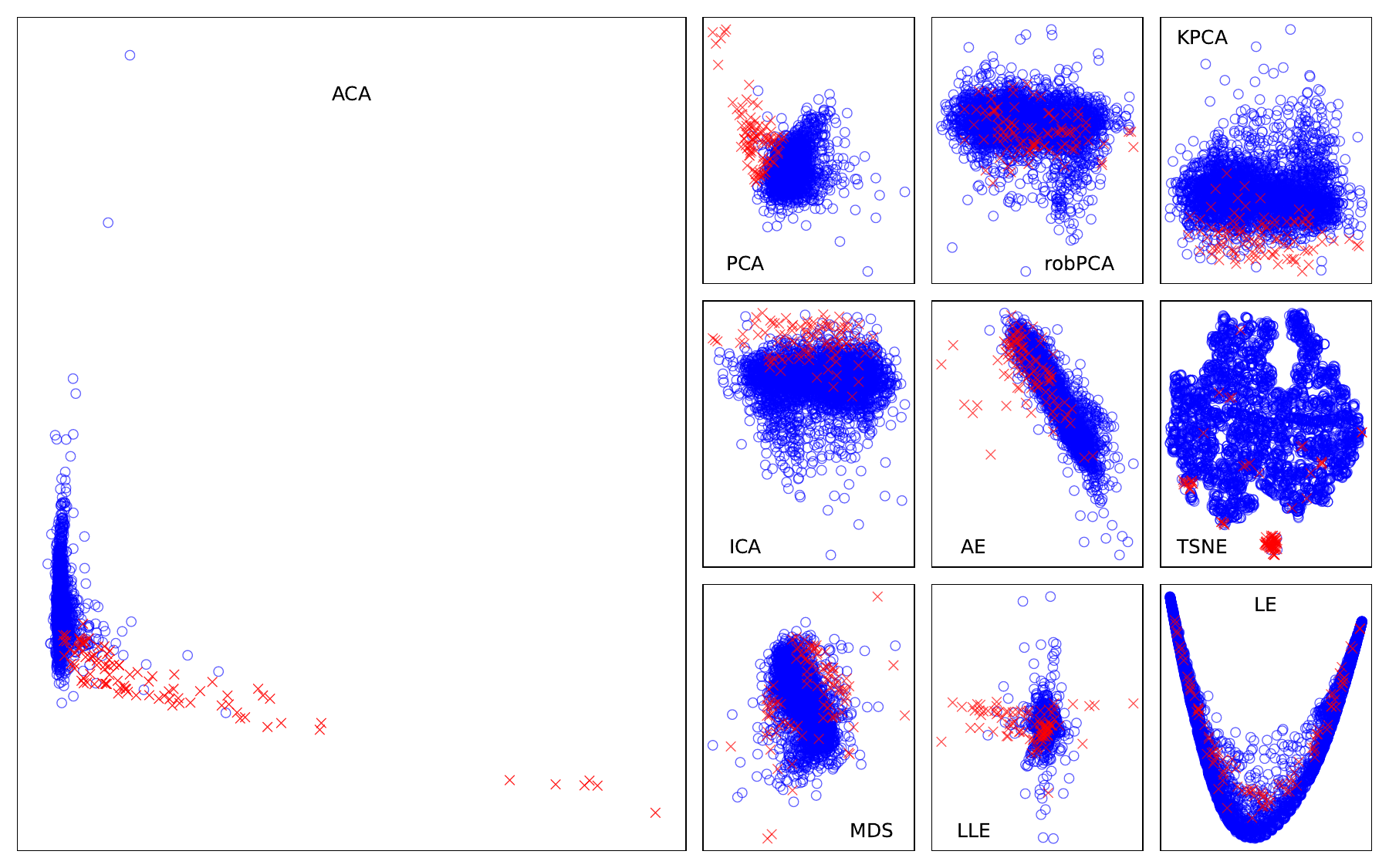}
\caption{Visualization of projection on two first components obtained by different methods for Thyroid data set.}\label{fig:realThyroid}
\end{figure}


\paragraph{Thyroid data set}For the Thyroid data set, we used $6$ continuous variables where the hyperfunction class is taken as abnormal. Here, visually (Figure~\ref{fig:realThyroid}), ACA outperforms the rest of the methods in addition to being the only one that highlights variable `2' (Triiodothyronine (T3) test) as the main one to explain anomalies, further giving much importance to variable `6' (TBG blood test); see also Table~\ref{tab:realThyroidVars}.

\section{Conclusion}
\label{sec:conc}


While today a universe of dimension-reduction methods is at practitioner's disposal, covering different conceptual approaches and application domains, these methods (be it a simple classical method like PCA (or its robust version) or more advanced techniques, \textit{e.g.}, $t$-SNE) are primarily aimed at finding a relevant representation space for the whole empirical data distribution and not at identifying anomalies. Even if in some case general dimension-reduction methods allow to detect or visualize some of them, this happens rather by chance as witnessed in the simulations of Section~\ref{sec:comparevisu} and real-data examples of Section~\ref{sec:realdata}.
The only way to get around anomalies is to search for them directly. ACA constitutes an attempt to fill this gap, aiming at representation of anomalies in a linear subspace of the original space $\mathbb{R}^d$.

ACA is easy to implement and mainly leverages standard existing (depth-computation and some further) tools with restriction of the direction $\bmu$'s search space to a lower-dimensional basis $\bmB$ being the only exception.
As we discuss in Section~\ref{ssec:algorithm}, ACA can be implemented with sufficient precision and polynomial complexity in both data set size $n$ and space dimension $d$. Numerous examples of this article prove reasonable approximation of such implementation in practice. Furthermore, involvement of $\bmB$ can be avoided by simply projecting the data on the orthogonal complement of the most recent found component; this would yield a somewhat different procedure though.

Positioning in the anomaly detection framework, ACA constructs an orthonormal basis of a pre-defined dimension, which---hopefully---provides insights on the location of anomalies from the training data set. When projecting new (out-of-sample) data in the same basis, different situations can arise, depending on the contamination model. In case of Huber model~\eqref{equ:HTCM}, the representation should be normally sufficient for gaining insights about anomalies, which should not be necessarily the case for others, in particular adversarial contamination~\citep[see, \textit{e.g.},][]{DiakonikolasKKMS16,BateniD20} that can appear in any part of $\mathbb{R}^d$ and be hidden by the normal data in all $2$- and $3$-dimensional projections. A possible strategy to act is then to compute the depth/outlyingness of suspected (or all) observations, and if these values indicate potential abnormality, (re-)run ACA on a data set including these observations.

ACA is not restricted to (asymmetric) projection depth only, and can be readily employed with other depth notions that satisfy~\eqref{equ:prjprop}, as well as those minimizable over projections, as it is the case for, \textit{e.g.}, weighted halfspace depth of~\cite{KotikH17}. Moreover, any further procedure that provides an anomaly score based on a univariate data's projection can be adapted to ACA framework as well. Furthermore, ACA can be extended to data sets in spaces where linear combination of variables is expected to provide a reasonable explanation of anomalies and the corresponding search procedure can be constructed.

It is important to mention that ACA's implementation depends on the equilibrium between computational cost and search precision. This can be particularly important for higher dimensions, where the direction-seeking routing might need more iterations to converge. In the case when data span lower-dimensional subspace of a high-dimensional space, either proper dimension-reduction should be performed, or Nelder-Mead algorithm should be privileged. If, on the other hand, data are spanned by a lower-dimensional non-linear manifold, the utility of ACA can be limited, as we show in the Supplementary Material.



With the most relevant information and illustrations incorporated in the body of the article, the Supplementary Materials to this article contain: Section~\ref{sup:sec:ndirs}: additional information concerning the choice of parameters from Section~\ref{ssec:parameters}, Section~\ref{sup:sec:simset}: additional simulations and settings information from Section~\ref{ssec:simset}, Section~\ref{sup:sec:comps}: full 10 coordinates constituting abnormal components for visualizations from Section~\ref{ssec:vissimresults}, Section~\ref{sup:sec:compDDC} and Section~\ref{sup:sec:compDIFFI}: comparisons with DDC and DIFFI respectively concerning explainability, Section~\ref{sup:sec:realdata}: visualization and components’ data on the 30 remaining real data sets, Section~\ref{sup:sec:nonlinear}: ACA applied to non-linear structures and Section~\ref{sup:sec:nm}: Nelder-Mead algorithm used for depth approximation.

\section*{Acknowledgments}

The authors greatly acknowledge the support of the CIFRE grant n° 2021/1739.






\bibliographystyle{chicago}
\bibliography{sources.bib}

\clearpage
\setcounter{section}{0}

\pagebreak
\begin{center}
{\LARGE Supplementary materials to the article\\

  ``Abnormal component analysis''}

\vspace{.5cm}
  
\author{    Romain Valla \\    {\small LTCI, Télécom Paris, Institut Polytechnique de Paris}\\
    Pavlo Mozharovskyi \\    {\small LTCI, Télécom Paris, Institut Polytechnique de Paris}\\
    Florence d'Alché-Buc \\    {\small LTCI, Télécom Paris, Institut Polytechnique de Paris}}
    
    \vspace{.5cm}
    
\date{June 05, 2025}
\maketitle
\end{center}

These supplementary materials contain additional information on the study of the performance of the abnormal component analysis (ACA) procedure. Section~\ref{sup:sec:ndirs} proposes information about choice of parameters (Section~\ref{ssec:parameters}), the settings used in the visualisation comparison (Section~\ref{sup:sec:simset}), and detailed results (mentioning every component) concerning all the methods involved in experiments (Section~\ref{sup:sec:comps}). Further, experiments concerning explainability are presented, where are ACA is compared to DDC (Section~\ref{sup:sec:compDDC}) and DIFFI (Section~\ref{sup:sec:compDIFFI}). Then, other real datasets are used to apply ACA for visualisation and explanation purpose (Section~\ref{sup:sec:realdata}). While it is expected that ACA’s results are not always better, they are usually comparable to the commonly employed techniques and illustrate the relevance of ACA. Finally Section~\ref{sup:sec:nonlinear} present limitation of ACA when applied to non-linear structures and Section~\ref{sup:sec:nm} present the details of Nelder-Mead algorithm used for depth approximation.

\newpage

\section{Number of directions}\label{sup:sec:ndirs}
\begin{figure}[H]
\centering
\includegraphics[width=0.65\textwidth,trim=1cm 1.5cm 2cm 2.5cm,clip=true]{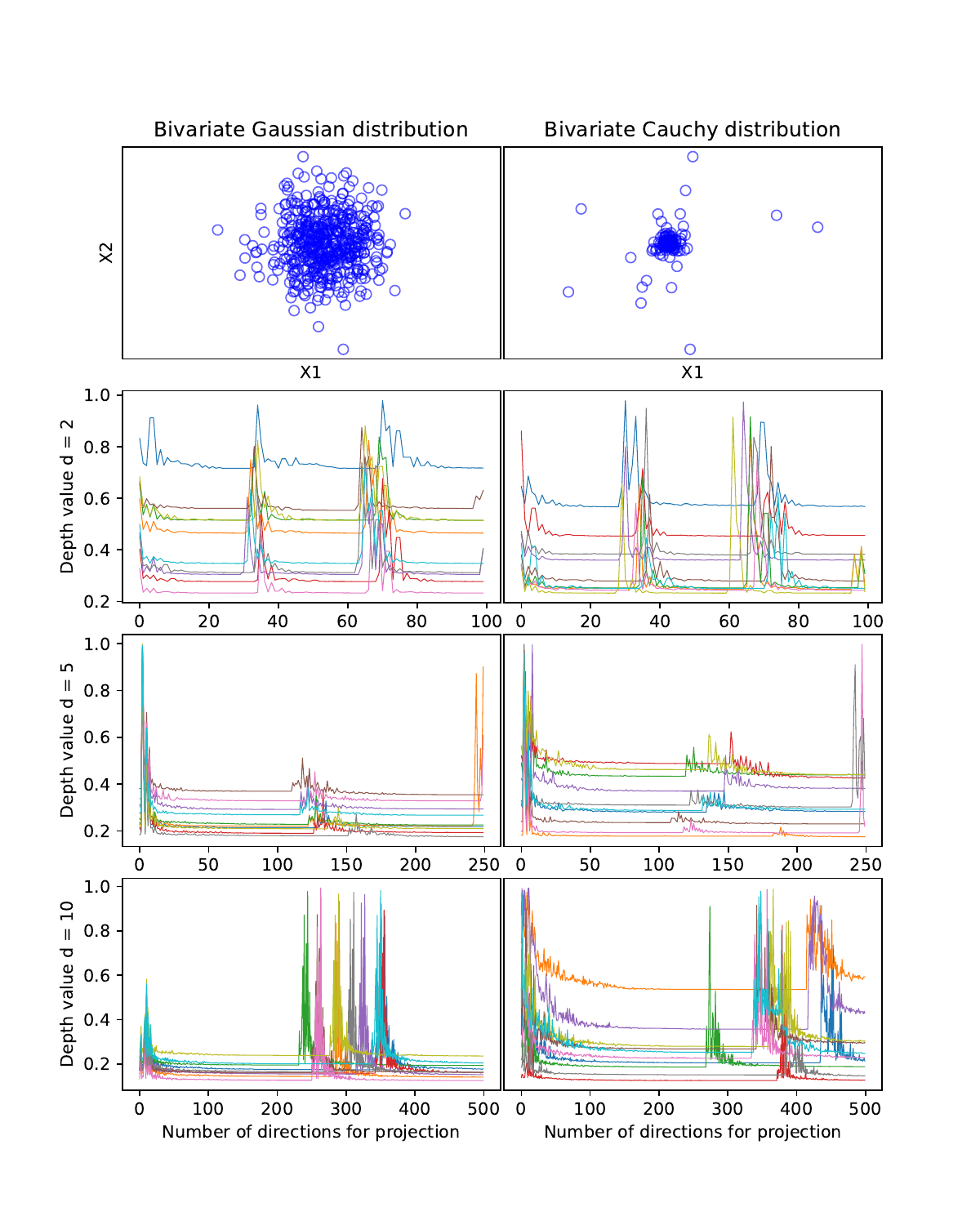}
\caption{Tuning process to determine the number of directions $k$. The depth value for $10$ points from a sample stemming from (non-correlated) Gaussian (left) and Cauchy (right) distributions, in dimensions $d=2,5,10$.}\label{fig:tunek}
\end{figure}

To decide on the number of directions in the application at hand, we suggest the following simple verification following the very principle of the class of depths satisfying the projection property~\eqref{equ:prjprop}: ``the smaller the approximated depth value the better it is''. That being said, even without knowing the true depth value, it is reasonable to choose the method and number of directions delivering the smallest depth value. A simple way to study the optimization behavior is the visual inspection of the development of the (minimal) depth value throughout the optimization iterations, for at least several points from the data set. For a sample from Gaussian and Cauchy distribution (with varying dimension), this is illustrated in Figure~\ref{fig:tunek}, where the repeating jumps of the depth value indicate re-starting the optimization routine in hope to avoid local minima.

\section{Simulation settings}\label{sup:sec:simset}

Having described the data generation parameters in Section~\ref{ssec:simset}, here is a view of the two covariance matrices used in setting MVN(A09) and MVN(hCN).

\begin{figure}[h]
\centering
$\begin{bmatrix}
1. & -0.9 & 0.81 & -0.73 & 0.66 & -0.59 & 0.53 & -0.48 & 0.43 & -0.39\\
-0.9 & 1. & -0.9 & 0.81 & -0.73 & 0.66 & -0.59 & 0.53 & -0.48 & 0.43\\
0.81 & -0.9 & 1. & -0.9 & 0.81 & -0.73 & 0.66 & -0.59 & 0.53 & -0.48\\
-0.73 & 0.81 & -0.9 & 1. & -0.9 & 0.81 & -0.73 & 0.66 & -0.59 & 0.53\\
0.66 & -0.73 & 0.81 & -0.9 & 1. & -0.9 & 0.81 & -0.73 & 0.66 & -0.59\\
-0.59 & 0.66 & -0.73 & 0.81 & -0.9 & 1. & -0.9 & 0.81 & -0.73 & 0.66\\
0.53 & -0.59 & 0.66 & -0.73 & 0.81 & -0.9 & 1. & -0.9 & 0.81 & -0.73\\
-0.48 & 0.53 & -0.59 & 0.66 & -0.73 & 0.81 & -0.9 & 1. & -0.9 & 0.81\\
0.43 & -0.48 & 0.53 & -0.59 & 0.66 & -0.73 & 0.81 & -0.9 & 1. & -0.9\\
-0.39 & 0.43 & -0.48 & 0.53 & -0.59 & 0.66 & -0.73 & 0.81 & -0.9 & 1.\\
\end{bmatrix}$
\caption{Covariance matrix $\boldsymbol{\Sigma}_{A09(d)}$ used in settings MVN(A09) and ELL($t(5)$) when $d=10$.}
\end{figure}

\begin{figure}[h]
\centering
$\begin{bmatrix}
1. & -0.03 & 0.14 & -0.12 & -0.33 & -0.01 & 0.5 & 0.01 & -0.23 & -0.07\\
-0.03 & 1. & 0.37 & 0.41 & 0.19 & 0.27 & 0.19 & -0.43 & -0.17 & 0.12\\
0.14 & 0.37 & 1. & 0.26 & 0.04 & 0.09 & 0.14 & -0.09 & -0.13 & 0.33\\
-0.12 & 0.41 & 0.26 & 1. & -0.25 & 0.41 & -0.01 & -0.28 & 0.42 & -0.02\\
-0.33 & 0.19 & 0.04 & -0.25 & 1. & -0.13 & -0.08 & -0.24 & 0.07 & 0.43\\
-0.01 & 0.27 & 0.09 & 0.41 & -0.13 & 1. & -0.25 & -0.3 & 0.53 & -0.31\\
0.5 & 0.19 & 0.14 & -0.01 & -0.08 & -0.25 & 1. & -0.18 & -0.09 & -0.09\\
0.01 & -0.43 & -0.09 & -0.28 & -0.24 & -0.3 & -0.18 & 1. & 0.07 & -0.35\\
-0.23 & -0.17 & -0.13 & 0.42 & 0.07 & 0.53 & -0.09 & 0.07 & 1. & -0.18\\
-0.07 & 0.12 & 0.33 & -0.02 & 0.43 & -0.31 & -0.09 & -0.35 & -0.18 & 1.\\
\end{bmatrix}$
\caption{Random covariance matrix with a high condition number~\citep[$=100$, following][see Section~4 for the exact matrix-generating procedure]{AGOSTINELLI15} used in setting MVN(hCN) when $d=10$.}
\end{figure}

\begin{figure}[H]
    \centering
    \includegraphics[width=0.85\textwidth]{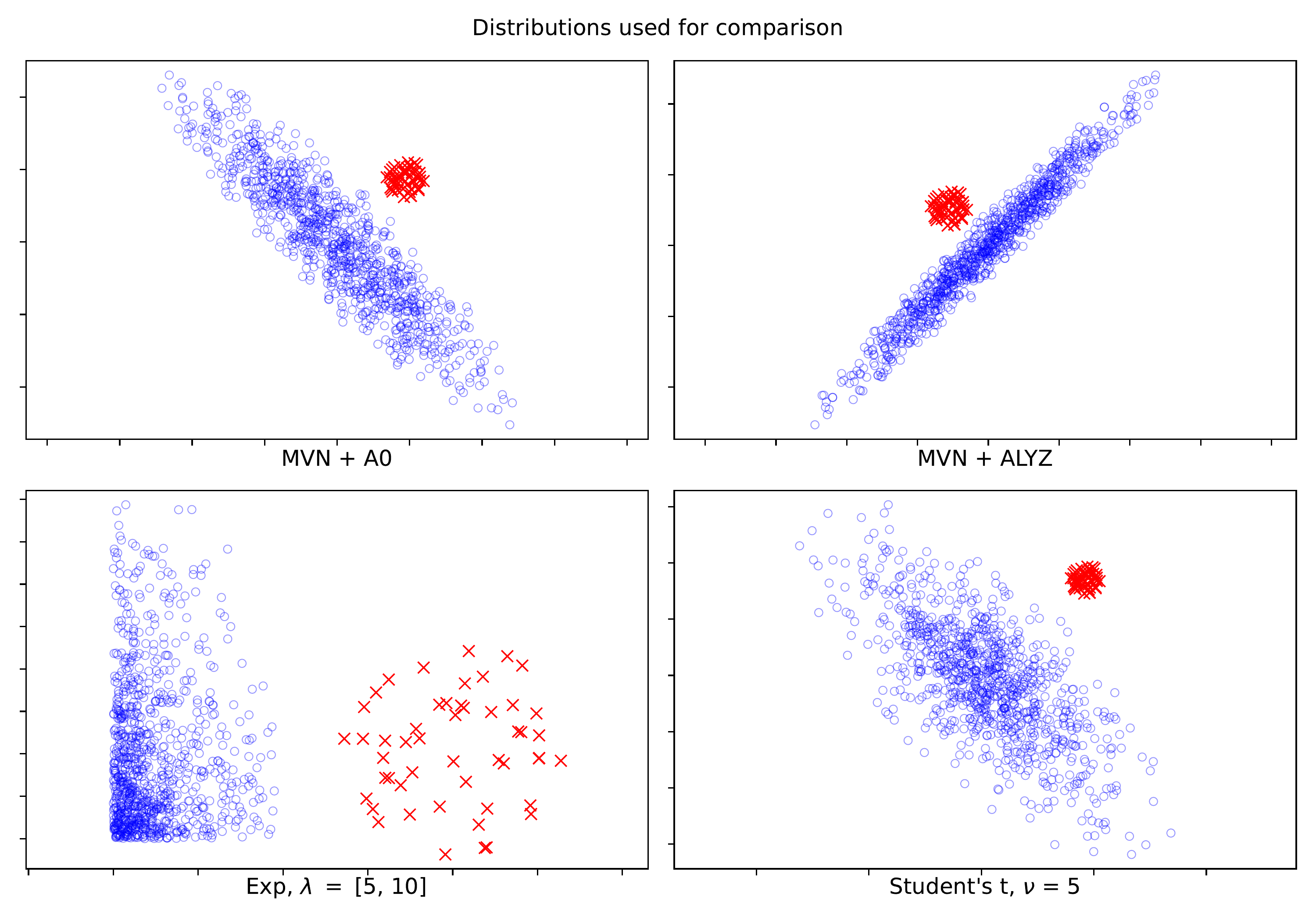}
    \caption{Example (in $\mathbb{R}^2$) of a data sample for four contaminated settings: MVN(A09) (top, left), MVN(hCN) (top, right), EXP (bottom, left), ELL($t(5)$) (bottom, right).}
    \label{fig:vis2d}
\end{figure}

For illustrative purposes, we plot a data sample from each of the four contaminated settings in $\mathbb{R}^2$ in Figure~\ref{fig:vis2d}.

\subsection{Additional simulation setting}\label{sup:ssec:simset}
\begin{figure}[!t]
\centering
    \begin{tabular}{cc}
        \hspace{1cm}robPCA for MVN(A09) & \hspace{1cm}robPCA for MVN(hCN) \\
        \includegraphics[width=0.4\textwidth, page=1, trim=0 0.25cm 0 0.75cm, clip]{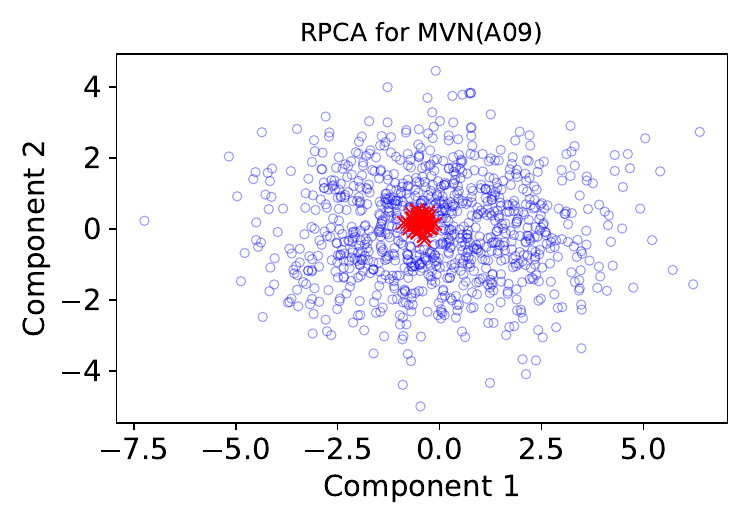} &
        \includegraphics[width=0.4\textwidth, page=2, trim=0 0.25cm 0 0.75cm, clip]{figures/settings/by_model/rpca.pdf}\\
        \hspace{1cm}robPCA for ELL($t(5)$) & \hspace{1cm}robPCA for EXP \\
        \includegraphics[width=0.4\textwidth, page=4, trim=0 0.25cm 0 0.75cm, clip]{figures/settings/by_model/rpca.pdf} &
        \includegraphics[width=0.4\textwidth, page=3, trim=0 0.25cm 0 0.75cm, clip]{figures/settings/by_model/rpca.pdf}\\
    \end{tabular}
\caption{Projection of data set $\bmY$ on first two components obtained by robPCA.}\label{fig:robpca2d}
\end{figure}

\medskip

\paragraph{robPCA} Projection of $\bmY$ on two first components obtained by robPCA, for the four mentioned above contaminated settings, is depicted in Figure~\ref{fig:robpca2d}. Using robust MCD estimates for the mean and covariance matrix (with the standard value for parameter $\alpha=(n+d+1)/2n$ (portion or anomalies in all our experiments never exceeds this parameter); see also \cite{ROUSSEEUW99} for the fast randomized algorithm), the group of anomalies is being ignored, and principal components well approximate the variance-maximizing directions (\textit{e.g.}, for MVN and ELL they are close to the axes of the ellipsoids defined by the population (=true) covariance matrix). With anomalies not necessarily lying on the first (or in general no) such axes, they are not readily identifiable/explainable from the visualization.

\begin{figure}[H]
    \centering
    \begin{tabular}{cccc}
         & PCA & ICA & Autoencoder\\
        \raisebox{1.75\normalbaselineskip}[0pt][0pt]{\rotatebox[origin=c]{90}{MVN(A09)}} & \includegraphics[width=0.3\textwidth, page=1, trim=1cm 0.95cm 0 0.75cm, clip]{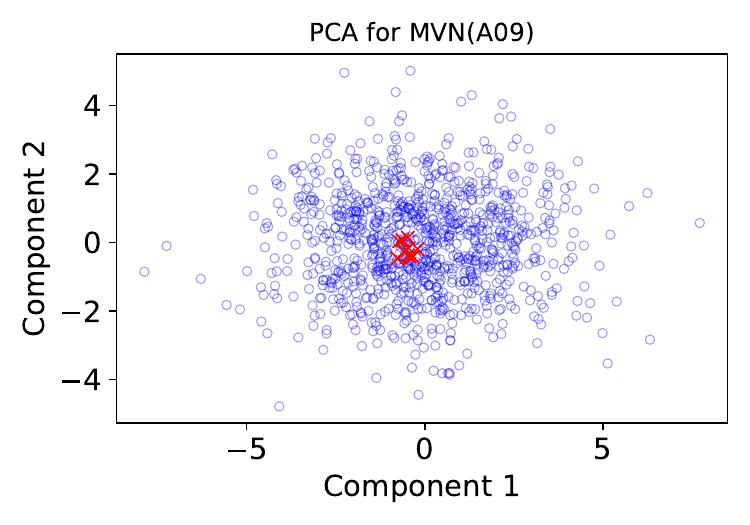} & \includegraphics[width=0.3\textwidth, page=1, trim=1cm 0.95cm 0 0.75cm, clip]{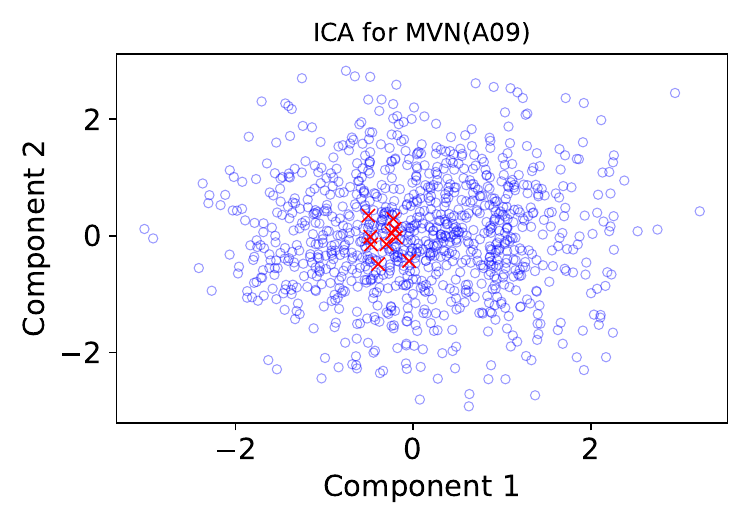} & \includegraphics[width=0.3\textwidth, page=1,trim=1cm 0.95cm 0 0.75cm, clip]{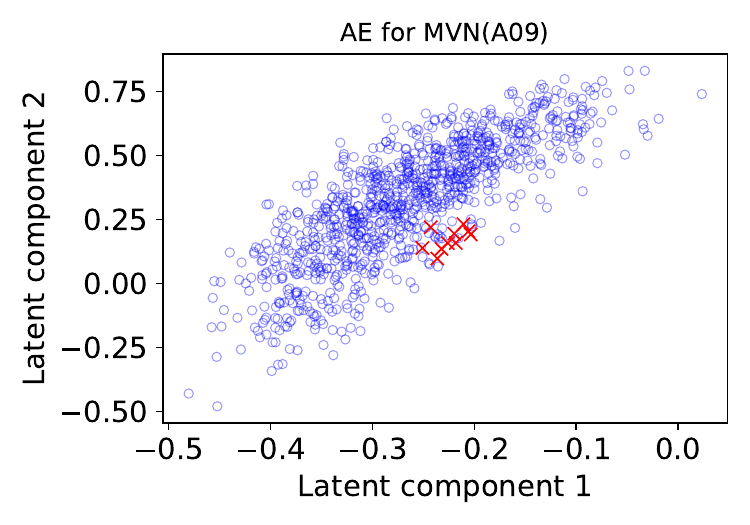}\\
        \raisebox{1.75\normalbaselineskip}[0pt][0pt]{\rotatebox[origin=c]{90}{MVN(hCN)}} & \includegraphics[width=0.3\textwidth, page=2, trim=1cm 0.95cm 0 0.75cm, clip]{figures/settings/by_model/pca_01.pdf} & \includegraphics[width=0.3\textwidth, page=2, trim=1cm 0.95cm 0 0.75cm, clip]{figures/settings/by_model/ica_01.pdf} & \includegraphics[width=0.3\textwidth, page=2, trim=1cm 0.95cm 0 0.75cm, clip]{figures/settings/by_model/ae_01.pdf}\\
        \raisebox{1.75\normalbaselineskip}[0pt][0pt]{\rotatebox[origin=c]{90}{ELL($t(5)$)}} & \includegraphics[width=0.3\textwidth, page=4, trim=1cm 0.95cm 0 0.75cm, clip]{figures/settings/by_model/pca_01.pdf} & \includegraphics[width=0.3\textwidth,  page=4, trim=1cm 0.95cm 0 0.75cm, clip]{figures/settings/by_model/ica_01.pdf} & \includegraphics[width=0.3\textwidth, page=4, trim=1cm 0.95cm 0 0.75cm, clip]{figures/settings/by_model/ae_01.pdf}\\
        \raisebox{1.75\normalbaselineskip}[0pt][0pt]{\rotatebox[origin=c]{90}{EXP}} & \includegraphics[width=0.3\textwidth, page=3, trim=1cm 0.95cm 0 0.75cm, clip]{figures/settings/by_model/pca_01.pdf} & \includegraphics[width=0.3\textwidth, page=3, trim=1cm 0.95cm 0 0.75cm, clip]{figures/settings/by_model/ica_01.pdf} & \includegraphics[width=0.3\textwidth, page=3, trim=1cm 0.95cm 0 0.75cm, clip]{figures/settings/by_model/ae_01.pdf}\\
    \end{tabular}
    \caption{Two-dimensional plots for PCA, ICA and Autoencoder applied to each setting when $\epsilon=0.01$. Axis are the 2 first components for PCA and ICA. Concerning the Autoencoder we visualise the $2d$ latent space.}
    \label{fig:sim2d:0.01}
\end{figure}

\begin{figure}[H]
\centering
    \begin{tabular}{cc}
        \hspace{1cm}ACA for MVN(A09) & \hspace{1cm}ACA for MVN(hCN) \\
        \includegraphics[width=0.4\textwidth, page=1, trim=0 0.25cm 0 0.75cm, clip]{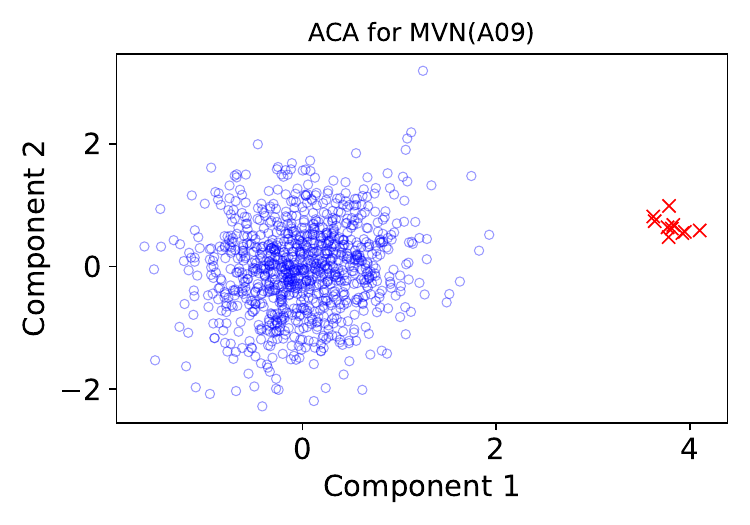} &
        \includegraphics[width=0.4\textwidth, page=2, trim=0 0.25cm 0 0.75cm, clip]{figures/settings/by_model/aca_01.pdf}\\
        \hspace{1cm}ACA for ELL($t(5)$) & \hspace{1cm}ACA for EXP \\
        \includegraphics[width=0.4\textwidth, page=4, trim=0 0.25cm 0 0.75cm, clip]{figures/settings/by_model/aca_01.pdf} &
        \includegraphics[width=0.4\textwidth, page=3, trim=0 0.25cm 0 0.75cm, clip]{figures/settings/by_model/aca_01.pdf}\\
    \end{tabular}
    \caption{Projection of data set $\bmY$ on first two components obtained by ACA in settings with $\epsilon=0.01$.}
\end{figure}

\begin{figure}[H]
\centering
    \begin{tabular}{cc}
        \hspace{1cm}ACA for MVN(A09) & \hspace{1cm}ACA for MVN(hCN) \\
        \includegraphics[width=0.4\textwidth, page=1, trim=0 0.25cm 0 0.75cm, clip]{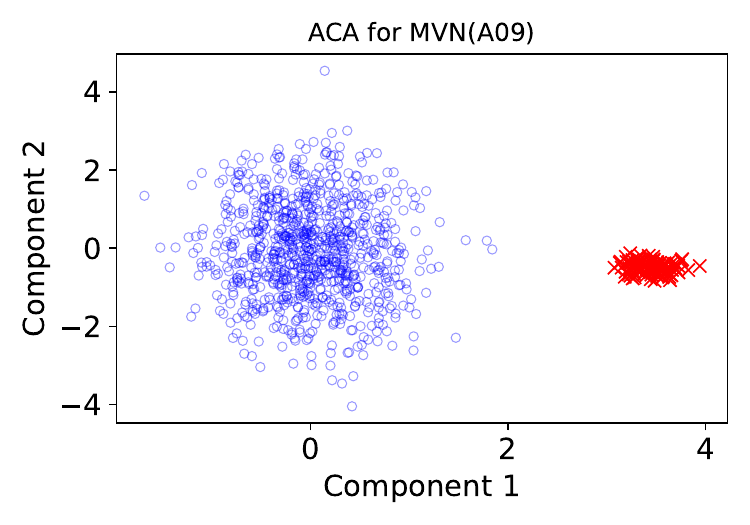} &
        \includegraphics[width=0.4\textwidth, page=2, trim=0 0.25cm 0 0.75cm, clip]{figures/settings/by_model/aca_10.pdf}\\
        \hspace{1cm}ACA for ELL($t(5)$) & \hspace{1cm}ACA for EXP \\
        \includegraphics[width=0.4\textwidth, page=4, trim=0 0.25cm 0 0.75cm, clip]{figures/settings/by_model/aca_10.pdf} &
        \includegraphics[width=0.4\textwidth, page=3, trim=0 0.25cm 0 0.75cm, clip]{figures/settings/by_model/aca_10.pdf}\\
    \end{tabular}
    \caption{Projection of data set $\bmY$ on first two components obtained by ACA in settings with $\epsilon=0.1$.}
\end{figure}

\begin{figure}[H]
    \centering
    \begin{tabular}{cccc}
         & PCA & ICA & Autoencoder\\
        \raisebox{3.5\normalbaselineskip}[0pt][0pt]{\rotatebox[origin=c]{90}{ELL($t(2)$)}} & \includegraphics[width=0.3\textwidth, page=4, trim=1cm 0.95cm 0 0.75cm, clip]{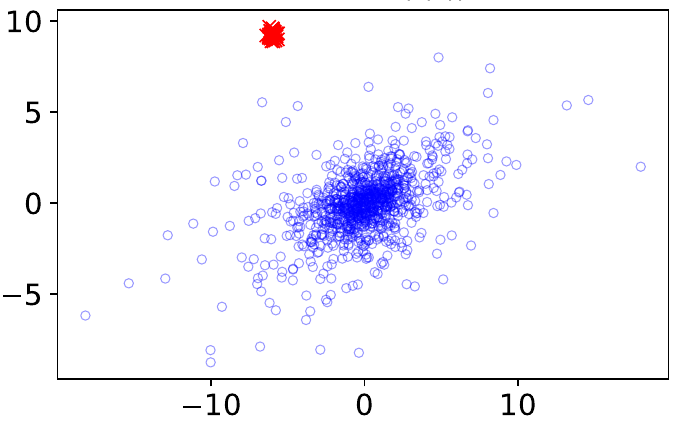} & \includegraphics[width=0.3\textwidth,  page=4, trim=1cm 0.95cm 0 0.75cm, clip]{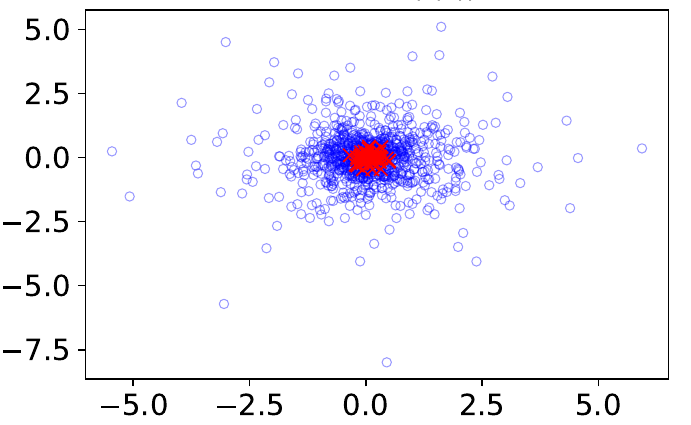} & \includegraphics[width=0.3\textwidth, page=4, trim=1cm 0.95cm 0 0.75cm, clip]{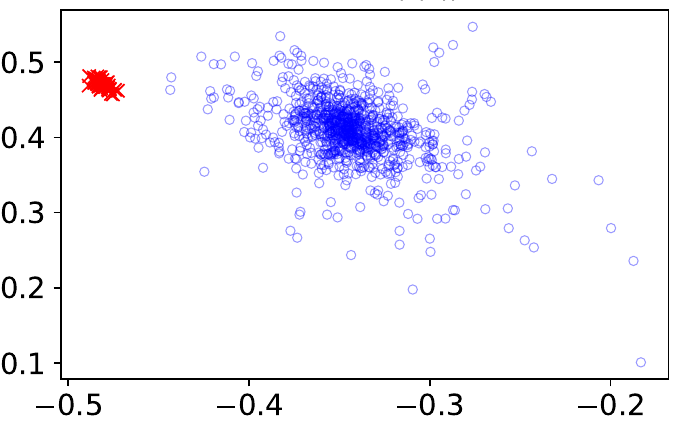}\\
        \raisebox{3.5\normalbaselineskip}[0pt][0pt]{\rotatebox[origin=c]{90}{ELL($t(3)$)}} & \includegraphics[width=0.3\textwidth, page=4, trim=1cm 0.95cm 0 0.75cm, clip]{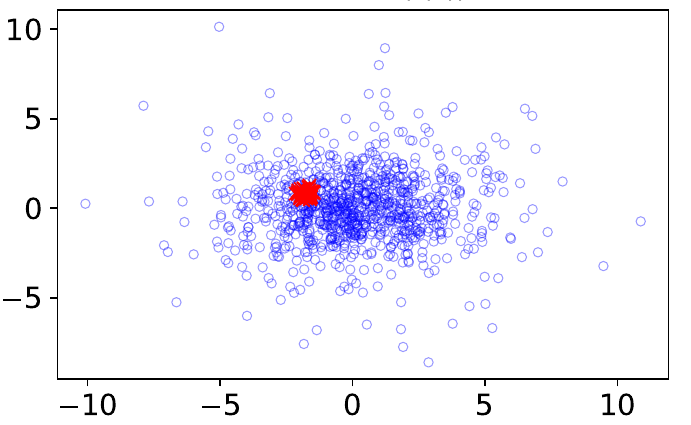} & \includegraphics[width=0.3\textwidth,  page=4, trim=1cm 0.95cm 0 0.75cm, clip]{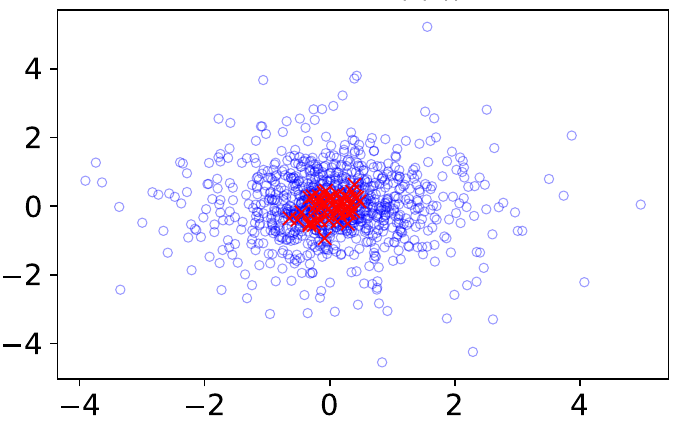} & \includegraphics[width=0.3\textwidth, page=4, trim=1cm 0.95cm 0 0.75cm, clip]{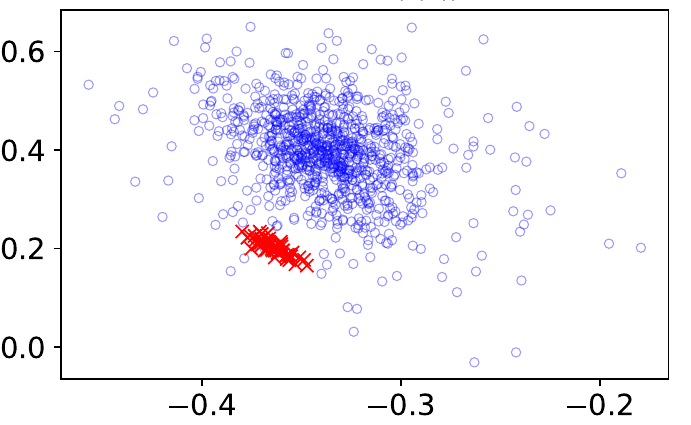}\\
    \end{tabular}
    \caption{Two-dimensional plots for PCA, ICA and Autoencoder applied to settings ELL($t(2)$) and ELL($t(3)$) when $\epsilon=0.05$. Axis are the 2 first components for PCA and ICA. Concerning the Autoencoder we visualise the $2d$ latent space.}
    \label{fig:sim2d:ell}
\end{figure}

\begin{figure}[H]
\centering
    \begin{tabular}{cc}
        \hspace{1cm}ACA for ELL($t(2)$) & \hspace{1cm}ACA for ELL($t(3)$) \\
        \includegraphics[width=0.4\textwidth, page=4, trim=0 0.25cm 0 0.75cm, clip]{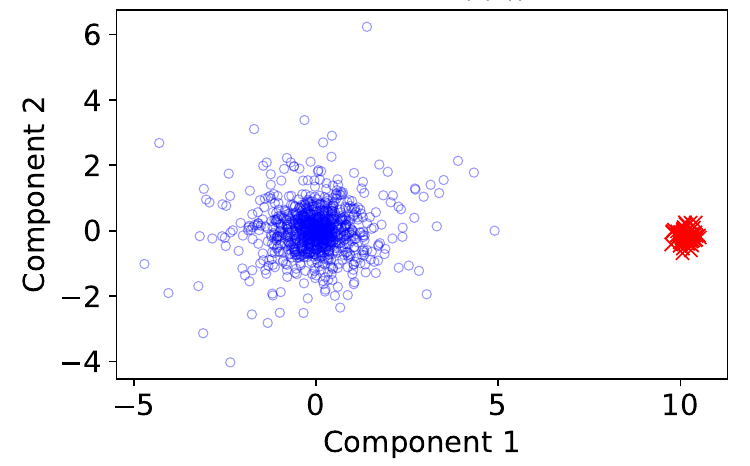} &
        \includegraphics[width=0.4\textwidth, page=4, trim=0 0.25cm 0 0.75cm, clip]{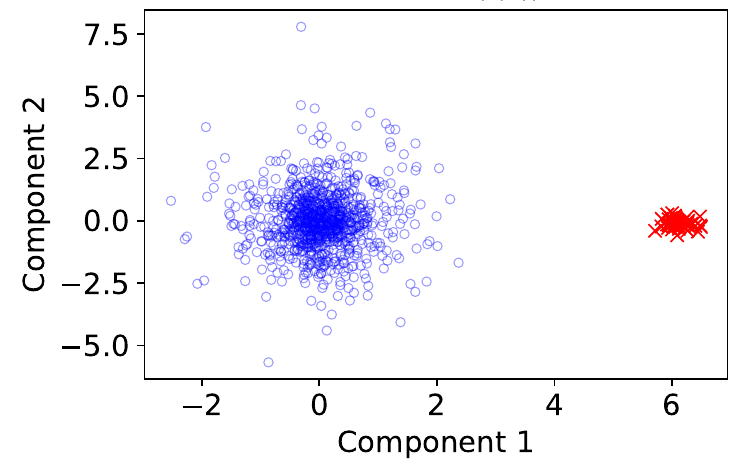}\\
    \end{tabular}
    \caption{Projection of data set $\bmY$ on first two components obtained by ACA in settings ELL($t(2)$) and ELL($t(3)$) with $\epsilon=0.05$.}
\end{figure}

\bigskip

\section{Components' coordinates}\label{sup:sec:comps}

Following visualisations in Section~\ref{ssec:vissimresults}, we show the component for every simulation setting and every method (except for the auto-encoder because of its non-linearity).

\begin{table}[H]
\centering
\begin{tabular}{|cc|cc|cc|cc|}
\hline
\multicolumn{2}{|c|}{MVN(A09)} & \multicolumn{2}{c|}{MVN(hCN)} & \multicolumn{2}{c|}{ELL($t(5)$)} & \multicolumn{2}{c|}{EXP} \\ \hline
PC1   & PC2   & PC1   & PC2   & PC1   & PC2   & PC1   & PC2    \\ \hline
0.19  & -0.41 & -0.17 & 0.56  & 0.2   & -0.36 & 0.    & 0.     \\
-0.26 & 0.43  & -0.38 & 0.02  & -0.3  & 0.44  & 0.01  & -0.    \\
0.28  & -0.36 & -0.37 & -0.21 & 0.3   & -0.28 & 0.01  & 0.01   \\
-0.32 & 0.26  & 0.24  & -0.05 & -0.31 & 0.21  & -0.   & 0.01   \\
0.37  & -0.14 & 0.41  & -0.   & 0.39  & -0.18 & 0.01  & -0.02  \\
-0.39 & -0.03 & 0.04  & 0.46  & -0.39 & -0.03 & -0.05 & -0.07  \\
0.37  & 0.24  & -0.62 & -0.23 & 0.37  & 0.24  & 0.    & -0.01  \\
-0.37 & -0.34 & 0.1   & -0.41 & -0.33 & -0.35 & 0.07  & 0.61   \\
0.3   & 0.37  & 0.01  & 0.3   & 0.29  & 0.44  & -0.03 & 0.79   \\
-0.26 & -0.34 & -0.26 & 0.35  & -0.21 & -0.39 & 1.    & -0.02  \\ \hline
\end{tabular}
\caption{First and second principal components for all settings.}
\end{table}

\begin{table}[H]
\centering
\begin{tabular}{|cc|cc|cc|cc|}
\hline
\multicolumn{2}{|c|}{MVN(A09)} & \multicolumn{2}{c|}{MVN(hCN)} & \multicolumn{2}{c|}{ELL($t(5)$)} & \multicolumn{2}{c|}{EXP} \\ \hline
robPC1 & robPC2 & robPC1 & robPC2 & robPC1 & robPC2 & robPC1 & robPC2 \\ \hline
-0.18  & 0.4    & -0.03  & 0.58   & -0.2   & 0.34   & -0.    & 0.     \\
0.25   & -0.43  & -0.37  & 0.13   & 0.28   & -0.39  & -0.    & -0.02  \\
-0.29  & 0.38   & -0.38  & -0.18  & -0.35  & 0.34   & -0.01  & 0.03   \\
0.33   & -0.29  & 0.19   & -0.04  & 0.35   & -0.28  & -0.02  & 0.     \\
-0.35  & 0.13   & 0.42   & -0.15  & -0.37  & 0.11   & -0.02  & -0.    \\
0.38   & 0.04   & 0.14   & 0.47   & 0.36   & 0.08   & -0.02  & -0.04  \\
-0.38  & -0.23  & -0.65  & -0.08  & -0.37  & -0.25  & 0.02   & -0.04  \\
0.36   & 0.36   & -0.05  & -0.32  & 0.32   & 0.36   & -0.02  & -0.11  \\
-0.31  & -0.36  & 0.11   & 0.22   & -0.29  & -0.42  & -0.09  & 0.99   \\
0.26   & 0.33   & -0.2   & 0.46   & 0.21   & 0.39   & 0.99   & 0.09   \\ \hline
\end{tabular}
\caption{First and second robust principal components for all settings.}
\end{table}

\begin{table}[H]
\centering
\begin{tabular}{|cc|cc|cc|cc|}
\hline
\multicolumn{2}{|c|}{MVN(A09)} & \multicolumn{2}{c|}{MVN(hCN)} & \multicolumn{2}{c|}{ELL($t(5)$)} & \multicolumn{2}{c|}{EXP} \\ \hline
IC1   & IC2   & IC1   & IC2   & IC1   & IC2   & IC1   & IC2   \\ \hline
0.2   & 0.21  & -0.07 & 0.71  & 0.11  & -0.28 & 0.05  & 0.07  \\
0.51  & 0.27  & 0.23  & 0.31  & 0.06  & -0.22 & 0.06  & 0.1   \\
-0.02 & -0.05 & 0.42  & -0.19 & 0.01  & -0.17 & -0.06 & -0.06 \\
0.59  & -0.03 & 0.48  & -0.05 & 0.31  & -0.12 & -0.14 & -0.   \\
-0.25 & 0.33  & 0.24  & 0.13  & -0.38 & 0.16  & 0.98  & -0.04 \\
-0.19 & 0.26  & 0.4   & -0.25 & 0.2   & 0.14  & -0.04 & -0.08 \\
0.43  & -0.27 & 0.05  & 0.12  & 0.02  & -0.24 & 0.04  & 0.99  \\
-0.02 & -0.35 & -0.01 & 0.2   & -0.82 & 0.04  & -0.   & 0.03  \\
-0.22 & -0.5  & 0.5   & 0.05  & -0.16 & 0.75  & 0.04  & 0.01  \\
-0.16 & -0.51 & 0.27  & -0.48 & -0.05 & 0.41  & 0.    & -0.04 \\ \hline
\end{tabular}
\caption{First and second independent components for all settings.}
\end{table}

\begin{table}[H]
\centering
\begin{tabular}{|cc|cc|cc|cc|}
\hline
\multicolumn{2}{|c|}{MVN(A09)} & \multicolumn{2}{c|}{MVN(hCN)} & \multicolumn{2}{c|}{ELL($t(5)$)} & \multicolumn{2}{c|}{EXP} \\ \hline
AC1   & AC2   & AC1   & AC2   & AC1   & AC2   & AC1   & AC2   \\ \hline
0.05  & -0.01 & 0.11  & -0.45 & -0.04 & -0.31 & -0.99 & 0.02  \\
-0.39 & -0.59 & 0.38  & 0.25  & -0.44 & -0.33 & 0.11  & 0.2   \\
-0.31 & 0.32  & 0.27  & -0.06 & -0.42 & -0.31 & -0.03 & -0.21 \\
0.42  & 0.15  & 0.27  & 0.19  & 0.35  & -0.36 & -0.02 & 0.05  \\
0.47  & -0.02 & 0.42  & 0.14  & 0.54  & -0.04 & -0.01 & -0.05 \\
-0.12 & 0.61  & -0.15 & -0.11 & -0.19 & 0.12  & -0.01 & -0.24 \\
-0.46 & 0.24  & -0.16 & 0.46  & -0.27 & -0.17 & 0.01  & 0.19  \\
-0.34 & -0.08 & 0.48  & -0.54 & -0.17 & -0.22 & -0.01 & 0.9   \\
0.02  & 0.16  & 0.42  & 0.4   & -0.24 & 0.62  & -0.   & -0.06 \\
0.12  & -0.26 & 0.27  & 0.    & -0.15 & 0.3   & 0.01  & -0.06 \\ \hline
\end{tabular}
\caption{First and second abnormal components for all settings.}
\end{table}

\section{On comparison with DDC}\label{sup:sec:compDDC}

Following the principle of ACA, in this section, we shall explore the general capacity of (asymmetric) projection depth to highlight variables responsible for observations' abnormality, in two comparative simulation studies.

\paragraph{Comparison with DDC} \cite{ROUSSEEUW18} proposed a method for detecting (and imputing) deviating data cells (DDC) in the dataset, which can be also seen as explaining anomaly-detection tool. (Under a cell-wise anomaly one understands a normal observation whose one or more coordinates are contaminated with other variables remaining intact.) In what follows, we shall compare (asymmetric) projection depth with DDC in detecting such deviating cells. With DDC exploiting correlation, we use the following challenging anomaly-detection setting: we generate $\bmY$ according to MV-Sk (see Section~\ref{ssec:simset}) contaminated with $10\%$ of anomalies from Section~\ref{ssec:vissimresults} restricting $\boldsymbol{\tilde\mu}$ to the set $\gamma M_{max} \bmu\,:\,\{\bmu^{(1)}>0,\,\|\bmu^{(2)}\| > \bmu^{(1)},\,\bmu\in\mathbb{S}^{d-1}\}$ (with $\bmu^{(i)}$ denoting $i$th coordinate of $\bmu$ and $M_{max}$ standing for the maximal Mahalanobis distance among normal data); we set $\gamma=0.8,0.9,1,1.5$, $n=1000$ and $d=2$. Thus, $\bmY$ is not only asymmetric, but as well all anomalies are trivially explained by the second variable. 

For $i$th variable of $\bmy$, the depth-based anomaly score is constructed as:
\begin{equation*}
	s^{(i)}(\bmy|\bmY) = \|\bmu_{pd}^{(i)}(\bmy|\bmY)\|\,\cdot\, \bigl(1 / D^{apd}(\bmy|\bmY) - 1\bigr)\,,
\end{equation*}
where $\bmu_{pd}(\bmy|\bmY) \in \argmin_{\bmu\in\mathbb{S}^{d-1}} D^{pd}(\bmy|\bmY)$.
For DDC, to each observation's variable we attribute the cell's standardized residuals as anomaly score \citep[see][for more details]{ROUSSEEUW18}. Using these scores to order observations in $\bmY$, we compare them to true anomalies by the area under the receiver operating characteristic (AUC); these are indicated in Figure~\ref{fig:DDCwithACAone}. We observe that, under asymmetric deviation from elliptical contours (perfectly described by correlation) DDC is outperformed by the non-parametric depth-based approach in this setting. For a broader setting, \textit{i.e.}, when letting more freedom for $\boldsymbol{\tilde\mu}$, DDC and depth perform comparably.

\subsection{Setting delivering equal performance}\label{sup:ssec:DDDequal}

\begin{figure}[H]
\centering
\includegraphics[width=0.7\textwidth]{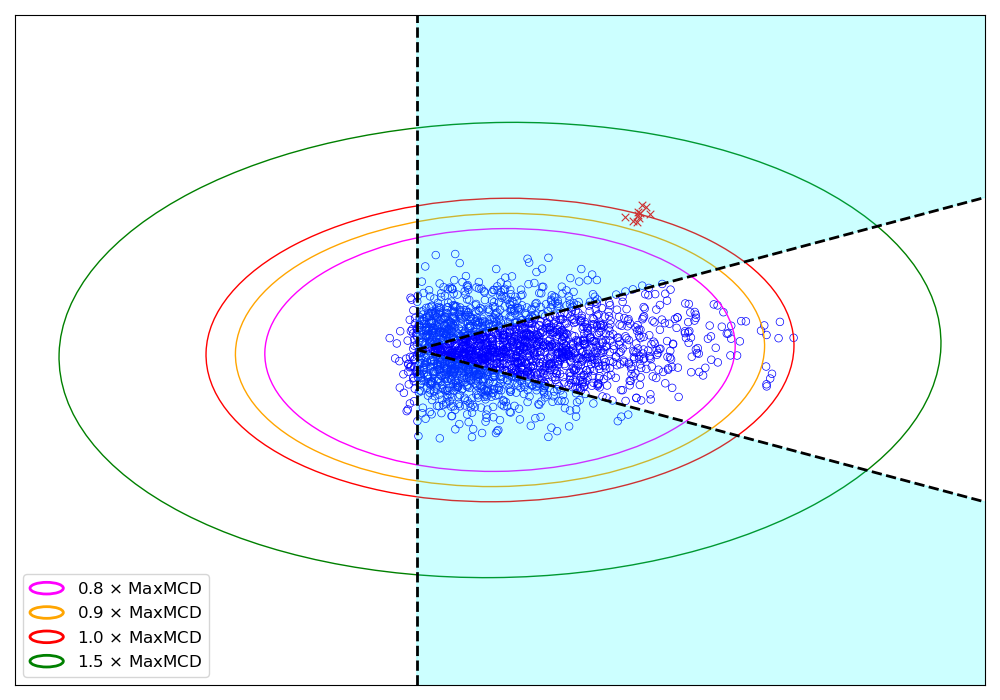}
\caption{First setting for comparison between ACA and DDC. Two symmetrical areas of $75^\circ$ from the ordinate in cyan, in which anomalies are placed at various Mahalanobis distances (represented by colored ellipses) from the center. Scatter for the Mahalanobis distance is estimated using minimum covariance determinant~\citep{ROUSSEEUW99} and location is in $(0,0)^\top$.}\label{fig:samesett}
\end{figure}

\begin{figure}[H]
\centering
\includegraphics[width=0.7\textwidth]{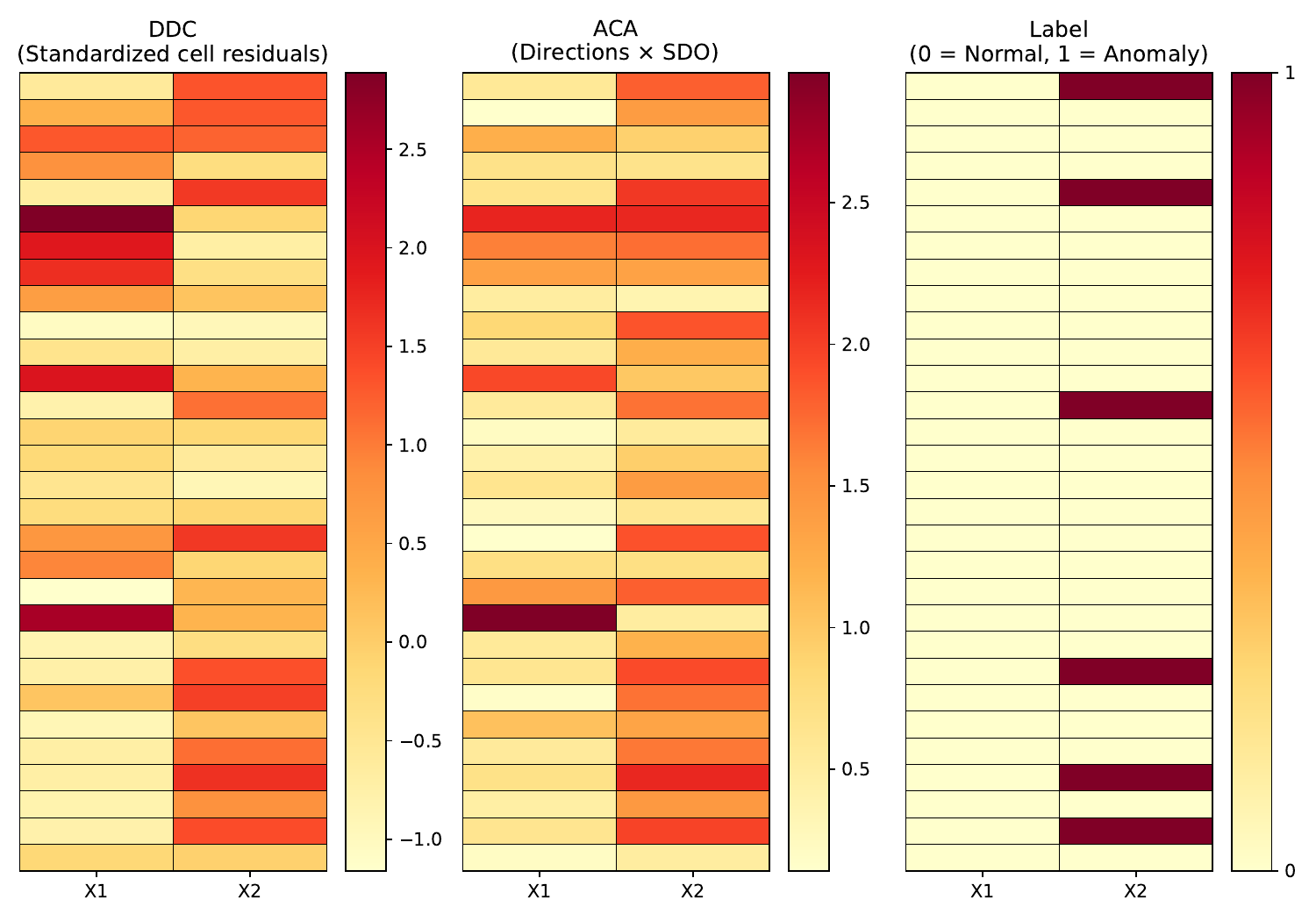}
\caption{Features obtained by DDC (left) and ACA (middle) to detect and explain anomalies compared to true labels (right).}\label{fig:features}
\end{figure}

\begin{figure}[H]
    \centering
    \begin{tabular}{cc}
    \small $\gamma=0.8$ & \small $\gamma=0.9$ \\
    \includegraphics[width=0.49\textwidth, page=1, trim=0cm 0cm 1cm 1cm, clip]{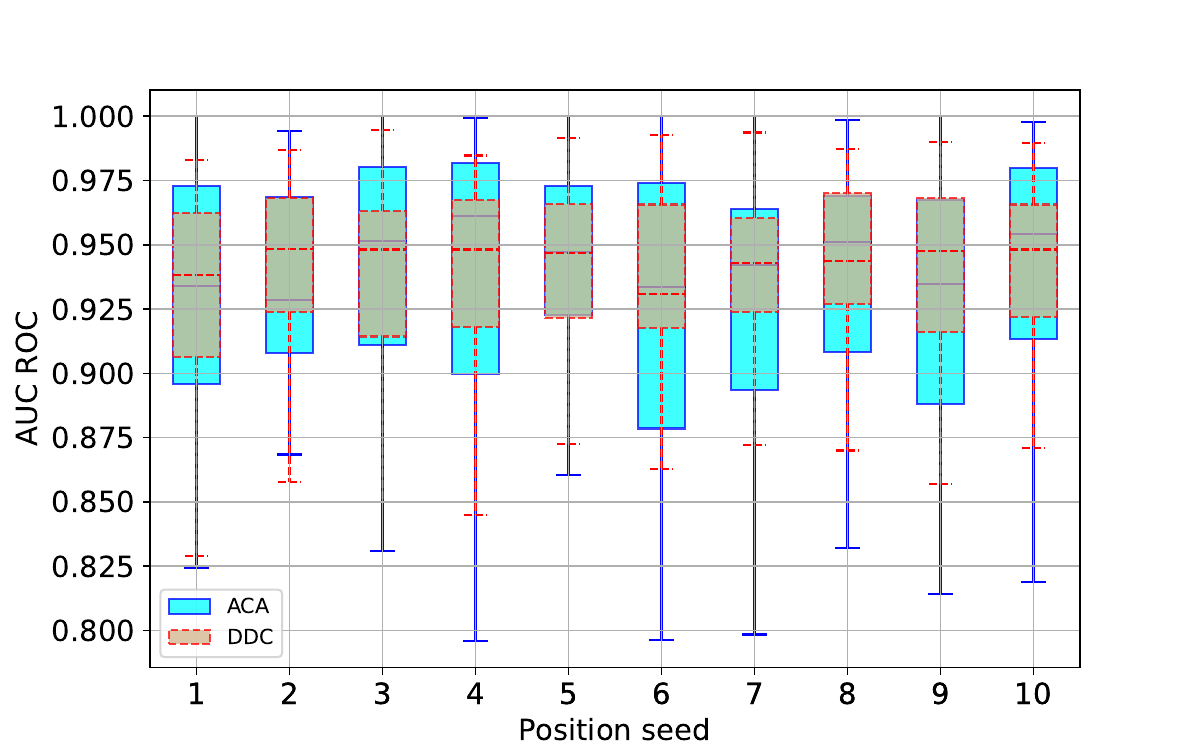} & \includegraphics[width=0.49\textwidth, page=2, trim=0cm 0cm 1cm 1cm, clip]{figures/DDC/ddc_aca.pdf} \\
    \small $\gamma=1$ & \small $\gamma=1.5$ \\
    \includegraphics[width=0.49\textwidth, page=3, trim=0cm 0cm 1cm 1cm, clip]{figures/DDC/ddc_aca.pdf} & \includegraphics[width=0.49\textwidth, page=4, trim=0cm 0cm 1cm 1cm, clip]{figures/DDC/ddc_aca.pdf}
    \end{tabular}
    \caption{AUC computed using respective features for both methods DDC and ACA in the setting where they have approximately equal performance; 50 simulations are performed for each of $10$ position seed values.}\label{fig:boxplots}
\end{figure}

First setting allows for anomalies' location in a wider area such that they can be more explained using the first variable $X_1$ than the second $X_2$. This results in approximately equal performance for both methods. The setting is explained in Figure~\ref{fig:samesett} and we display the features in Figure~\ref{fig:features} obtained with DDC and ACA which are used to compute AUC indicated in Figure~\ref{fig:boxplots}.

\subsection{Setting delivering different performance}\label{sup:ssec:DDDbetter}

\begin{figure}[H]
\centering
\includegraphics[width=0.7\textwidth]{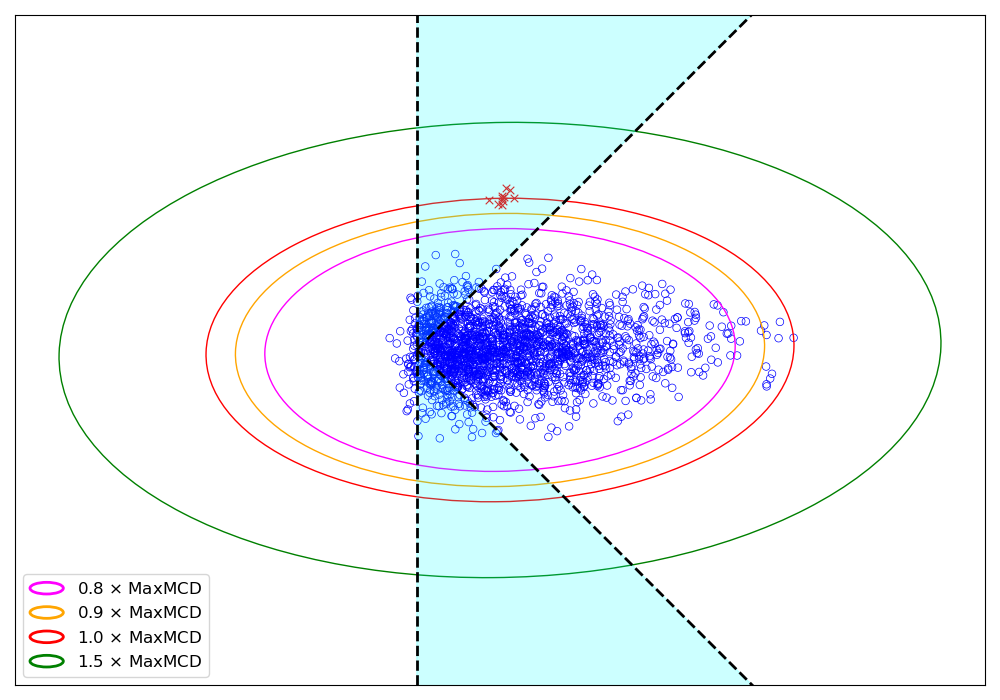}
\caption{Second setting for comparison between ACA and DDC. Two symmetric areas of $45^\circ$ from the ordinate in cyan, in which anomalies are placed at various Mahalanobis distances (represented by colored ellipses) from the center. Scatter for the Mahalanobis distance is estimated using minimum covariance determinant~\citep{ROUSSEEUW99} and location is in $(0,0)^\top$.}\label{fig:bettersett}
\end{figure}

\begin{figure}[H]
\centering
\includegraphics[width=0.7\textwidth]{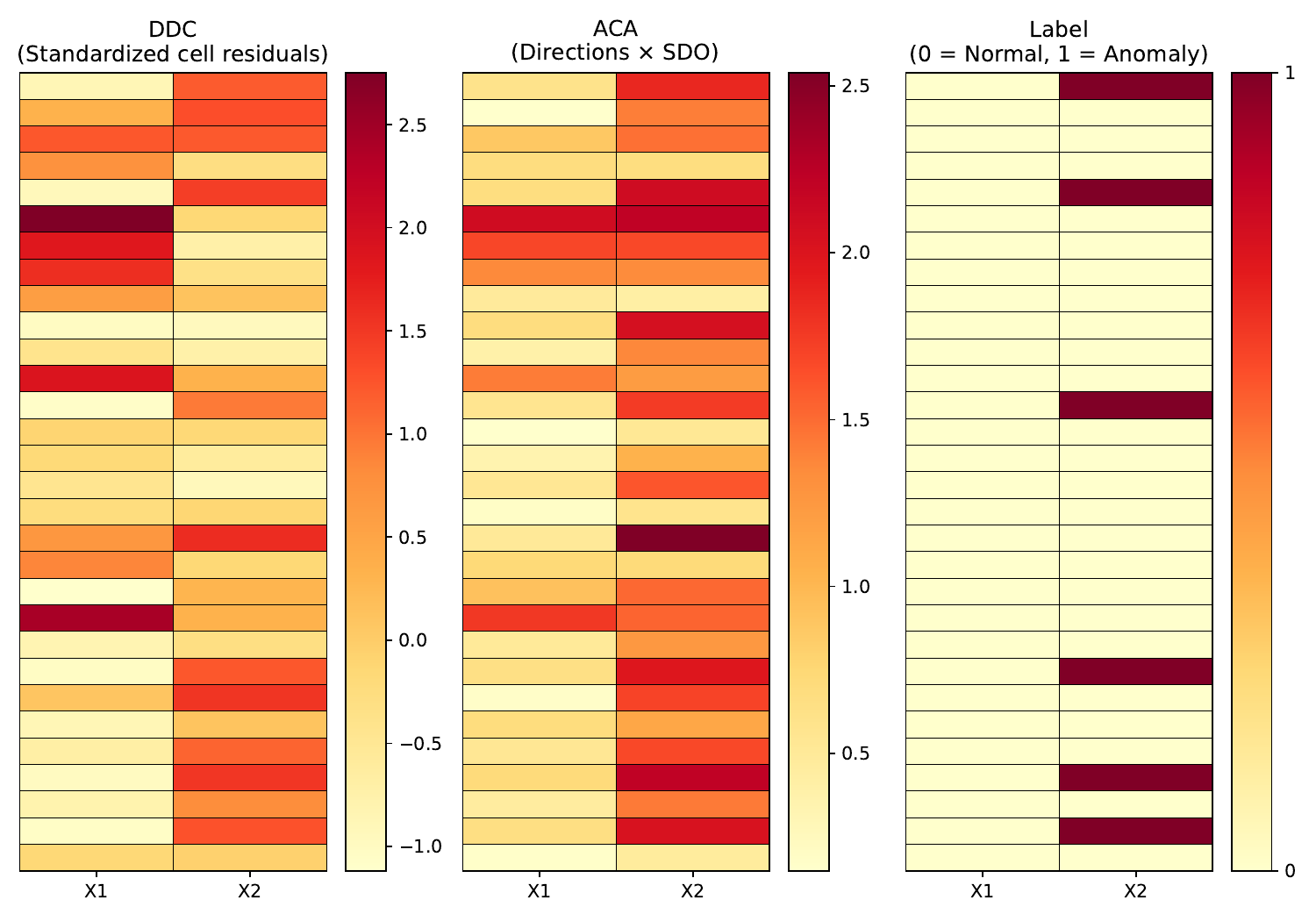}
\caption{Features obtained by DDC (left) and ACA (middle) to detect and explain anomalies compared to true labels (right).}\label{fig:features2}
\end{figure}

\begin{figure}[!h]
    \centering
    \begin{tabular}{cc}
    \small $\gamma=0.8$ & \small $\gamma=0.9$ \\
    \includegraphics[width=0.4\textwidth, page=5, trim=0cm 0cm 2cm 1cm, clip]{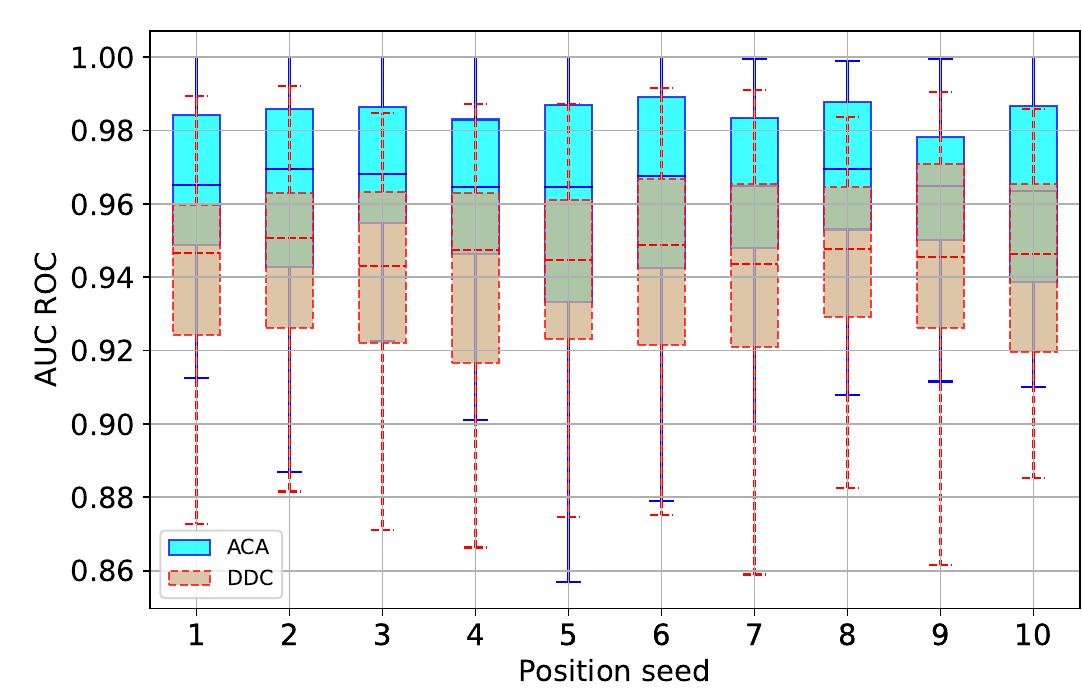} & \includegraphics[width=0.4\textwidth, page=6, trim=0cm 0cm 2cm 1cm, clip]{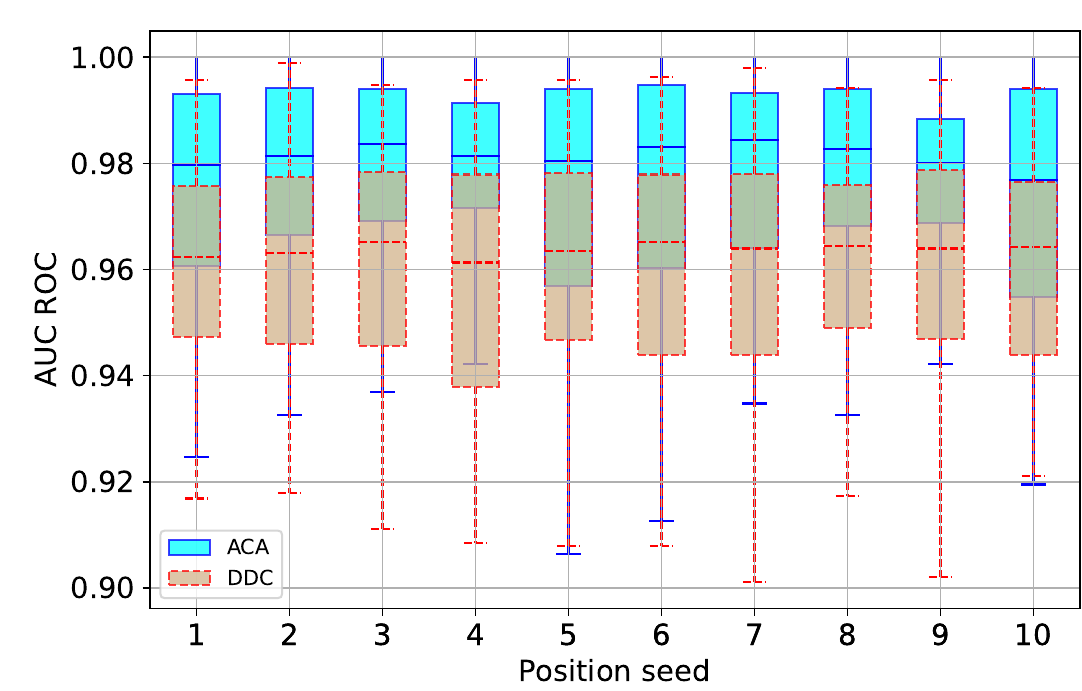} \\
    \small $\gamma=1$ & \small $\gamma=1.5$ \\
    \includegraphics[width=0.4\textwidth, page=7, trim=0cm 0cm 2cm 1cm, clip]{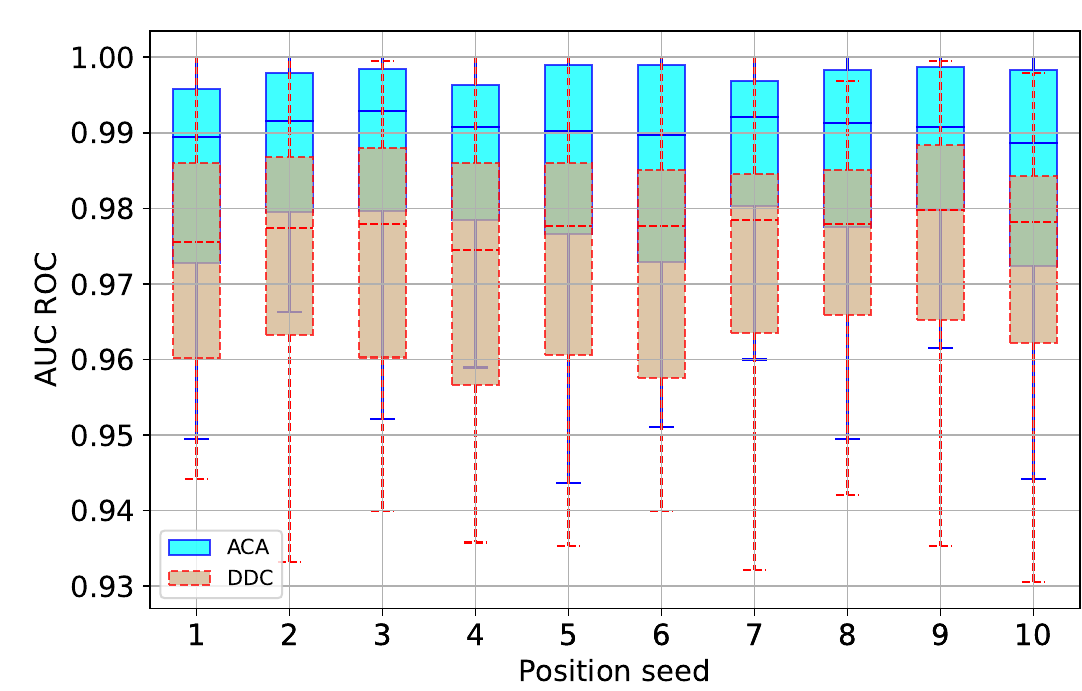} & \includegraphics[width=0.4\textwidth, page=8, trim=0cm 0cm 2cm 1cm, clip]{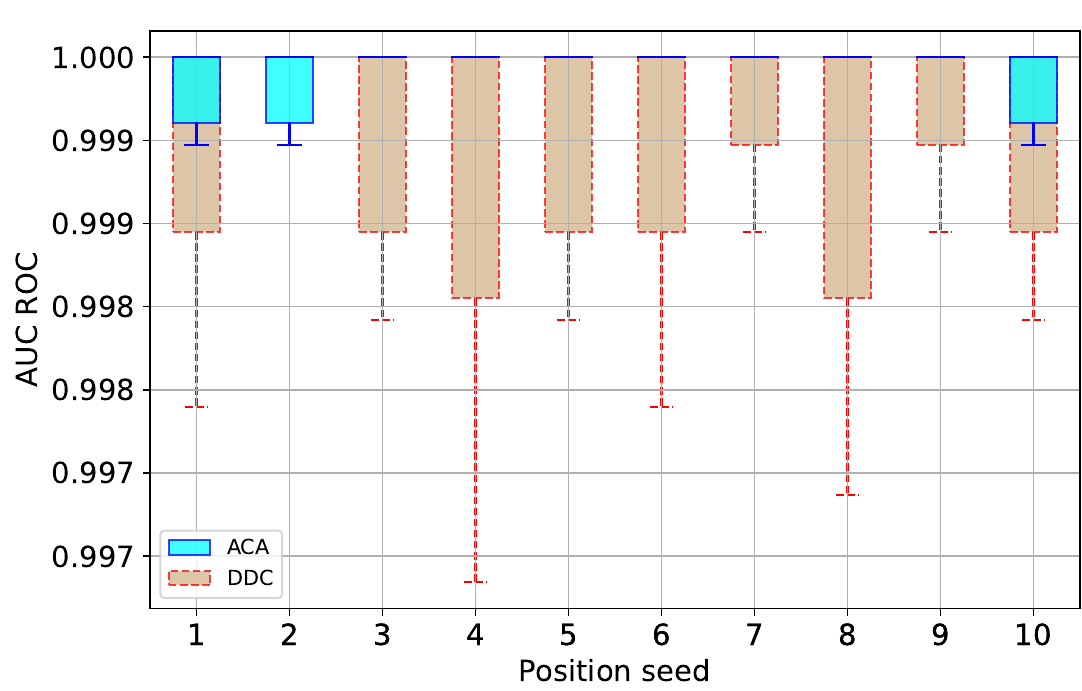}
    \end{tabular}
    \caption{AUC of cells' anomaly scores attributed by DDC and ACA's directions for 1000 bivariate observations drawn from MV-Sk setting; 50 simulations are performed for each of $10$ position seed values.}\label{fig:DDCwithACAone}
\end{figure}

This second setting where anomalies are placed in the same way as before with $X_2$ being more `responsible' for abnormality than $X_1$. This results in better performance for ACA. The setting is explained in Figure~\ref{fig:bettersett} and we display the features in Figure~\ref{fig:features2} obtained with DDC and ACA which are used to compute AUR indicated in Figure~\ref{fig:DDCwithACAone}.

\section{On comparison with DIFFI}\label{sup:sec:compDIFFI}

\begin{figure}[!h]
\centering
    \begin{tabular}{cc}
        \includegraphics[width=0.4\textwidth]{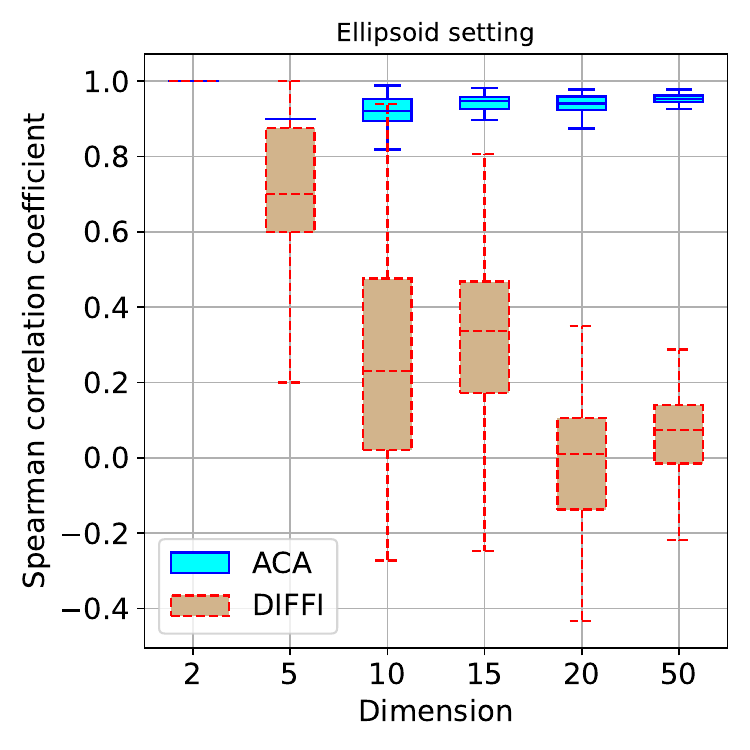} & \includegraphics[width=0.4\textwidth]{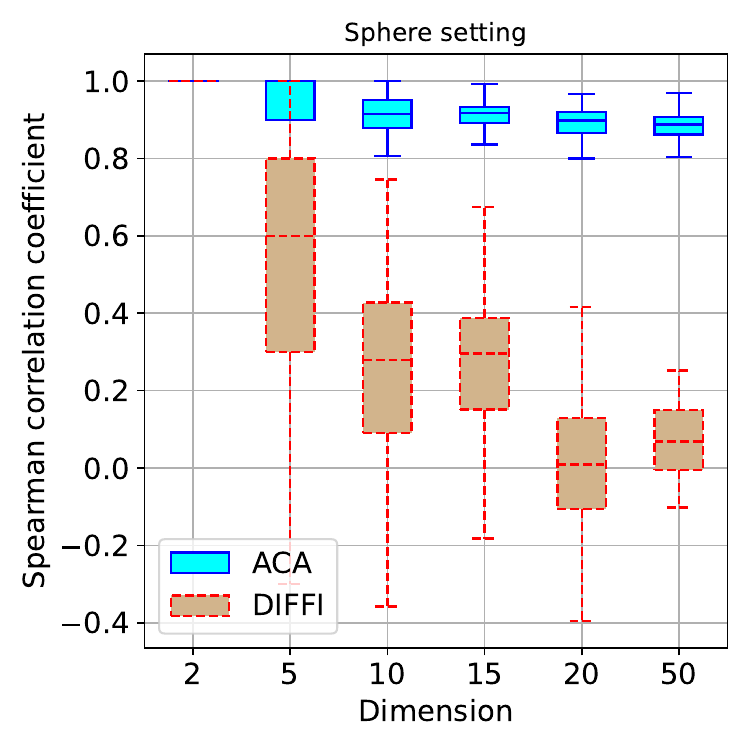}
    \end{tabular}
    \caption{Spearman correlation coefficient (over $50$ repetitions) of variables' importance ranking (with the population order) for ACA and DIFFI for elliptical (MVN(A09), left) and spherical (right) standard normal distribution.}\label{fig:diffi}
\end{figure}

\cite{CARLETTI23} introduced depth-based isolation forest feature importance (DIFFI) that allows to evaluate contribution of each variable to the abnormality of an observation, and thus constitutes a natural candidate for comparison. We generate data from MVN(A09) and contaminate them with $10\%$ of anomalies generated from from $\mathcal{N}(\boldsymbol{\tilde\mu},\boldsymbol{I}_d/1000)$ with
\begin{equation}\label{equ:diffivars}
	\boldsymbol{\tilde\mu} = \bigl( (2 \cdot d)^2, (2 \cdot (d - 1))^2, ..., (2 \cdot 1)^2 \bigr)^\top\,,
\end{equation}
\textit{i.e.}, variables' importance decreases in their literal order. For ACA, the variable importance is derived from the variable's contribution (in absolute value) to the first AC. The resulting order correlation (with the correct order induced by~\eqref{equ:diffivars}) is indicated in Figure~\ref{fig:diffi} (left), for varying space dimension $d$. One observes that ACA preserves good correlation level when $d$ increases. Not to disadvantage DIFFI, we also attempt the spherical (standard) normal distribution instead of MVN(A09); see Figure~\ref{fig:diffi} (right).

\section{Real data}\label{sup:sec:realdata}

Following datasets are explored in the same manner as in Section~\ref{sec:realdata}. Obtained visualisations and explanations are competitive to those delivered by the methods of the state of the art. We also add tables with variables' contributions for components for ACA, PCA, robPCA and ICA to interpret anomalies location.

\begin{table}[H]
\centering
\begin{tabular}{l|rrccc}
\textbf{Name} & $n$ & $d$ & \textbf{\# Anomalies} & \textbf{\% Anomalies} & \textbf{Category} \\ \hline
ALOI                & -    & 27  & 1508 & 3.04  & Image         \\
Annthyroid          & 7200 & 6   & 534  & 7.42  & Healthcare    \\
Breastw             & 683  & 9   & 239  & 34.99 & Healthcare    \\
Cardio              & 1831 & 21  & 176  & 9.61  & Healthcare    \\
Cardiotocography    & 2114 & 21  & 466  & 22.04 & Healthcare    \\
Celeba              & -    & 39  & 4547 & 2.24  & Image         \\
Cover               & -    & 10  & 2747 & 0.96  & Botany        \\
\end{tabular}
\caption{Datasets description (- signifies datasets with $n>10\,000$ and randomly sampled to $n=10\,000$)}
\end{table}

 \begin{table}[H]
\centering
\begin{tabular}{c|cccccccc|}
\cline{2-9}
 & \multicolumn{8}{c|}{Musk data set} \\ \cline{2-9} 
 & \multicolumn{2}{c|}{PCA} & \multicolumn{2}{c|}{robPCA} & \multicolumn{2}{c|}{ICA} & \multicolumn{2}{c|}{ACA} \\ \cline{2-9} 
 & PC1 & \multicolumn{1}{c|}{PC2} & robPC1 & \multicolumn{1}{c|}{robPC2} & IC1 & \multicolumn{1}{c|}{IC2} & AC1 & AC2 \\ \hline
\multicolumn{1}{|c|}{Var1} & 114 (1\%) & \multicolumn{1}{c|}{102 (2\%)} & 165 (13\%) & \multicolumn{1}{c|}{3 (3\%)} & 30 (2\%) & \multicolumn{1}{c|}{62 (6\%)} & 44 (3\%) & 5 (6\%) \\
\multicolumn{1}{|c|}{Var2} & 113 (1\%) & \multicolumn{1}{c|}{155 (2\%)} & 163 (12\%) & \multicolumn{1}{c|}{68 (2\%)} & 128 (2\%) & \multicolumn{1}{c|}{32 (6\%)} & 105 (3\%) & 155 (2\%) \\
\multicolumn{1}{|c|}{Var3} & 99 (1\%) & \multicolumn{1}{c|}{96 (2\%)} & 164 (4\%) & \multicolumn{1}{c|}{41 (2\%)} & 64 (2\%) & \multicolumn{1}{c|}{125 (5\%)} & 14 (2\%) & 46 (2\%) \\ \hline
\end{tabular}
\caption{Contributions of three most important (for each component) variables for PCA, robPCA, ICA, and ACA for Musk data set.}\label{tab:realMuskVars}
\end{table}

\begin{table}[H]
\centering
\begin{tabular}{c|cccccccc|}
\cline{2-9}
 & \multicolumn{8}{c|}{Satellite image (2) data set} \\ \cline{2-9} 
 & \multicolumn{2}{c|}{PCA} & \multicolumn{2}{c|}{robPCA} & \multicolumn{2}{c|}{ICA} & \multicolumn{2}{c|}{ACA} \\ \cline{2-9} 
 & PC1 & \multicolumn{1}{c|}{PC2} & robPC1 & \multicolumn{1}{c|}{robPC2} & IC1 & \multicolumn{1}{c|}{IC2} & AC1 & AC2 \\ \hline
\multicolumn{1}{|c|}{Var1} & 19 (3\%) & \multicolumn{1}{c|}{17 (5\%)} & 24 (7\%) & \multicolumn{1}{c|}{25 (6\%)} & 19 (4\%) & \multicolumn{1}{c|}{13 (4\%)} & 18 (8\%) & 12 (9\%) \\
\multicolumn{1}{|c|}{Var2} & 15 (3\%) & \multicolumn{1}{c|}{13 (5\%)} & 35 (7\%) & \multicolumn{1}{c|}{13 (6\%)} & 23 (4\%) & \multicolumn{1}{c|}{17 (4\%)} & 2 (8\%) & 10 (8\%) \\
\multicolumn{1}{|c|}{Var3} & 18 (3\%) & \multicolumn{1}{c|}{29 (5\%)} & 15 (6\%) & \multicolumn{1}{c|}{18 (6\%)} & 31 (4\%) & \multicolumn{1}{c|}{5 (4\%)} & 26 (6\%) & 8 (7\%) \\ \hline
\end{tabular}
\caption{Contributions of three most important (for each component) variables for PCA, robPCA, ICA, and ACA for Satellite image (2) data set.}\label{tab:realSatimageVars}
\end{table}

\begin{table}[H]
\centering
\begin{tabular}{c|cccccccc|}
\cline{2-9}
 & \multicolumn{8}{c|}{Thyroid data set} \\ \cline{2-9} 
 & \multicolumn{2}{c|}{PCA} & \multicolumn{2}{c|}{robPCA} & \multicolumn{2}{c|}{ICA} & \multicolumn{2}{c|}{ACA} \\ \cline{2-9} 
 & PC1 & \multicolumn{1}{c|}{PC2} & robPC1 & \multicolumn{1}{c|}{robPC2} & IC1 & \multicolumn{1}{c|}{IC2} & AC1 & AC2 \\ \hline
\multicolumn{1}{|c|}{Var1} & 4 (28\%) & \multicolumn{1}{c|}{5 (28\%)} & 1 (90\%) & \multicolumn{1}{c|}{5 (91\%)} & 1 (73\%) & \multicolumn{1}{c|}{4 (33\%)} & 2 (75\%) & 6 (64\%) \\
\multicolumn{1}{|c|}{Var2} & 6 (22\%) & \multicolumn{1}{c|}{1 (21\%)} & 3 (4\%) & \multicolumn{1}{c|}{6 (8\%)} & 4 (11\%) & \multicolumn{1}{c|}{5 (26\%)} & 6 (12\%) & 4 (18\%) \\
\multicolumn{1}{|c|}{Var3} & 3 (22\%) & \multicolumn{1}{c|}{6 (20\%)} & 4 (4\%) & \multicolumn{1}{c|}{4 (1\%)} & 6 (6\%) & \multicolumn{1}{c|}{3 (22\%)} & 4 (8\%) & 2 (12\%) \\ \hline
\end{tabular}
\caption{Contributions of three most important (for each component) variables for PCA, robPCA, ICA, and ACA for Satellite image (2) data set.}\label{tab:realThyroidVars}
\end{table}

\begin{figure}[H]
\centering
\includegraphics[width=\textwidth]{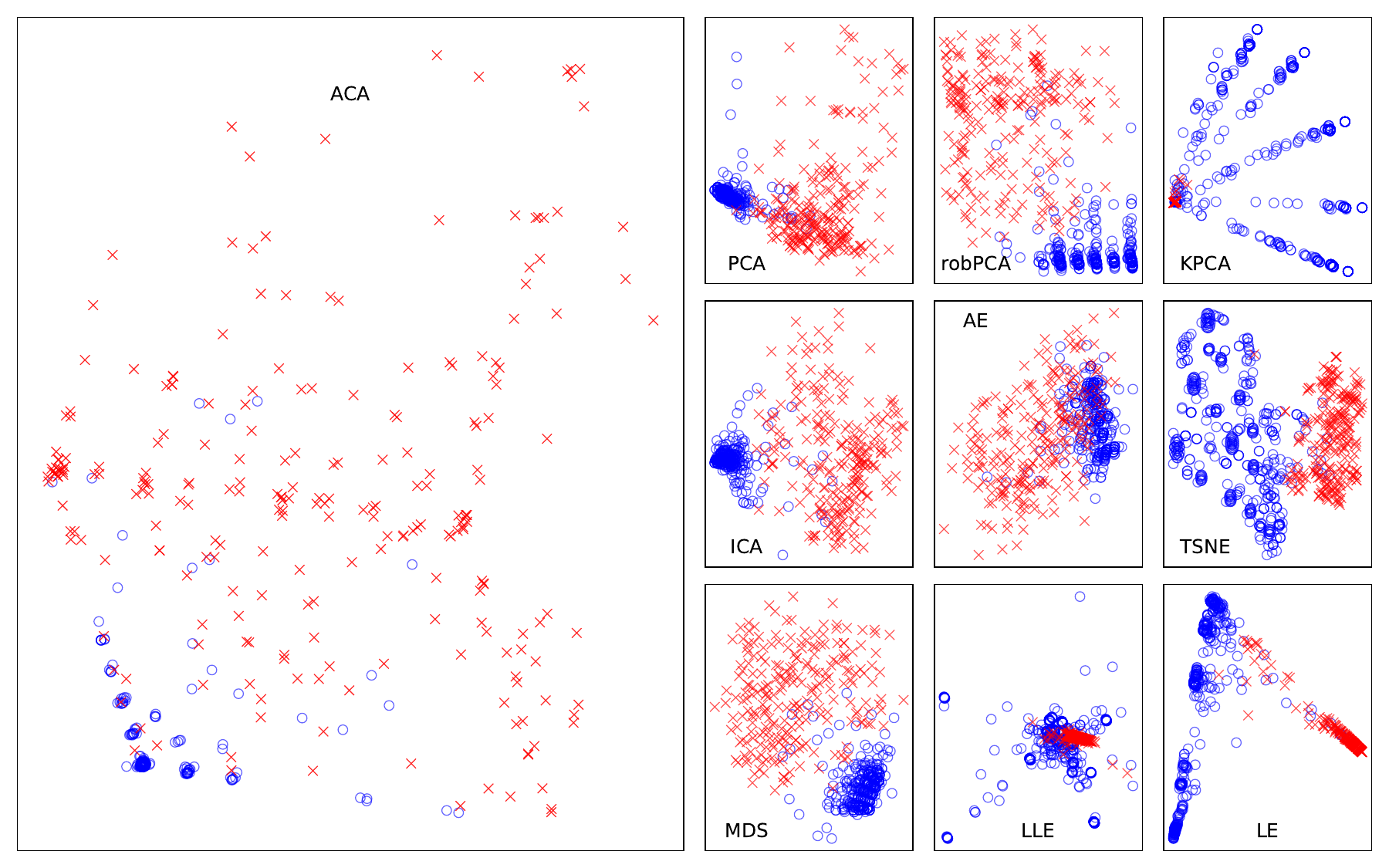}
\caption{Visualization of projection on two first components obtained by different methods for breastw dataset.}
\end{figure}

\begin{table}[H]
\centering
\begin{tabular}{l|cccccccc|}
\cline{2-9}
\multicolumn{1}{c|}{} & \multicolumn{8}{c|}{breastw} \\ \cline{2-9} 
\multicolumn{1}{c|}{} & \multicolumn{2}{c|}{PCA} & \multicolumn{2}{c|}{robPCA} & \multicolumn{2}{c|}{ICA} & \multicolumn{2}{c|}{ACA} \\ \cline{2-9} 
\multicolumn{1}{c|}{} & PC1 & \multicolumn{1}{c|}{PC2} & robPC1 & \multicolumn{1}{c|}{robPC2} & IC1 & \multicolumn{1}{c|}{IC2} & AC1 & AC2 \\ \hline
\multicolumn{1}{|c|}{Var1} & 2 (13\%) & \multicolumn{1}{c|}{9 (47\%)} & 1 (80\%) & \multicolumn{1}{c|}{6 (51\%)} & 6 (32\%) & \multicolumn{1}{c|}{6 (33\%)} & 8 (62\%) & 9 (54\%) \\
\multicolumn{1}{|c|}{Var2} & 3 (13\%) & \multicolumn{1}{c|}{6 (14\%)} & 4 (10\%) & \multicolumn{1}{c|}{5 (18\%)} & 4 (14\%) & \multicolumn{1}{c|}{8 (22\%)} & 9 (18\%) & 6 (34\%) \\
\multicolumn{1}{|c|}{Var3} & 7 (12\%) & \multicolumn{1}{c|}{7 (12\%)} & 3 (4\%) & \multicolumn{1}{c|}{3 (14\%)} & 1 (12\%) & \multicolumn{1}{c|}{2 (11\%)} & 6 (14\%) & 8 (8\%) \\
\multicolumn{1}{|l|}{Var4} & 5 (11\%) & \multicolumn{1}{c|}{5 (9\%)} & 5 (2\%) & \multicolumn{1}{c|}{4 (8\%)} & 3 (11\%) & \multicolumn{1}{c|}{5 (10\%)} & 5 (3\%) & 4 (2\%) \\
\multicolumn{1}{|l|}{Var5} & 8 (11\%) & \multicolumn{1}{c|}{1 (7\%)} & 2 (2\%) & \multicolumn{1}{c|}{7 (4\%)} & 7 (11\%) & \multicolumn{1}{c|}{9 (9\%)} & 7 (1\%) & 2 (1\%) \\
\hline
\end{tabular}
\caption{Most important variables for every method applied to Breastw dataset.}
\end{table}

\begin{figure}[H]
\centering
\includegraphics[width=\textwidth]{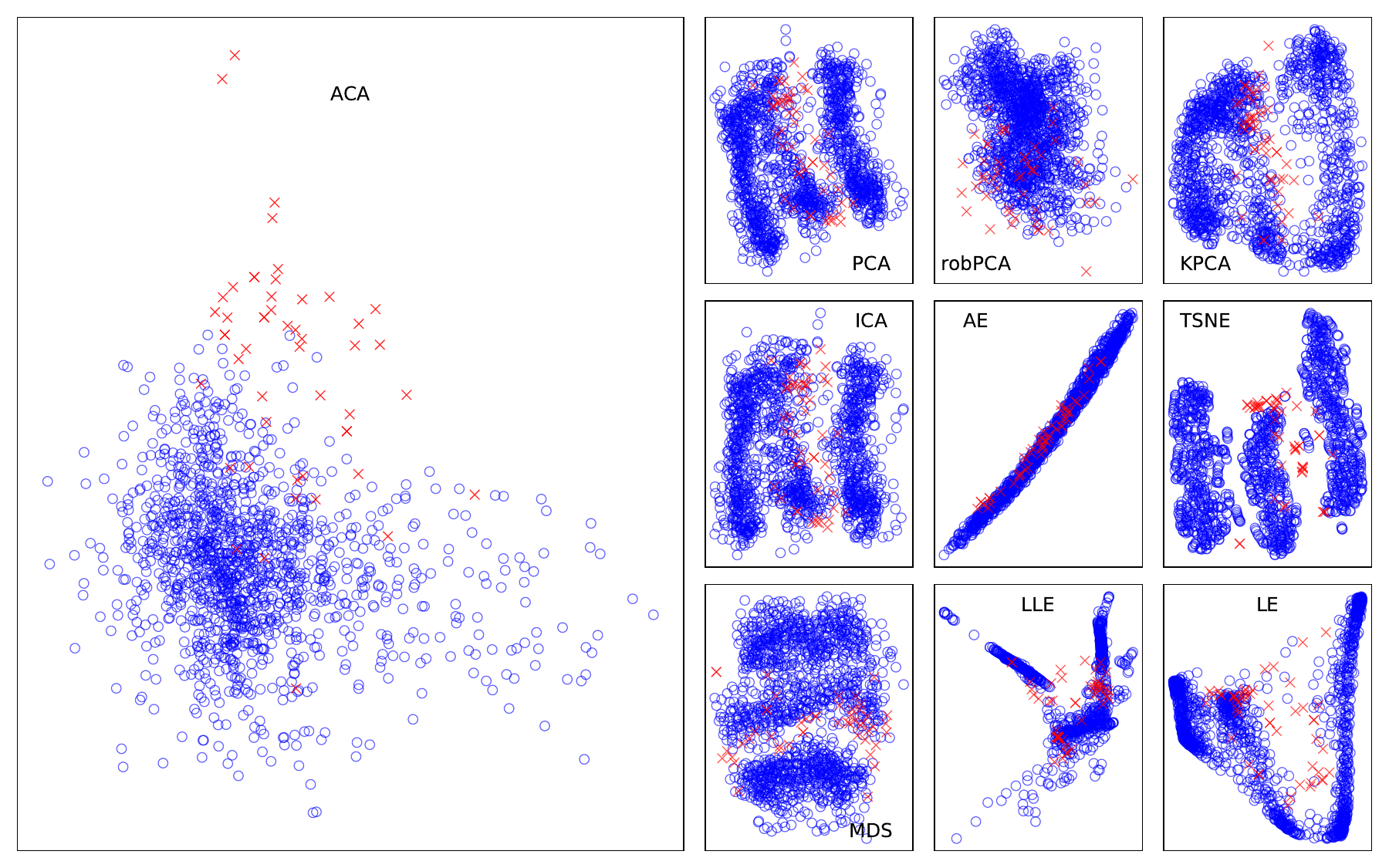}
\caption{Visualization of projection on two first components obtained by different methods for vowels dataset.}
\end{figure}

\begin{table}[H]
\centering
\begin{tabular}{c|cccccccc|}
\cline{2-9}
 & \multicolumn{8}{c|}{vowels} \\ \cline{2-9} 
 & \multicolumn{2}{c|}{PCA} & \multicolumn{2}{c|}{robPCA} & \multicolumn{2}{c|}{ICA} & \multicolumn{2}{c|}{ACA} \\ \cline{2-9} 
 & PC1 & \multicolumn{1}{c|}{PC2} & robPC1 & \multicolumn{1}{c|}{robPC2} & IC1 & \multicolumn{1}{c|}{IC2} & AC1 & AC2 \\ \hline
\multicolumn{1}{|c|}{Var1} & 2 (15\%) & \multicolumn{1}{c|}{5 (15\%)} & 5 (29\%) & \multicolumn{1}{c|}{9 (20\%)} & 9 (16\%) & \multicolumn{1}{c|}{4 (15\%)} & 3 (15\%) & 1 (28\%) \\
\multicolumn{1}{|c|}{Var2} & 3 (15\%) & \multicolumn{1}{c|}{4 (14\%)} & 2 (15\%) & \multicolumn{1}{c|}{10 (14\%)} & 3 (15\%) & \multicolumn{1}{c|}{8 (14\%)} & 4 (13\%) & 7 (13\%) \\
\multicolumn{1}{|c|}{Var3} & 9 (14\%) & \multicolumn{1}{c|}{8 (14\%)} & 6 (9\%) & \multicolumn{1}{c|}{6 (13\%)} & 2 (15\%) & \multicolumn{1}{c|}{5 (12\%)} & 11 (11\%) & 11 (11\%) \\
\multicolumn{1}{|c|}{Var4} & 11 (9\%) & \multicolumn{1}{c|}{1 (13\%)} & 8 (9\%) & \multicolumn{1}{c|}{12 (10\%)} & 5 (13\%) & \multicolumn{1}{c|}{6 (12\%)} & 1 (10\%) & 9 (9\%) \\
\multicolumn{1}{|c|}{Var5} & 5 (8\%) & \multicolumn{1}{c|}{7 (12\%)} & 7 (8\%) & \multicolumn{1}{c|}{4 (9\%)} & 1 (11\%) & \multicolumn{1}{c|}{1 (11\%)} & 10 (10\%) & 4 (9\%) \\
\hline
\end{tabular}
\caption{Most important variables for every method applied to Vowels dataset.}
\end{table}

\begin{figure}[H]
\centering
\includegraphics[width=\textwidth]{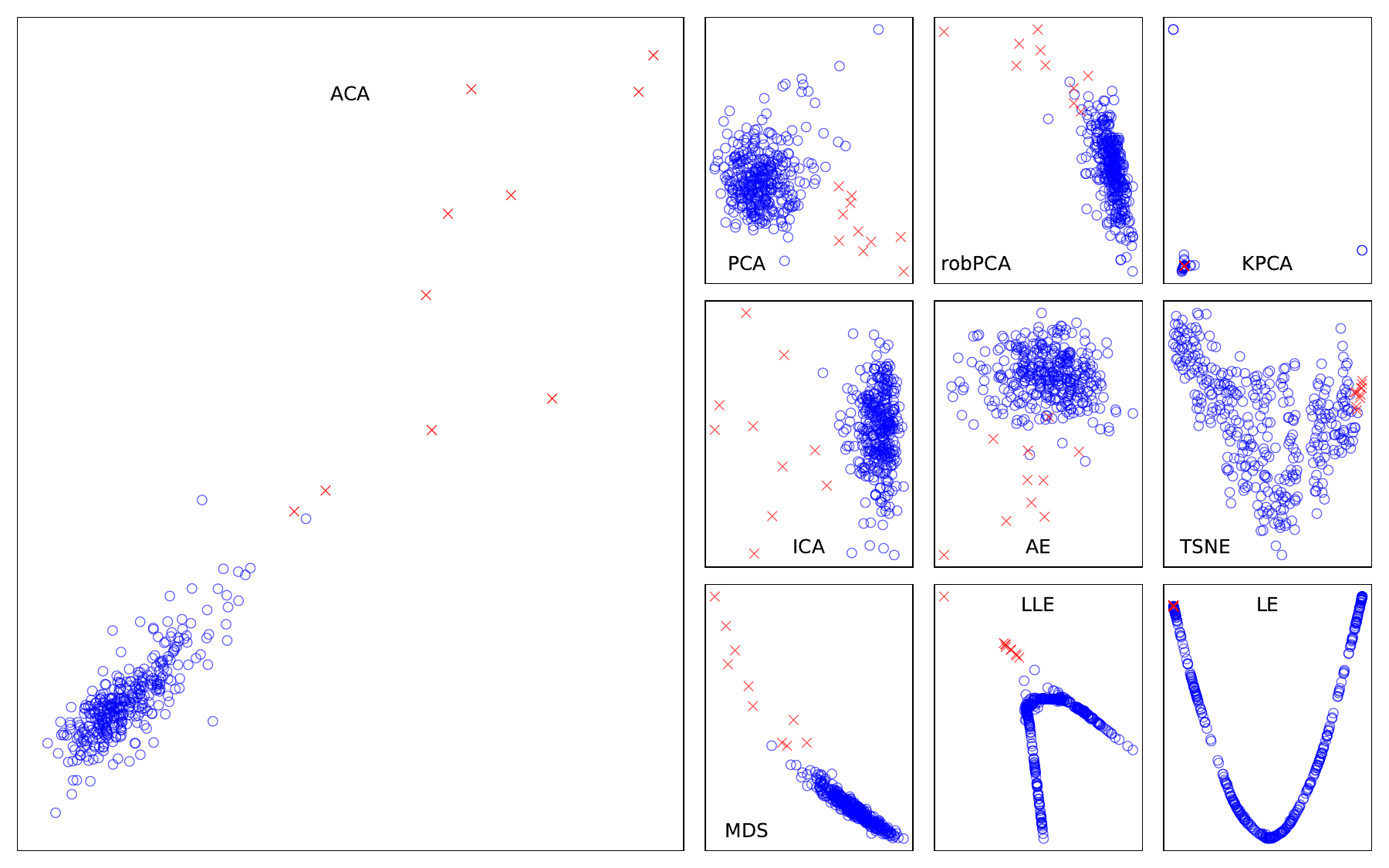}
\caption{Visualization of projection on two first components obtained by different methods for WDBC dataset.}
\end{figure}

\begin{table}[H]
\centering
\begin{tabular}{c|cccccccc|}
\cline{2-9}
 & \multicolumn{8}{c|}{WDBC} \\ \cline{2-9} 
 & \multicolumn{2}{c|}{PCA} & \multicolumn{2}{c|}{robPCA} & \multicolumn{2}{c|}{ICA} & \multicolumn{2}{c|}{ACA} \\ \cline{2-9} 
 & PC1 & \multicolumn{1}{c|}{PC2} & robPC1 & \multicolumn{1}{c|}{robPC2} & IC1 & \multicolumn{1}{c|}{IC2} & AC1 & AC2 \\ \hline
\multicolumn{1}{|c|}{Var1} & 8 (6\%) & \multicolumn{1}{c|}{10 (7\%)} & 13 (22\%) & \multicolumn{1}{c|}{6 (48\%)} & 4 (43\%) & \multicolumn{1}{c|}{4 (52\%)} & 4 (8\%) & 16 (8\%) \\
\multicolumn{1}{|c|}{Var2} & 7 (6\%) & \multicolumn{1}{c|}{1 (6\%)} & 12 (12\%) & \multicolumn{1}{c|}{3 (35\%)} & 24 (40\%) & \multicolumn{1}{c|}{24 (30\%)} & 24 (8\%) & 19 (8\%) \\
\multicolumn{1}{|c|}{Var3} & 28 (6\%) & \multicolumn{1}{c|}{4 (6\%)} & 9 (12\%) & \multicolumn{1}{c|}{8 (6\%)} & 14 (7\%) & \multicolumn{1}{c|}{14 (7\%)} & 6 (7\%) & 23 (7\%) \\
\multicolumn{1}{|c|}{Var4} & 6 (5\%) & \multicolumn{1}{c|}{21 (6\%)} & 15 (10\%) & \multicolumn{1}{c|}{7 (3\%)} & 3 (5\%) & \multicolumn{1}{c|}{3 (6\%)} & 13 (7\%) & 28 (6\%) \\
\multicolumn{1}{|c|}{Var5} & 27 (5\%) & \multicolumn{1}{c|}{3 (6\%)} & 14 (8\%) & \multicolumn{1}{c|}{9 (2\%)} & 22 (2\%) & \multicolumn{1}{c|}{22 (2\%)} & 16 (6\%) & 30 (6\%) \\
\hline
\end{tabular}
\caption{Most important variables for every method applied to WDBC dataset.}
\end{table}

\begin{figure}[H]
\centering
\includegraphics[width=\textwidth]{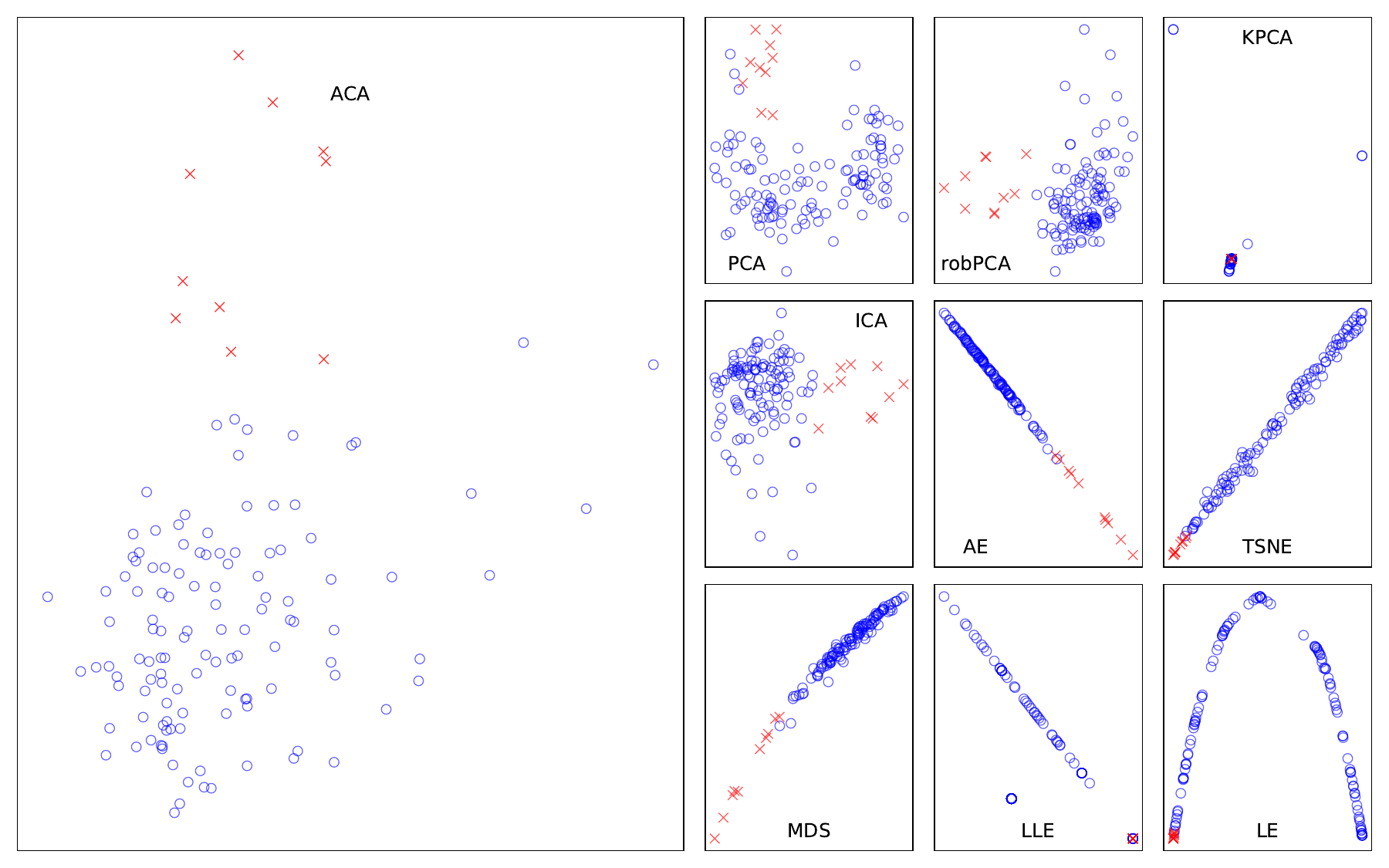}
\caption{Visualization of projection on two first components obtained by different methods for wine dataset.}
\end{figure}

\begin{table}[H]
\centering
\begin{tabular}{c|cccccccc|}
\cline{2-9}
 & \multicolumn{8}{c|}{wine} \\ \cline{2-9} 
 & \multicolumn{2}{c|}{PCA} & \multicolumn{2}{c|}{robPCA} & \multicolumn{2}{c|}{ICA} & \multicolumn{2}{c|}{ACA} \\ \cline{2-9} 
 & PC1 & \multicolumn{1}{c|}{PC2} & robPC1 & \multicolumn{1}{c|}{robPC2} & IC1 & \multicolumn{1}{c|}{IC2} & AC1 & AC2 \\ \hline
\multicolumn{1}{|c|}{Var1} & 12 (13\%) & \multicolumn{1}{c|}{13 (16\%)} & 7 (62\%) & \multicolumn{1}{c|}{5 (46\%)} & 5 (70\%) & \multicolumn{1}{c|}{5 (93\%)} & 5 (34\%) & 9 (16\%) \\
\multicolumn{1}{|c|}{Var2} & 7 (13\%) & \multicolumn{1}{c|}{5 (14\%)} & 4 (12\%) & \multicolumn{1}{c|}{6 (16\%)} & 13 (26\%) & \multicolumn{1}{c|}{4 (2\%)} & 4 (17\%) & 3 (14\%) \\
\multicolumn{1}{|c|}{Var3} & 11 (11\%) & \multicolumn{1}{c|}{1 (13\%)} & 9 (9\%) & \multicolumn{1}{c|}{4 (12\%)} & 4 (2\%) & \multicolumn{1}{c|}{13 (2\%)} & 2 (10\%) & 13 (14\%) \\
\multicolumn{1}{|c|}{Var4} & 6 (11\%) & \multicolumn{1}{c|}{3 (11\%)} & 5 (8\%) & \multicolumn{1}{c|}{7 (7\%)} & 10 (1\%) & \multicolumn{1}{c|}{10 (1\%)} & 11 (10\%) & 8 (13\%) \\
\multicolumn{1}{|c|}{Var5} & 10 (10\%) & \multicolumn{1}{c|}{10 (11\%)} & 12 (2\%) & \multicolumn{1}{c|}{3 (6\%)} & 9 (1\%) & \multicolumn{1}{c|}{9 (1\%)} & 10 (7\%) & 12 (12\%) \\
\hline
\end{tabular}
\caption{Most important variables for every method applied to Wine dataset.}
\end{table}

\begin{figure}[!h]
\centering
\includegraphics[width=\textwidth]{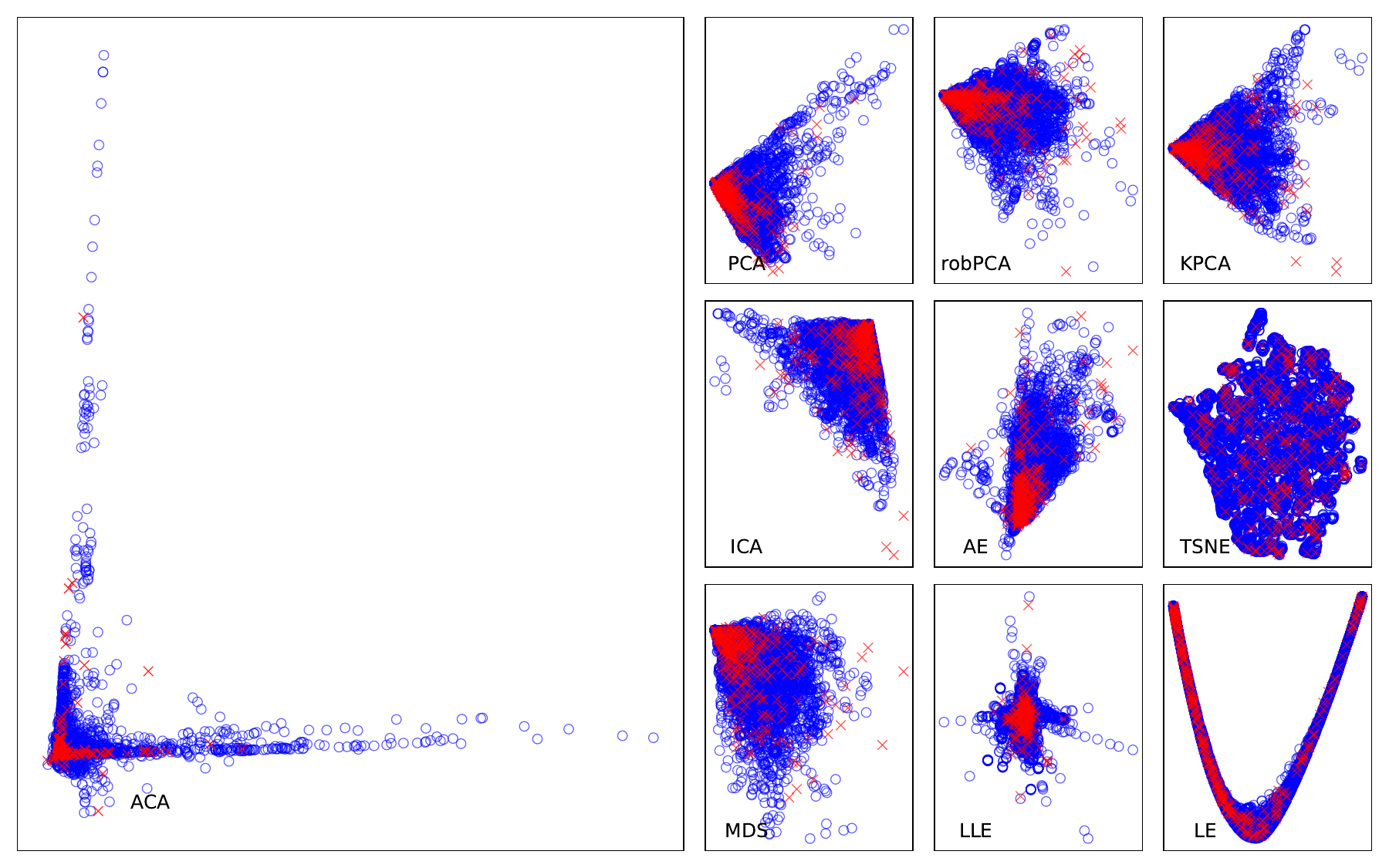}
\caption{Visualization of projection on two first components obtained by different methods for ALOI dataset.}\label{fig:realALOI}
\end{figure}

\begin{table}[H]
\centering
\begin{tabular}{c|cccccccc|}
\cline{2-9}
& \multicolumn{8}{c|}{ALOI}                                                                                                                                    \\ \cline{2-9} 
& \multicolumn{2}{c|}{PCA}                 & \multicolumn{2}{c|}{robPCA}               & \multicolumn{2}{c|}{ICA}                   & \multicolumn{2}{c|}{ACA} \\ \cline{2-9} 
& PC1      & \multicolumn{1}{c|}{PC2}      & robPC1   & \multicolumn{1}{c|}{robPC2}    & IC1       & \multicolumn{1}{c|}{IC2}       & AC1        & AC2         \\ \hline
\multicolumn{1}{|c|}{Var1}  & 1 (10\%) & \multicolumn{1}{c|}{1 (8\%)}  & 1 (40\%) & \multicolumn{1}{c|}{11 (29\%)} & 27 (30\%) & \multicolumn{1}{c|}{10 (22\%)} & 2 (39\%)   & 4 (30\%)    \\
\multicolumn{1}{|c|}{Var2}  & 5 (7\%)  & \multicolumn{1}{c|}{13 (8\%)} & 22 (7\%) & \multicolumn{1}{c|}{12 (20\%)} & 10 (20\%) & \multicolumn{1}{c|}{27 (22\%)} & 14 (5\%)   & 18 (12\%)   \\
\multicolumn{1}{|c|}{Var3}  & 15 (6\%) & \multicolumn{1}{c|}{26 (7\%)} & 11 (6\%) & \multicolumn{1}{c|}{14 (13\%)} & 1 (10\%)  & \multicolumn{1}{c|}{1 (18\%)}  & 12 (4\%)   & 12 (8\%)    \\
\multicolumn{1}{|c|}{Var4}  & 14 (6\%) & \multicolumn{1}{c|}{5 (7\%)}  & 24 (5\%) & \multicolumn{1}{c|}{13 (9\%)}  & 14 (10\%) & \multicolumn{1}{c|}{19 (10\%)} & 26 (4\%)   & 14 (4\%)    \\
\multicolumn{1}{|c|}{Var5}  & 18 (6\%) & \multicolumn{1}{c|}{22 (6\%)} & 10 (4\%) & \multicolumn{1}{c|}{22 (6\%)}  & 19 (9\%)  & \multicolumn{1}{c|}{22 (9\%)}  & 10 (4\%)   & 5 (4\%)     \\
\hline
\end{tabular}
\caption{Most important variables for every method applied to ALOI dataset.}
\end{table}

\begin{figure}[!h]
\centering
\includegraphics[width=\textwidth]{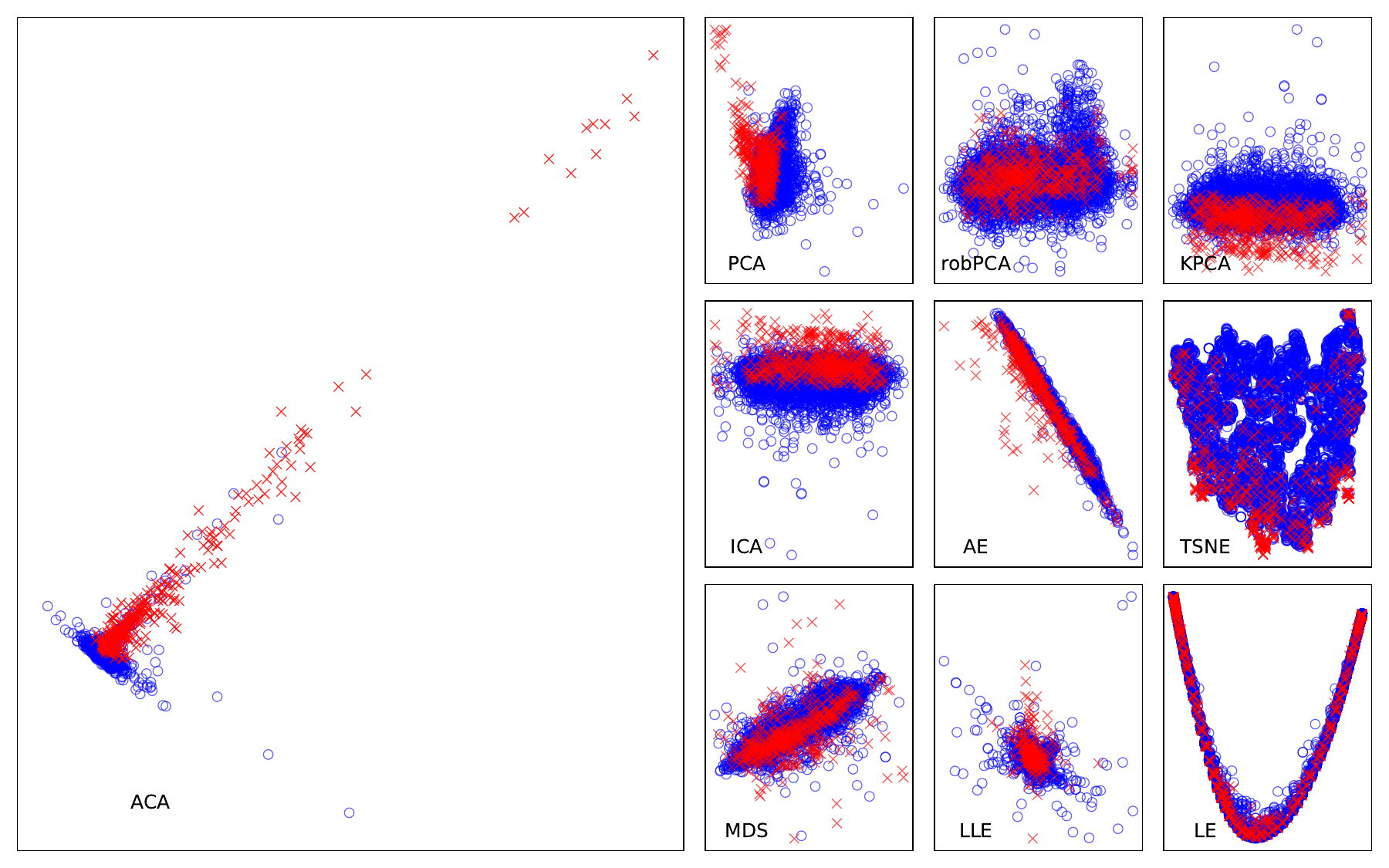}
\caption{Visualization of projection on two first components obtained by different methods for annthyroid dataset.}\label{fig:realannthyroid}
\end{figure}

\begin{table}[H]
\centering
\begin{tabular}{c|cccccccc|}
\cline{2-9}
& \multicolumn{8}{c|}{Annthyroid}                                                                                            \\ \cline{2-9} 
& \multicolumn{2}{c|}{PCA}                 & \multicolumn{2}{c|}{robPCA}              & \multicolumn{2}{c|}{ICA}                 & \multicolumn{2}{c|}{ACA} \\ \cline{2-9} 
& PC1      & \multicolumn{1}{c|}{PC2}      & robPC1   & \multicolumn{1}{c|}{robPC2}   & IC1      & \multicolumn{1}{c|}{IC2}      & AC1         & AC2        \\ \hline
\multicolumn{1}{|c|}{Var1} & 4 (28\%) & \multicolumn{1}{c|}{5 (31\%)} & 1 (99\%) & \multicolumn{1}{c|}{5 (94\%)} & 1 (87\%) & \multicolumn{1}{c|}{4 (42\%)} & 2 (32\%)    & 2 (37\%)   \\
\multicolumn{1}{|c|}{Var2} & 6 (24\%) & \multicolumn{1}{c|}{1 (22\%)} & 3 (1\%)  & \multicolumn{1}{c|}{6 (2\%)}  & 6 (6\%)  & \multicolumn{1}{c|}{6 (42\%)} & 5 (23\%)    & 5 (22\%)   \\
\multicolumn{1}{|c|}{Var3} & 3 (21\%) & \multicolumn{1}{c|}{6 (19\%)} & 4 (1\%)  & \multicolumn{1}{c|}{4 (2\%)}  & 4 (5\%)  & \multicolumn{1}{c|}{2 (10\%)} & 4 (22\%)    & 4 (17\%)   \\
\multicolumn{1}{|c|}{Var4} & 2 (13\%) & \multicolumn{1}{c|}{2 (16\%)} & 2 (0\%)  & \multicolumn{1}{c|}{2 (1\%)}  & 2 (2\%)  & \multicolumn{1}{c|}{3 (4\%)}  & 6 (21\%)    & 6 (16\%)   \\
\multicolumn{1}{|c|}{Var5} & 5 (9\%)  & \multicolumn{1}{c|}{3 (12\%)} & 5 (0\%)  & \multicolumn{1}{c|}{3 (1\%)}  & 5 (1\%)  & \multicolumn{1}{c|}{5 (2\%)}  & 3 (2\%)     & 3 (7\%)    \\
\multicolumn{1}{|c|}{Var6} & 1 (5\%)  & \multicolumn{1}{c|}{4 (0\%)}  & 6 (0\%)  & \multicolumn{1}{c|}{1 (0\%)}  & 3 (0\%)  & \multicolumn{1}{c|}{1 (0\%)}  & 1 (0\%)     & 1 (0\%)    \\ \hline
\end{tabular}
\caption{Most important variables for every method applied to Annthyroid dataset.}
\end{table}

\begin{figure}[!h]
\centering
\includegraphics[width=\textwidth]{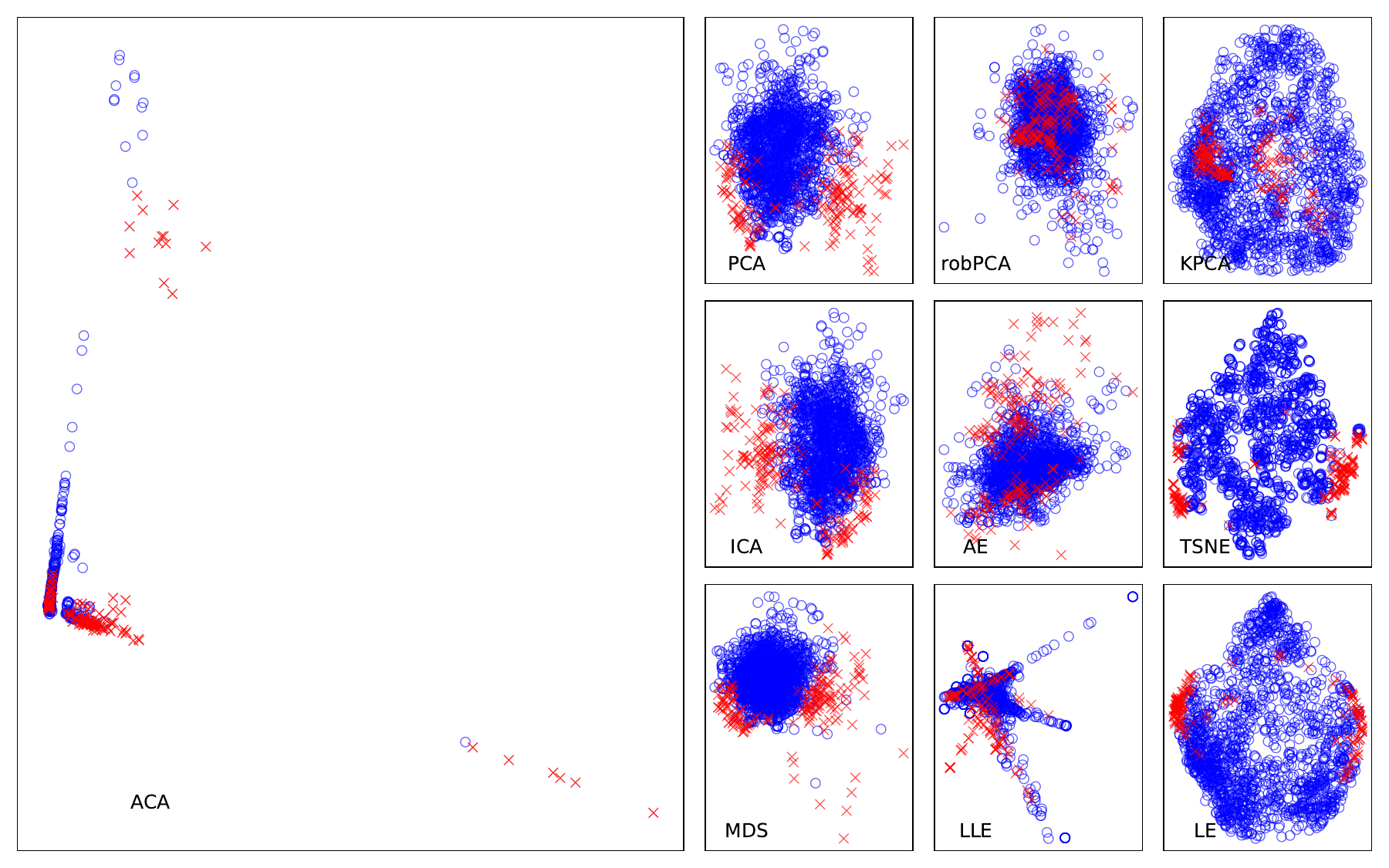}
\caption{Visualization of projection on two first components obtained by different methods for cardio dataset.}\label{fig:realcardio}
\end{figure}

\begin{table}[H]
\centering
\begin{tabular}{c|cccccccc|}
\cline{2-9}
& \multicolumn{8}{c|}{Cardio}                                                                                    \\ \cline{2-9} 
& \multicolumn{2}{c|}{PCA}                  & \multicolumn{2}{c|}{robPCA}               & \multicolumn{2}{c|}{ICA}                   & \multicolumn{2}{c|}{ACA} \\ \cline{2-9} 
& PC1      & \multicolumn{1}{c|}{PC2}       & robPC1    & \multicolumn{1}{c|}{robPC2}   & IC1       & \multicolumn{1}{c|}{IC2}       & AC1         & AC2        \\ \hline
\multicolumn{1}{|c|}{Var1}  & 13 (8\%) & \multicolumn{1}{c|}{14 (10\%)} & 1 (18\%)  & \multicolumn{1}{c|}{4 (19\%)} & 18 (12\%) & \multicolumn{1}{c|}{14 (10\%)} & 6 (56\%)    & 3 (60\%)   \\
\multicolumn{1}{|c|}{Var2}  & 18 (8\%) & \multicolumn{1}{c|}{19 (9\%)}  & 14 (18\%) & \multicolumn{1}{c|}{9 (13\%)} & 19 (11\%) & \multicolumn{1}{c|}{12 (10\%)} & 7 (22\%)    & 6 (12\%)   \\
\multicolumn{1}{|c|}{Var3}  & 12 (8\%) & \multicolumn{1}{c|}{17 (9\%)}  & 12 (15\%) & \multicolumn{1}{c|}{3 (10\%)} & 17 (11\%) & \multicolumn{1}{c|}{15 (8\%)}  & 3 (13\%)    & 7 (6\%)    \\
\multicolumn{1}{|c|}{Var4}  & 9 (8\%)  & \multicolumn{1}{c|}{18 (8\%)}  & 13 (10\%) & \multicolumn{1}{c|}{2 (9\%)}  & 1 (8\%)   & \multicolumn{1}{c|}{2 (7\%)}   & 18 (3\%)    & 13 (5\%)   \\
\multicolumn{1}{|c|}{Var5}  & 20 (7\%) & \multicolumn{1}{c|}{2 (8\%)}   & 4 (7\%)   & \multicolumn{1}{c|}{7 (9\%)}  & 7 (6\%)   & \multicolumn{1}{c|}{13 (7\%)}  & 19 (3\%)    & 12 (4\%)   \\
\hline
\end{tabular}
\caption{Most important variables for every method applied to Cardio dataset.}
\end{table}

\begin{figure}[!h]
\centering
\includegraphics[width=\textwidth]{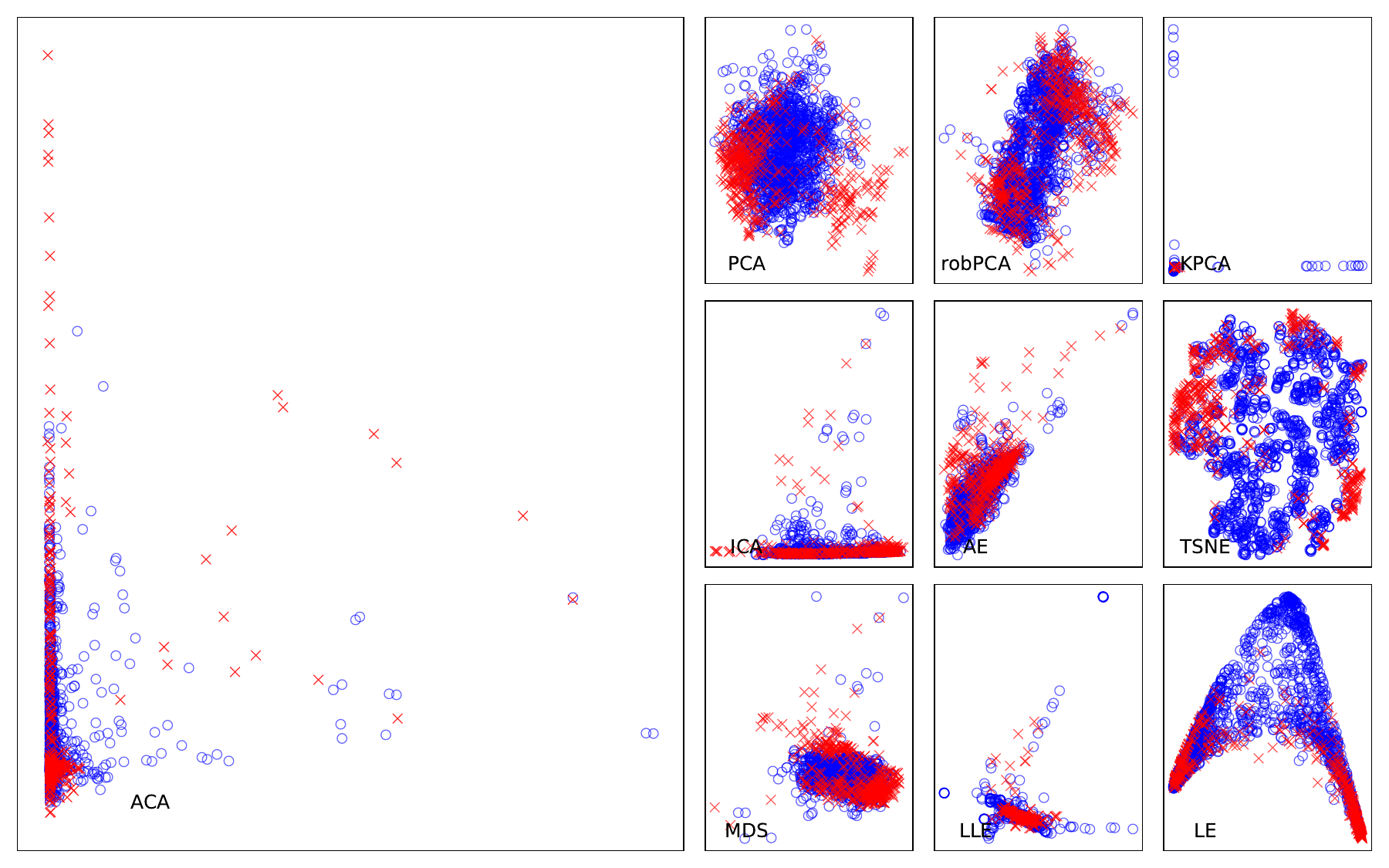}
\caption{Visualization of projection on two first components obtained by different methods for Cardiotocography dataset.}\label{fig:realCardiotocography}
\end{figure}

\begin{table}[H]
\centering
\begin{tabular}{c|cccccccc|}
\cline{2-9}
& \multicolumn{8}{c|}{Cardiotocography}                                                                                      \\ \cline{2-9} 
& \multicolumn{2}{c|}{PCA}                  & \multicolumn{2}{c|}{robPCA}                & \multicolumn{2}{c|}{ICA}                  & \multicolumn{2}{c|}{ACA} \\ \cline{2-9} 
& PC1      & \multicolumn{1}{c|}{PC2}       & robPC1    & \multicolumn{1}{c|}{robPC2}    & IC1       & \multicolumn{1}{c|}{IC2}      & AC1        & AC2         \\ \hline
\multicolumn{1}{|c|}{Var1}  & 13 (8\%) & \multicolumn{1}{c|}{19 (10\%)} & 9 (21\%)  & \multicolumn{1}{c|}{11 (26\%)} & 12 (25\%) & \multicolumn{1}{c|}{3 (83\%)} & 3 (65\%)   & 20 (36\%)   \\
\multicolumn{1}{|c|}{Var2}  & 12 (8\%) & \multicolumn{1}{c|}{14 (10\%)} & 7 (12\%)  & \multicolumn{1}{c|}{13 (15\%)} & 13 (18\%) & \multicolumn{1}{c|}{20 (3\%)} & 9 (6\%)    & 21 (6\%)    \\
\multicolumn{1}{|c|}{Var3}  & 6 (8\%)  & \multicolumn{1}{c|}{17 (9\%)}  & 10 (12\%) & \multicolumn{1}{c|}{7 (14\%)}  & 20 (14\%) & \multicolumn{1}{c|}{5 (3\%)}  & 6 (6\%)    & 16 (6\%)    \\
\multicolumn{1}{|c|}{Var4}  & 18 (8\%) & \multicolumn{1}{c|}{18 (8\%)}  & 2 (11\%)  & \multicolumn{1}{c|}{6 (11\%)}  & 3 (8\%)   & \multicolumn{1}{c|}{13 (2\%)} & 14 (3\%)   & 12 (6\%)    \\
\multicolumn{1}{|c|}{Var5}  & 20 (7\%) & \multicolumn{1}{c|}{1 (7\%)}   & 5 (10\%)  & \multicolumn{1}{c|}{10 (10\%)} & 14 (7\%)  & \multicolumn{1}{c|}{12 (2\%)} & 12 (3\%)   & 8 (6\%)     \\
\hline
\end{tabular}
\caption{Most important variables for every method applied to Cardiotocography dataset.}
\end{table}

\begin{figure}[!h]
\centering
\includegraphics[width=\textwidth]{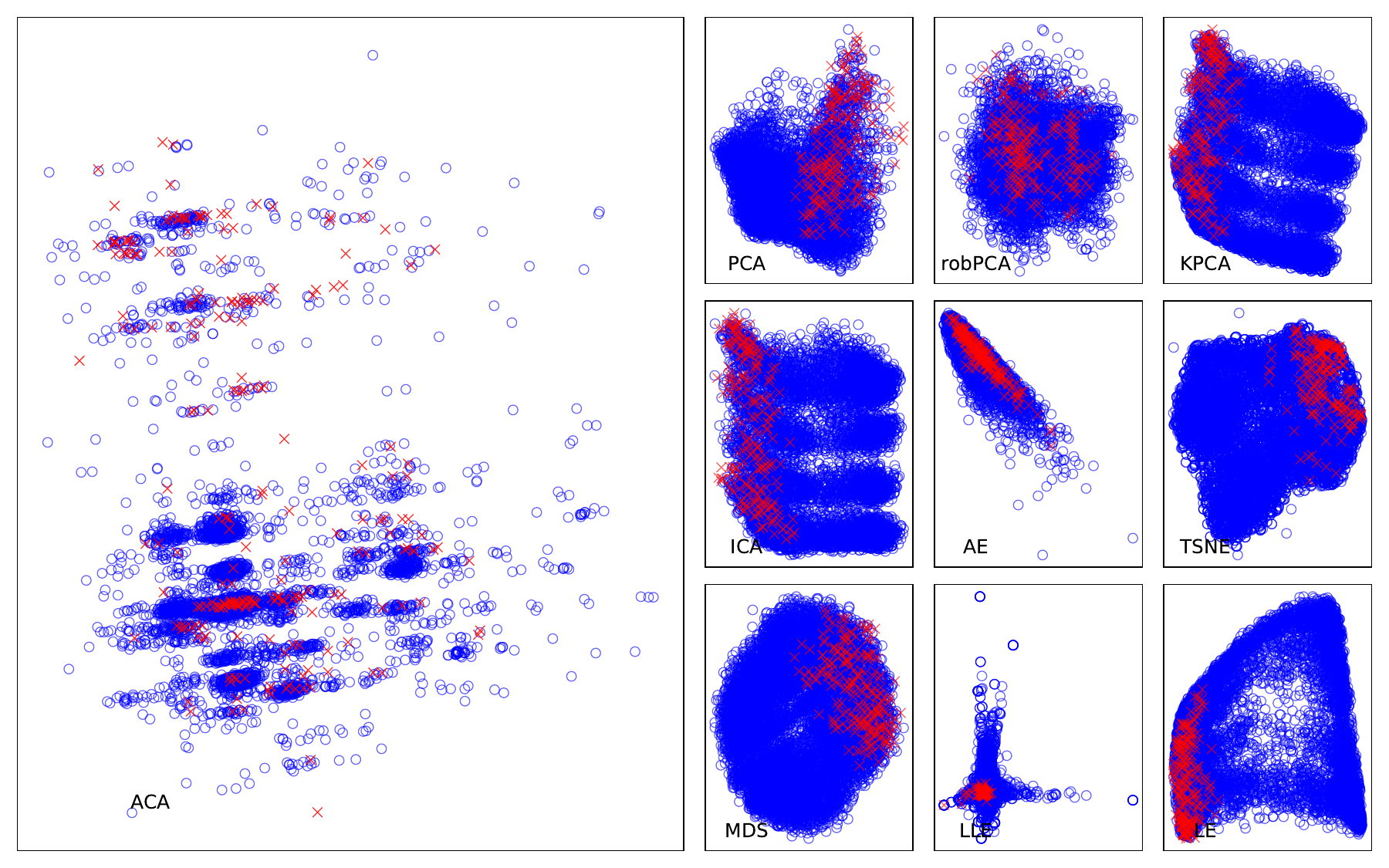}
\caption{Visualization of projection on two first components obtained by different methods for celeba dataset.}\label{fig:realceleba}
\end{figure}

\begin{table}[H]
\centering
\begin{tabular}{c|cccccccc|}
\cline{2-9}
& \multicolumn{8}{c|}{Celeba}                                                                                                \\ \cline{2-9} 
& \multicolumn{2}{c|}{PCA}                 & \multicolumn{2}{c|}{robPCA}                 & \multicolumn{2}{c|}{ICA}                  & \multicolumn{2}{c|}{ACA} \\ \cline{2-9} 
& PC1      & \multicolumn{1}{c|}{PC2}      & robPC1     & \multicolumn{1}{c|}{robPC2}    & IC1      & \multicolumn{1}{c|}{IC2}       & AC1         & AC2        \\ \hline
\multicolumn{1}{|c|}{Var1}  & 20 (6\%) & \multicolumn{1}{c|}{31 (8\%)} & 37 (100\%) & \multicolumn{1}{c|}{9 (13\%)}  & 36 (9\%) & \multicolumn{1}{c|}{31 (18\%)} & 15 (25\%)   & 35 (24\%)  \\
\multicolumn{1}{|c|}{Var2}  & 36 (6\%) & \multicolumn{1}{c|}{19 (8\%)} & 39 (0\%)   & \multicolumn{1}{c|}{12 (11\%)} & 20 (9\%) & \multicolumn{1}{c|}{19 (17\%)} & 13 (17\%)   & 22 (17\%)  \\
\multicolumn{1}{|c|}{Var3}  & 18 (6\%) & \multicolumn{1}{c|}{21 (7\%)} & 10 (0\%)   & \multicolumn{1}{c|}{13 (9\%)}  & 18 (8\%) & \multicolumn{1}{c|}{21 (16\%)} & 18 (11\%)   & 30 (15\%)  \\
\multicolumn{1}{|c|}{Var4}  & 24 (5\%) & \multicolumn{1}{c|}{14 (7\%)} & 17 (0\%)   & \multicolumn{1}{c|}{4 (9\%)}   & 3 (6\%)  & \multicolumn{1}{c|}{7 (6\%)}   & 14 (11\%)   & 17 (11\%)  \\
\multicolumn{1}{|c|}{Var5}  & 3 (4\%)  & \multicolumn{1}{c|}{13 (6\%)} & 16 (0\%)   & \multicolumn{1}{c|}{5 (8\%)}   & 7 (5\%)  & \multicolumn{1}{c|}{4 (5\%)}   & 36 (11\%)   & 16 (9\%)   \\
\hline
\end{tabular}
\caption{Most important variables for every method applied to Celeba dataset.}
\end{table}

\begin{figure}[!h]
\centering
\includegraphics[width=\textwidth]{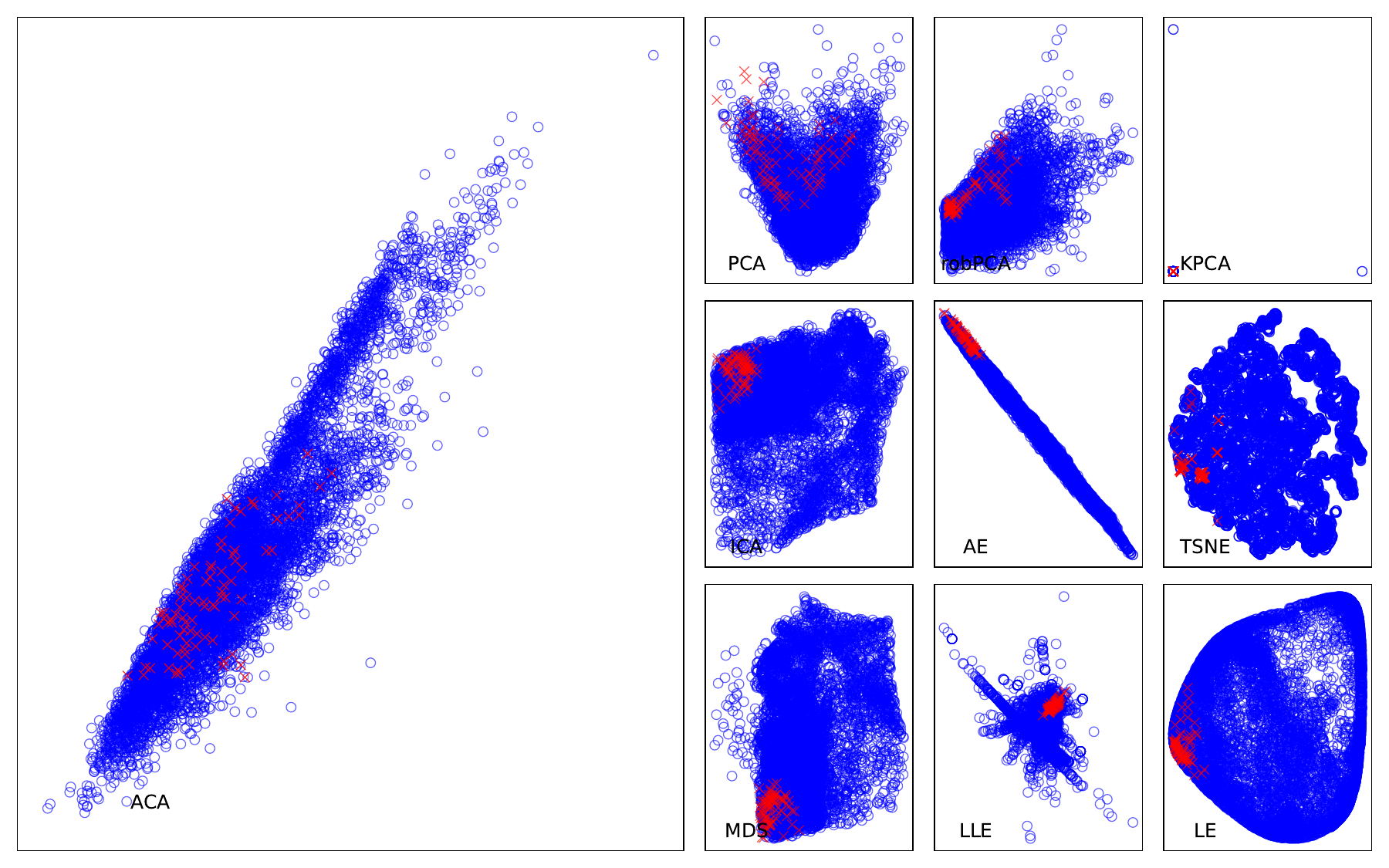}
\caption{Visualization of projection on two first components obtained by different methods for cover dataset.}\label{fig:realcover}
\end{figure}

\begin{table}[H]
\centering
\begin{tabular}{c|cccccccc|}
\cline{2-9}
& \multicolumn{8}{c|}{Cover}                                                                                                 \\ \cline{2-9} 
& \multicolumn{2}{c|}{PCA}                  & \multicolumn{2}{c|}{robPCA}              & \multicolumn{2}{c|}{ICA}                  & \multicolumn{2}{c|}{ACA} \\ \cline{2-9} 
& PC1      & \multicolumn{1}{c|}{PC2}       & robPC1   & \multicolumn{1}{c|}{robPC2}   & IC1      & \multicolumn{1}{c|}{IC2}       & AC1         & AC2        \\ \hline
\multicolumn{1}{|c|}{Var1}  & 9 (22\%) & \multicolumn{1}{c|}{3 (20\%)}  & 4 (50\%) & \multicolumn{1}{c|}{5 (67\%)} & 6 (92\%) & \multicolumn{1}{c|}{10 (73\%)} & 2 (32\%)    & 7 (48\%)   \\
\multicolumn{1}{|c|}{Var2}  & 2 (18\%) & \multicolumn{1}{c|}{6 (14\%)}  & 3 (41\%) & \multicolumn{1}{c|}{8 (14\%)} & 1 (4\%)  & \multicolumn{1}{c|}{6 (23\%)}  & 9 (16\%)    & 8 (30\%)   \\
\multicolumn{1}{|c|}{Var3}  & 7 (17\%) & \multicolumn{1}{c|}{10 (14\%)} & 5 (3\%)  & \multicolumn{1}{c|}{9 (6\%)}  & 4 (1\%)  & \multicolumn{1}{c|}{1 (2\%)}   & 5 (12\%)    & 3 (12\%)   \\
\multicolumn{1}{|c|}{Var4}  & 8 (15\%) & \multicolumn{1}{c|}{7 (13\%)}  & 2 (3\%)  & \multicolumn{1}{c|}{7 (4\%)}  & 10 (1\%) & \multicolumn{1}{c|}{2 (1\%)}   & 3 (8\%)     & 5 (3\%)    \\
\multicolumn{1}{|c|}{Var5}  & 1 (8\%)  & \multicolumn{1}{c|}{8 (13\%)}  & 1 (2\%)  & \multicolumn{1}{c|}{6 (4\%)}  & 2 (1\%)  & \multicolumn{1}{c|}{4 (1\%)}   & 8 (8\%)     & 2 (3\%)    \\
\hline
\end{tabular}
\caption{Most important variables for every method applied to Cover dataset.}
\end{table}

\begin{figure}[!h]
\centering
\includegraphics[width=\textwidth]{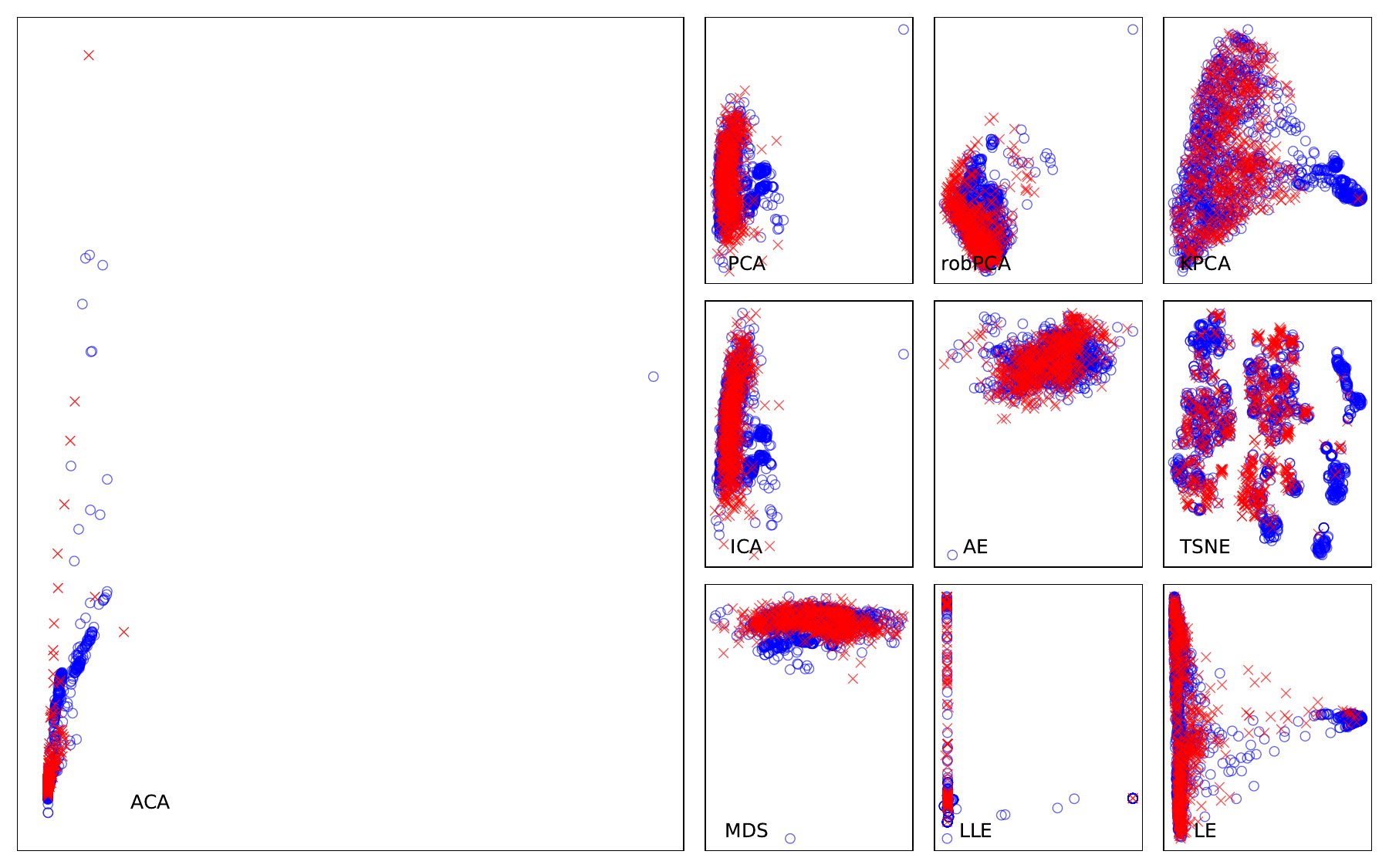}
\caption{Visualization of projection on two first components obtained by different methods for fault dataset.}\label{fig:realfault}
\end{figure}

\begin{table}[H]
\centering
\begin{tabular}{c|cccccccc|}
\cline{2-9}
& \multicolumn{8}{c|}{Fault}                                                                                                 \\ \cline{2-9} 
& \multicolumn{2}{c|}{PCA}                  & \multicolumn{2}{c|}{robPCA}                & \multicolumn{2}{c|}{ICA}                  & \multicolumn{2}{c|}{ACA} \\ \cline{2-9} 
& PC1      & \multicolumn{1}{c|}{PC2}       & robPC1    & \multicolumn{1}{c|}{robPC2}    & IC1      & \multicolumn{1}{c|}{IC2}       & AC1        & AC2         \\ \hline
\multicolumn{1}{|c|}{Var1}  & 22 (7\%) & \multicolumn{1}{c|}{25 (11\%)} & 25 (29\%) & \multicolumn{1}{c|}{25 (27\%)} & 24 (8\%) & \multicolumn{1}{c|}{25 (11\%)} & 7 (24\%)   & 18 (37\%)   \\
\multicolumn{1}{|c|}{Var2}  & 23 (7\%) & \multicolumn{1}{c|}{21 (10\%)} & 4 (22\%)  & \multicolumn{1}{c|}{4 (23\%)}  & 22 (7\%) & \multicolumn{1}{c|}{21 (10\%)} & 6 (24\%)   & 12 (9\%)    \\
\multicolumn{1}{|c|}{Var3}  & 8 (7\%)  & \multicolumn{1}{c|}{19 (9\%)}  & 1 (10\%)  & \multicolumn{1}{c|}{6 (10\%)}  & 27 (7\%) & \multicolumn{1}{c|}{19 (8\%)}  & 5 (18\%)   & 13 (9\%)    \\
\multicolumn{1}{|c|}{Var4}  & 5 (6\%)  & \multicolumn{1}{c|}{26 (8\%)}  & 6 (10\%)  & \multicolumn{1}{c|}{1 (9\%)}   & 8 (6\%)  & \multicolumn{1}{c|}{26 (7\%)}  & 8 (12\%)   & 27 (7\%)    \\
\multicolumn{1}{|c|}{Var5}  & 24 (6\%) & \multicolumn{1}{c|}{10 (7\%)}  & 9 (4\%)   & \multicolumn{1}{c|}{9 (5\%)}   & 5 (6\%)  & \multicolumn{1}{c|}{20 (7\%)}  & 1 (4\%)    & 6 (6\%)     \\
\hline
\end{tabular}
\caption{Most important variables for every method applied to Fault dataset.}
\end{table}

\begin{figure}[!h]
\centering
\includegraphics[width=\textwidth]{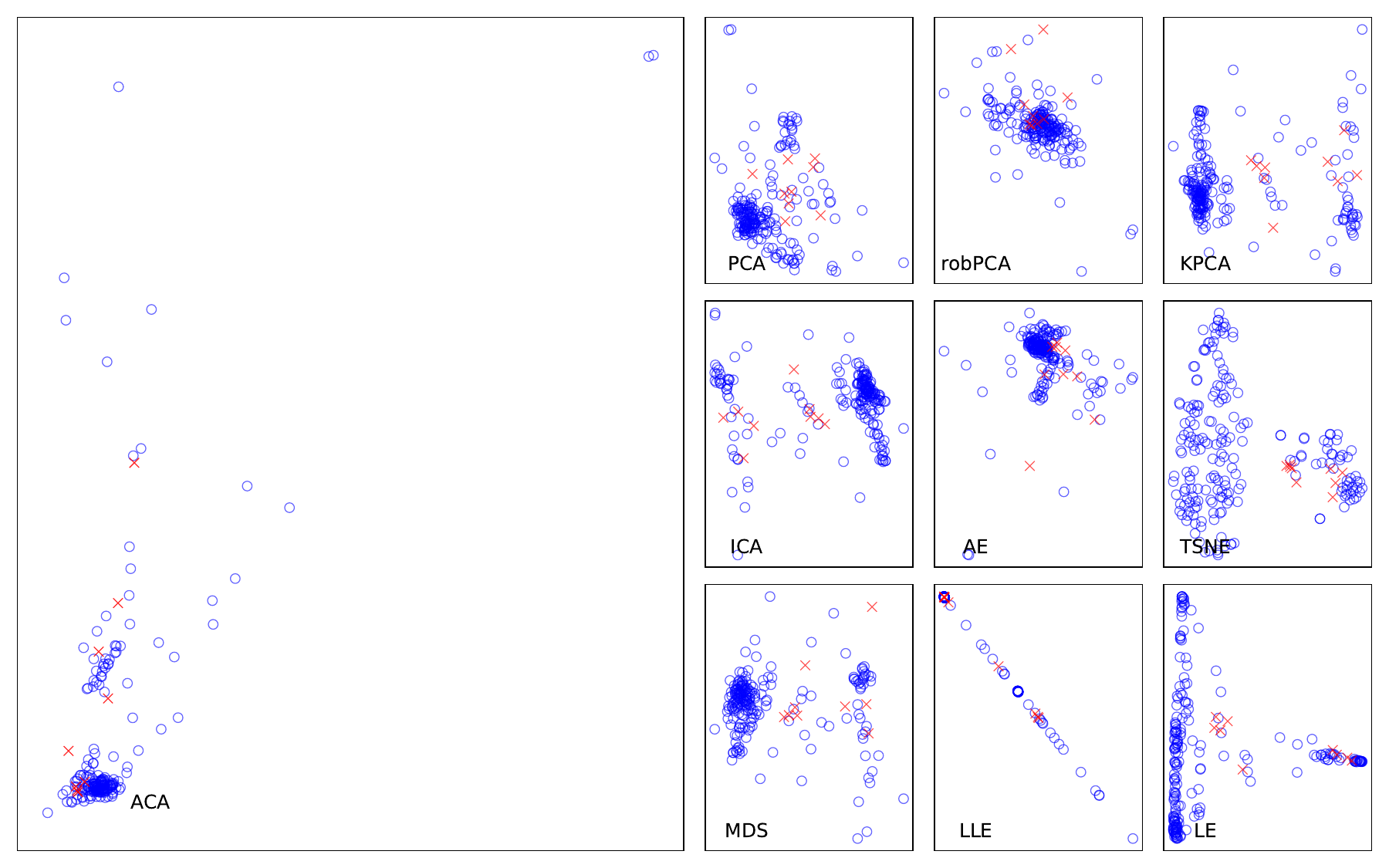}
\caption{Visualization of projection on two first components obtained by different methods for glass dataset.}\label{fig:realglass}
\end{figure}

\begin{table}[H]
\centering
\begin{tabular}{c|cccccccc|}
\cline{2-9}
& \multicolumn{8}{c|}{Glass}                                                                                                 \\ \cline{2-9} 
& \multicolumn{2}{c|}{PCA}                 & \multicolumn{2}{c|}{robPCA}              & \multicolumn{2}{c|}{ICA}                 & \multicolumn{2}{c|}{ACA} \\ \cline{2-9} 
& PC1      & \multicolumn{1}{c|}{PC2}      & robPC1   & \multicolumn{1}{c|}{robPC2}   & IC1      & \multicolumn{1}{c|}{IC2}      & AC1         & AC2        \\ \hline
\multicolumn{1}{|c|}{Var1} & 5 (28\%) & \multicolumn{1}{c|}{4 (31\%)} & 4 (26\%) & \multicolumn{1}{c|}{1 (33\%)} & 3 (55\%) & \multicolumn{1}{c|}{5 (25\%)} & 6 (58\%)    & 5 (34\%)   \\
\multicolumn{1}{|c|}{Var2} & 7 (23\%) & \multicolumn{1}{c|}{3 (23\%)} & 1 (24\%) & \multicolumn{1}{c|}{4 (32\%)} & 4 (19\%) & \multicolumn{1}{c|}{7 (24\%)} & 1 (11\%)    & 7 (31\%)   \\
\multicolumn{1}{|c|}{Var3} & 3 (17\%) & \multicolumn{1}{c|}{2 (14\%)} & 6 (23\%) & \multicolumn{1}{c|}{7 (25\%)} & 1 (8\%)  & \multicolumn{1}{c|}{4 (24\%)} & 2 (9\%)     & 2 (24\%)   \\
\multicolumn{1}{|c|}{Var4} & 6 (12\%) & \multicolumn{1}{c|}{6 (14\%)} & 5 (20\%) & \multicolumn{1}{c|}{2 (6\%)}  & 5 (7\%)  & \multicolumn{1}{c|}{1 (12\%)} & 3 (9\%)     & 3 (5\%)    \\
\multicolumn{1}{|c|}{Var5} & 2 (10\%) & \multicolumn{1}{c|}{7 (11\%)} & 2 (6\%)  & \multicolumn{1}{c|}{3 (2\%)}  & 2 (6\%)  & \multicolumn{1}{c|}{6 (12\%)} & 4 (7\%)     & 6 (4\%)    \\
\multicolumn{1}{|c|}{Var6} & 1 (8\%)  & \multicolumn{1}{c|}{1 (7\%)}  & 3 (1\%)  & \multicolumn{1}{c|}{5 (2\%)}  & 7 (3\%)  & \multicolumn{1}{c|}{2 (2\%)}  & 5 (3\%)     & 4 (2\%)    \\
\multicolumn{1}{|c|}{Var7} & 4 (2\%)  & \multicolumn{1}{c|}{5 (1\%)}  & 7 (0\%)  & \multicolumn{1}{c|}{6 (1\%)}  & 6 (3\%)  & \multicolumn{1}{c|}{3 (2\%)}  & 7 (2\%)     & 1 (1\%)    \\ \hline
\end{tabular}
\caption{Most important variables for every method applied to Glass dataset.}
\end{table}

\begin{figure}[!h]
\centering
\includegraphics[width=\textwidth]{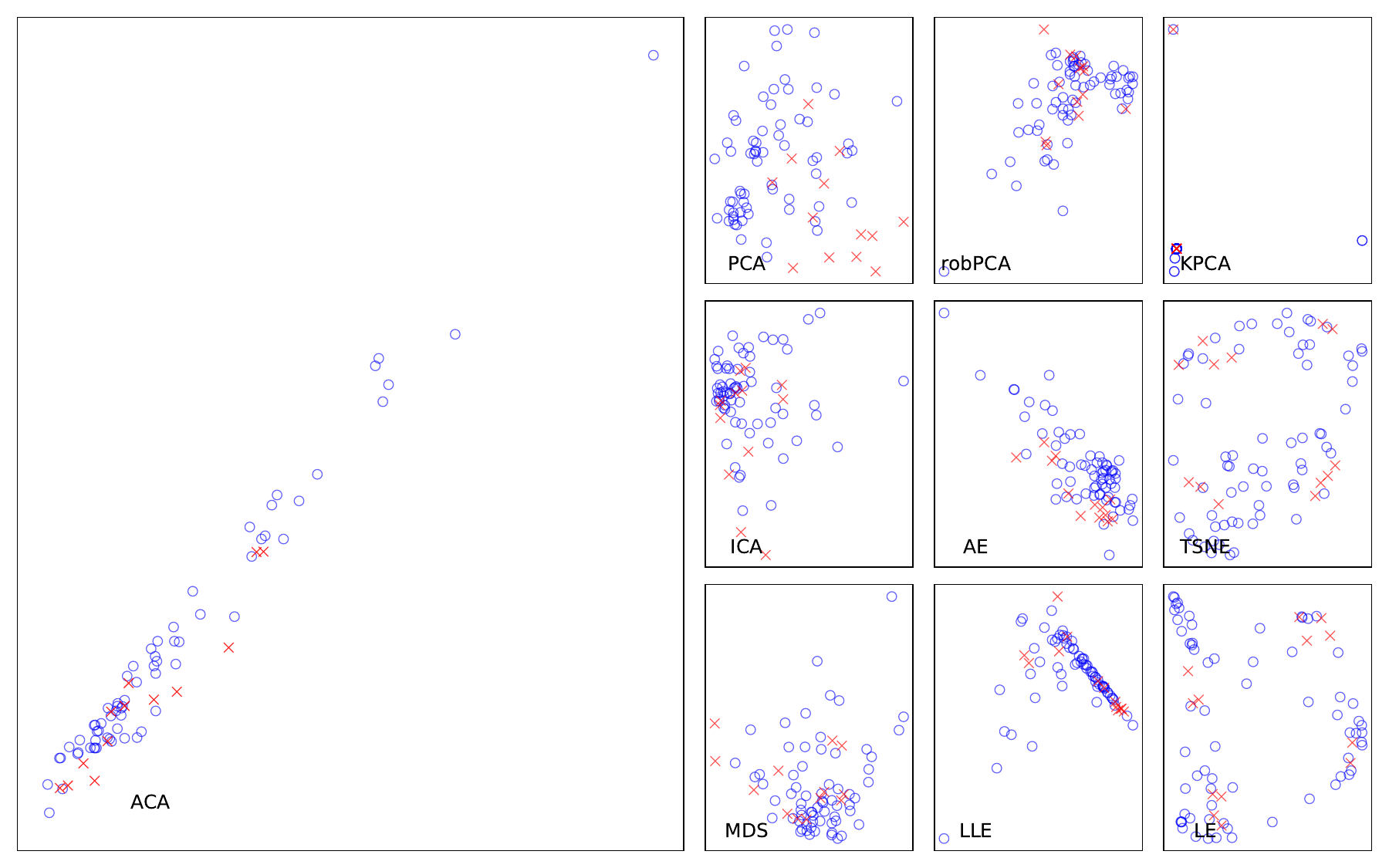}
\caption{Visualization of projection on two first components obtained by different methods for Hepatitis dataset.}\label{fig:realHepatitis}
\end{figure}

\begin{table}[H]
\centering
\begin{tabular}{c|cccccccc|}
\cline{2-9}
& \multicolumn{8}{c|}{Hepatitis}                                                                                             \\ \cline{2-9} 
& \multicolumn{2}{c|}{PCA}                 & \multicolumn{2}{c|}{robPCA}               & \multicolumn{2}{c|}{ICA}                   & \multicolumn{2}{c|}{ACA} \\ \cline{2-9} 
& PC1      & \multicolumn{1}{c|}{PC2}      & robPC1   & \multicolumn{1}{c|}{robPC2}    & IC1       & \multicolumn{1}{c|}{IC2}       & AC1         & AC2        \\ \hline
\multicolumn{1}{|c|}{Var1}  & 17 (9\%) & \multicolumn{1}{c|}{7 (11\%)} & 4 (79\%) & \multicolumn{1}{c|}{16 (25\%)} & 16 (83\%) & \multicolumn{1}{c|}{15 (67\%)} & 10 (18\%)   & 16 (26\%)  \\
\multicolumn{1}{|c|}{Var2}  & 11 (8\%) & \multicolumn{1}{c|}{8 (11\%)} & 3 (10\%) & \multicolumn{1}{c|}{17 (20\%)} & 15 (11\%) & \multicolumn{1}{c|}{16 (22\%)} & 7 (13\%)    & 18 (13\%)  \\
\multicolumn{1}{|c|}{Var3}  & 14 (8\%) & \multicolumn{1}{c|}{9 (8\%)}  & 2 (5\%)  & \multicolumn{1}{c|}{15 (13\%)} & 18 (3\%)  & \multicolumn{1}{c|}{18 (6\%)}  & 13 (11\%)   & 17 (9\%)   \\
\multicolumn{1}{|c|}{Var4}  & 12 (8\%) & \multicolumn{1}{c|}{6 (8\%)}  & 9 (1\%)  & \multicolumn{1}{c|}{18 (8\%)}  & 1 (2\%)   & \multicolumn{1}{c|}{1 (3\%)}   & 19 (8\%)    & 9 (8\%)    \\
\multicolumn{1}{|c|}{Var5}  & 19 (7\%) & \multicolumn{1}{c|}{5 (7\%)}  & 12 (1\%) & \multicolumn{1}{c|}{14 (7\%)}  & 14 (0\%)  & \multicolumn{1}{c|}{17 (0\%)}  & 16 (7\%)    & 1 (8\%)    \\
\hline
\end{tabular}
\caption{Most important variables for every method applied to Hepatitis dataset.}
\end{table}

\begin{figure}[!h]
\centering
\includegraphics[width=\textwidth]{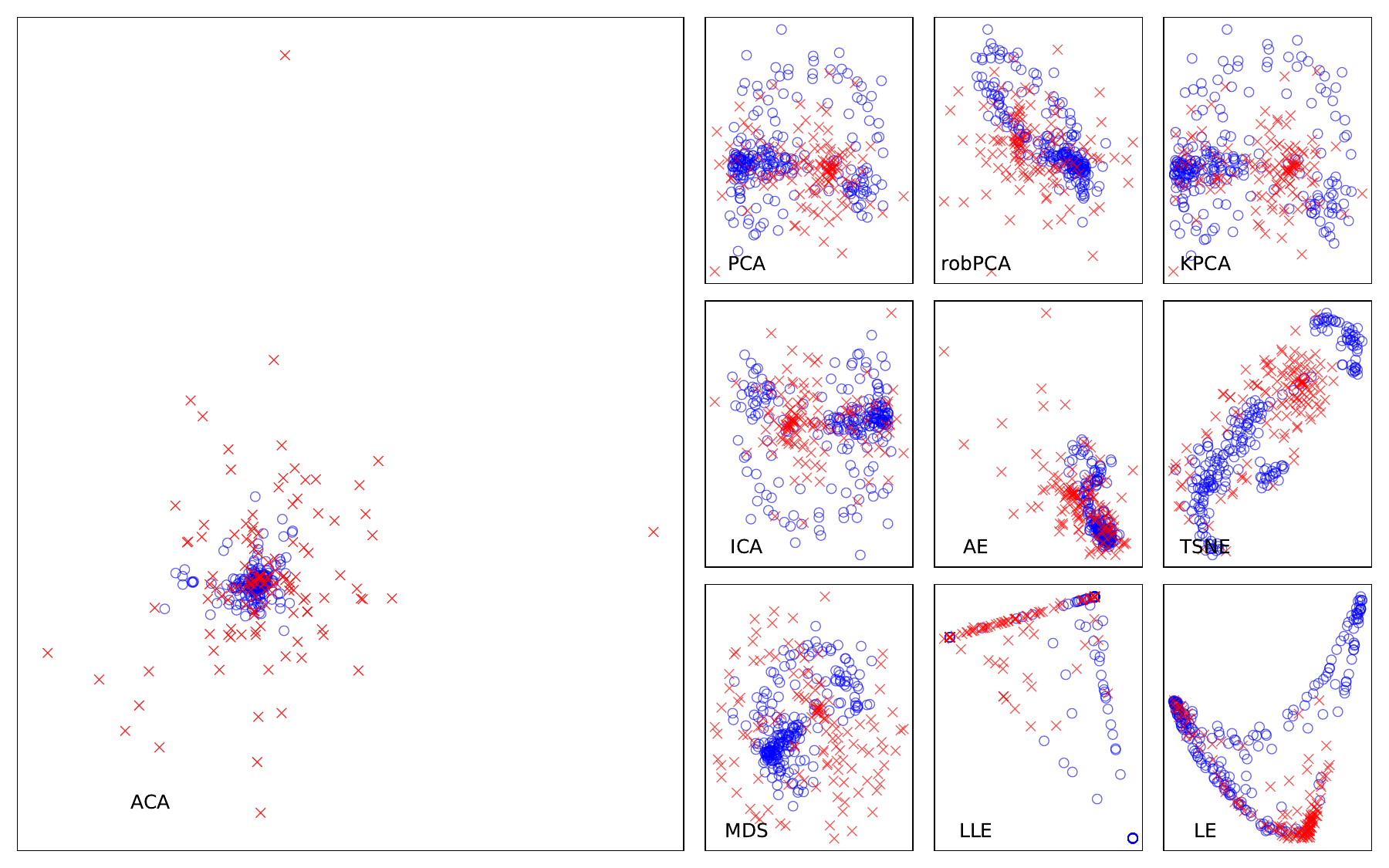}
\caption{Visualization of projection on two first components obtained by different methods for Ionosphere dataset.}\label{fig:realIonosphere}
\end{figure}

\begin{table}[H]
\centering
\begin{tabular}{c|cccccccc|}
\cline{2-9}
& \multicolumn{8}{c|}{Ionosphere}                                                                                            \\ \cline{2-9} 
& \multicolumn{2}{c|}{PCA}                 & \multicolumn{2}{c|}{robPCA}              & \multicolumn{2}{c|}{ICA}                 & \multicolumn{2}{c|}{ACA} \\ \cline{2-9} 
& PC1      & \multicolumn{1}{c|}{PC2}      & robPC1   & \multicolumn{1}{c|}{robPC2}   & IC1      & \multicolumn{1}{c|}{IC2}      & AC1         & AC2        \\ \hline
\multicolumn{1}{|c|}{Var1}  & 13 (6\%) & \multicolumn{1}{c|}{18 (8\%)} & 5 (10\%) & \multicolumn{1}{c|}{6 (10\%)} & 13 (7\%) & \multicolumn{1}{c|}{18 (8\%)} & 1 (6\%)     & 16 (7\%)   \\
\multicolumn{1}{|c|}{Var2}  & 19 (6\%) & \multicolumn{1}{c|}{20 (7\%)} & 4 (8\%)  & \multicolumn{1}{c|}{18 (9\%)} & 15 (6\%) & \multicolumn{1}{c|}{20 (7\%)} & 32 (6\%)    & 5 (6\%)    \\
\multicolumn{1}{|c|}{Var3}  & 15 (6\%) & \multicolumn{1}{c|}{26 (7\%)} & 17 (7\%) & \multicolumn{1}{c|}{8 (9\%)}  & 11 (6\%) & \multicolumn{1}{c|}{26 (7\%)} & 26 (5\%)    & 30 (6\%)   \\
\multicolumn{1}{|c|}{Var4}  & 17 (6\%) & \multicolumn{1}{c|}{24 (6\%)} & 9 (7\%)  & \multicolumn{1}{c|}{21 (8\%)} & 17 (6\%) & \multicolumn{1}{c|}{22 (7\%)} & 23 (5\%)    & 1 (6\%)    \\
\multicolumn{1}{|c|}{Var5}  & 11 (6\%) & \multicolumn{1}{c|}{22 (6\%)} & 12 (6\%) & \multicolumn{1}{c|}{19 (6\%)} & 19 (6\%) & \multicolumn{1}{c|}{24 (6\%)} & 7 (5\%)     & 11 (5\%)   \\
\hline
\end{tabular}
\caption{Most important variables for every method applied to Ionosphere dataset.}
\end{table}

\begin{figure}[!h]
\centering
\includegraphics[width=\textwidth]{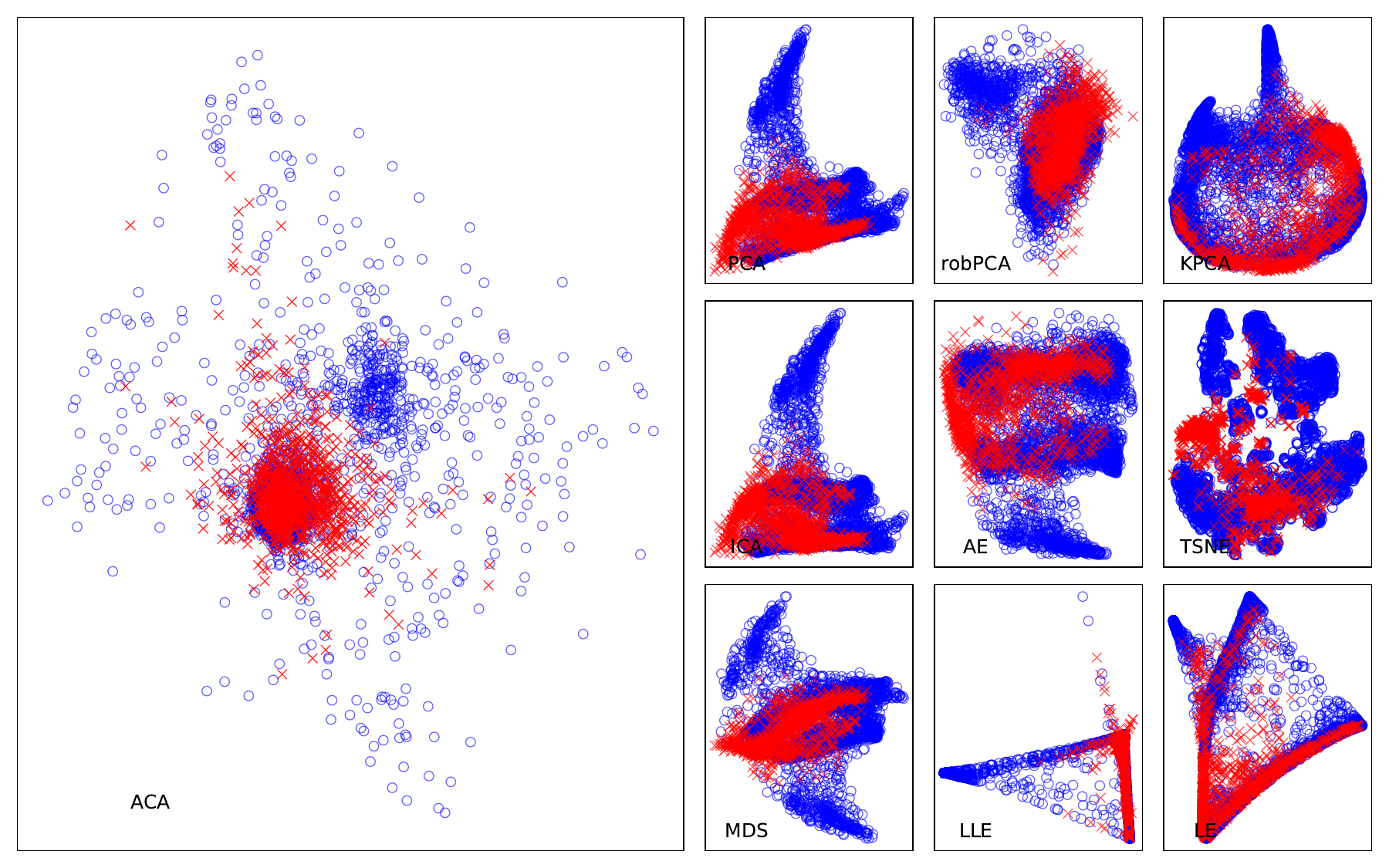}
\caption{Visualization of projection on two first components obtained by different methods for landsat dataset.}\label{fig:reallandsat}
\end{figure}

\begin{table}[H]
\centering
\begin{tabular}{c|cccccccc|}
\cline{2-9}
& \multicolumn{8}{c|}{Landsat}                                                                                               \\ \cline{2-9} 
& \multicolumn{2}{c|}{PCA}                 & \multicolumn{2}{c|}{robPCA}              & \multicolumn{2}{c|}{ICA}                 & \multicolumn{2}{c|}{ACA} \\ \cline{2-9} 
& PC1      & \multicolumn{1}{c|}{PC2}      & robPC1   & \multicolumn{1}{c|}{robPC2}   & IC1      & \multicolumn{1}{c|}{IC2}      & AC1         & AC2        \\ \hline
\multicolumn{1}{|c|}{Var1}  & 18 (3\%) & \multicolumn{1}{c|}{20 (4\%)} & 8 (10\%) & \multicolumn{1}{c|}{27 (7\%)} & 19 (4\%) & \multicolumn{1}{c|}{20 (4\%)} & 10 (9\%)    & 34 (8\%)   \\
\multicolumn{1}{|c|}{Var2}  & 14 (3\%) & \multicolumn{1}{c|}{24 (4\%)} & 16 (8\%) & \multicolumn{1}{c|}{35 (6\%)} & 15 (4\%) & \multicolumn{1}{c|}{24 (4\%)} & 24 (8\%)    & 36 (7\%)   \\
\multicolumn{1}{|c|}{Var3}  & 22 (3\%) & \multicolumn{1}{c|}{16 (4\%)} & 12 (8\%) & \multicolumn{1}{c|}{30 (6\%)} & 23 (4\%) & \multicolumn{1}{c|}{16 (3\%)} & 12 (7\%)    & 30 (5\%)   \\
\multicolumn{1}{|c|}{Var4}  & 30 (3\%) & \multicolumn{1}{c|}{32 (4\%)} & 2 (7\%)  & \multicolumn{1}{c|}{29 (6\%)} & 7 (4\%)  & \multicolumn{1}{c|}{32 (3\%)} & 28 (6\%)    & 6 (5\%)    \\
\multicolumn{1}{|c|}{Var5}  & 6 (3\%)  & \multicolumn{1}{c|}{8 (4\%)}  & 21 (7\%) & \multicolumn{1}{c|}{34 (6\%)} & 31 (4\%) & \multicolumn{1}{c|}{8 (3\%)}  & 22 (6\%)    & 22 (4\%)   \\
\hline
\end{tabular}
\caption{Most important variables for every method applied to Landsat dataset.}
\end{table}

\begin{figure}[!h]
\centering
\includegraphics[width=\textwidth]{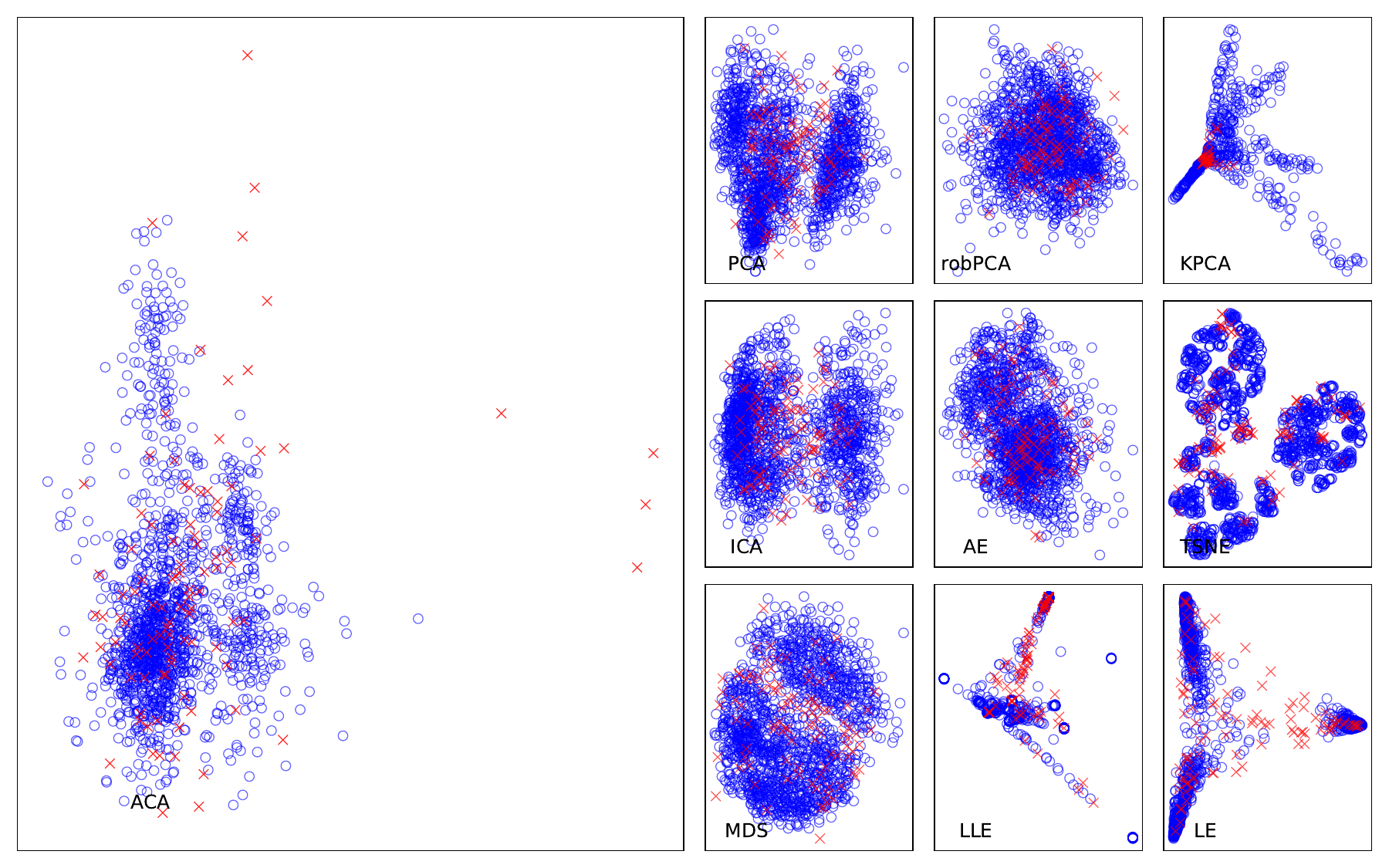}
\caption{Visualization of projection on two first components obtained by different methods for letter dataset.}\label{fig:realletter}
\end{figure}

\begin{table}[H]
\centering
\begin{tabular}{c|cccccccc|}
\cline{2-9}
& \multicolumn{8}{c|}{Letter}                                                                                                \\ \cline{2-9} 
& \multicolumn{2}{c|}{PCA}                 & \multicolumn{2}{c|}{robPCA}               & \multicolumn{2}{c|}{ICA}                   & \multicolumn{2}{c|}{ACA} \\ \cline{2-9} 
& PC1      & \multicolumn{1}{c|}{PC2}      & robPC1    & \multicolumn{1}{c|}{robPC2}   & IC1       & \multicolumn{1}{c|}{IC2}       & AC1         & AC2        \\ \hline
\multicolumn{1}{|c|}{Var1}  & 13 (6\%) & \multicolumn{1}{c|}{17 (6\%)} & 32 (12\%) & \multicolumn{1}{c|}{4 (14\%)} & 13 (10\%) & \multicolumn{1}{c|}{18 (13\%)} & 22 (13\%)   & 24 (17\%)  \\
\multicolumn{1}{|c|}{Var2}  & 29 (6\%) & \multicolumn{1}{c|}{19 (6\%)} & 21 (9\%)  & \multicolumn{1}{c|}{5 (9\%)}  & 29 (9\%)  & \multicolumn{1}{c|}{19 (9\%)}  & 26 (12\%)   & 20 (6\%)   \\
\multicolumn{1}{|c|}{Var3}  & 14 (5\%) & \multicolumn{1}{c|}{1 (5\%)}  & 28 (6\%)  & \multicolumn{1}{c|}{19 (9\%)} & 5 (6\%)   & \multicolumn{1}{c|}{20 (8\%)}  & 27 (6\%)    & 21 (6\%)   \\
\multicolumn{1}{|c|}{Var4}  & 30 (5\%) & \multicolumn{1}{c|}{3 (5\%)}  & 31 (6\%)  & \multicolumn{1}{c|}{3 (8\%)}  & 3 (6\%)   & \multicolumn{1}{c|}{21 (8\%)}  & 23 (6\%)    & 9 (6\%)    \\
\multicolumn{1}{|c|}{Var5}  & 9 (5\%)  & \multicolumn{1}{c|}{16 (5\%)} & 30 (5\%)  & \multicolumn{1}{c|}{20 (6\%)} & 24 (5\%)  & \multicolumn{1}{c|}{17 (8\%)}  & 4 (6\%)     & 30 (6\%)   \\
\hline
\end{tabular}
\caption{Most important variables for every method applied to Letter dataset.}
\end{table}

\begin{figure}[!h]
\centering
\includegraphics[width=\textwidth]{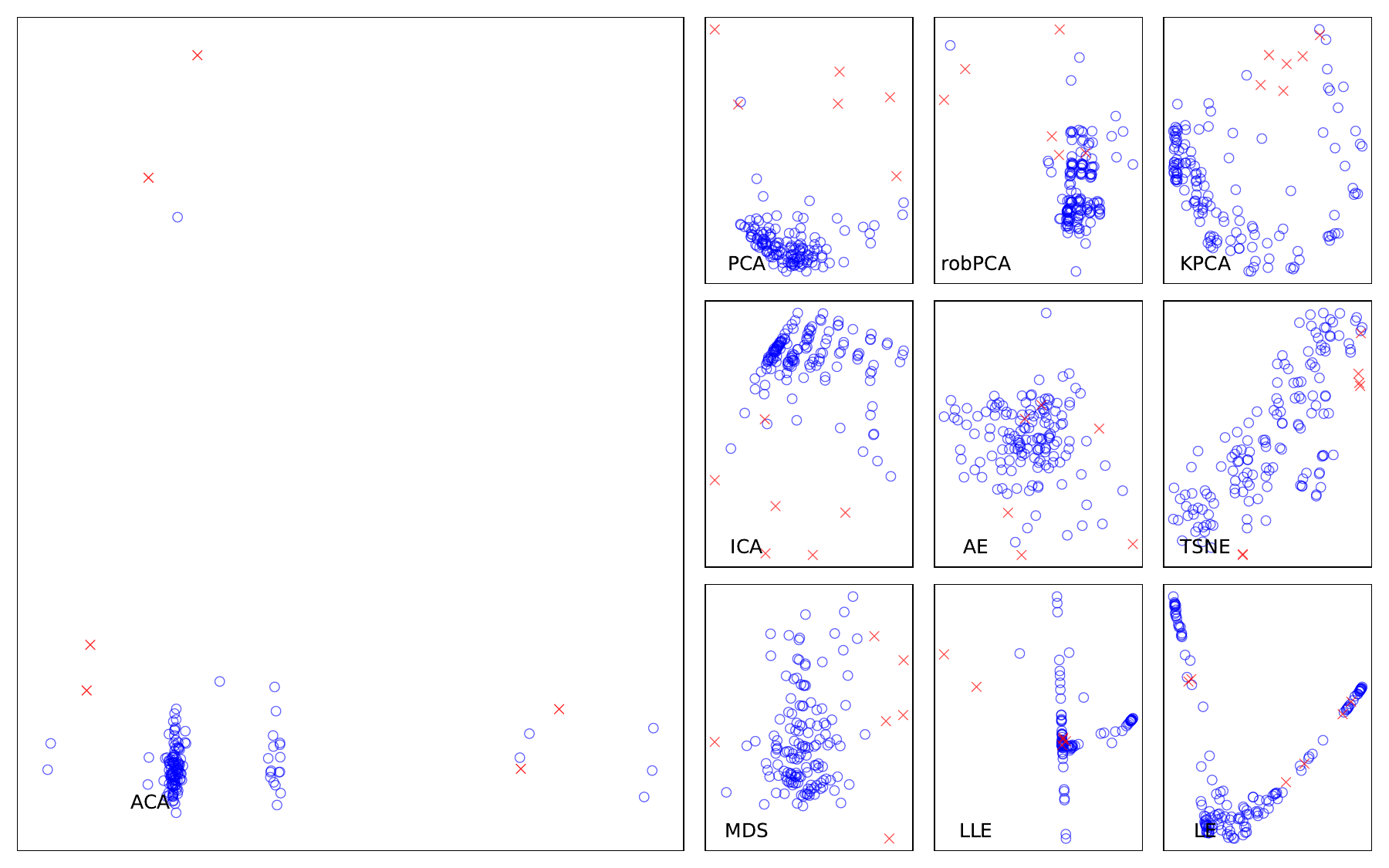}
\caption{Visualization of projection on two first components obtained by different methods for Lymphography dataset.}\label{fig:realLymphography}
\end{figure}

\begin{table}[H]
\centering
\begin{tabular}{c|cccccccc|}
\cline{2-9}
& \multicolumn{8}{c|}{Lymphography}                                                                                          \\ \cline{2-9} 
& \multicolumn{2}{c|}{PCA}                   & \multicolumn{2}{c|}{robPCA}                & \multicolumn{2}{c|}{ICA}                   & \multicolumn{2}{c|}{ACA} \\ \cline{2-9} 
& PC1       & \multicolumn{1}{c|}{PC2}       & robPC1    & \multicolumn{1}{c|}{robPC2}    & IC1       & \multicolumn{1}{c|}{IC2}       & AC1        & AC2         \\ \hline
\multicolumn{1}{|c|}{Var1}  & 18 (10\%) & \multicolumn{1}{c|}{11 (13\%)} & 12 (34\%) & \multicolumn{1}{c|}{13 (57\%)} & 18 (22\%) & \multicolumn{1}{c|}{10 (16\%)} & 4 (37\%)   & 12 (38\%)   \\
\multicolumn{1}{|c|}{Var2}  & 5 (10\%)  & \multicolumn{1}{c|}{12 (10\%)} & 9 (12\%)  & \multicolumn{1}{c|}{14 (19\%)} & 10 (20\%) & \multicolumn{1}{c|}{7 (12\%)}  & 3 (20\%)   & 2 (13\%)    \\
\multicolumn{1}{|c|}{Var3}  & 4 (9\%)   & \multicolumn{1}{c|}{9 (10\%)}  & 10 (9\%)  & \multicolumn{1}{c|}{2 (4\%)}   & 11 (8\%)  & \multicolumn{1}{c|}{11 (10\%)} & 7 (15\%)   & 1 (10\%)    \\
\multicolumn{1}{|c|}{Var4}  & 3 (9\%)   & \multicolumn{1}{c|}{13 (9\%)}  & 5 (8\%)   & \multicolumn{1}{c|}{7 (3\%)}   & 7 (6\%)   & \multicolumn{1}{c|}{4 (9\%)}   & 6 (9\%)    & 11 (8\%)    \\
\multicolumn{1}{|c|}{Var5}  & 7 (8\%)   & \multicolumn{1}{c|}{1 (8\%)}   & 6 (8\%)   & \multicolumn{1}{c|}{11 (3\%)}  & 15 (6\%)  & \multicolumn{1}{c|}{13 (7\%)}  & 11 (5\%)   & 13 (7\%)    \\
\hline
\end{tabular}
\caption{Most important variables for every method applied to Lymphography dataset.}
\end{table}

\begin{figure}[!h]
\centering
\includegraphics[width=\textwidth]{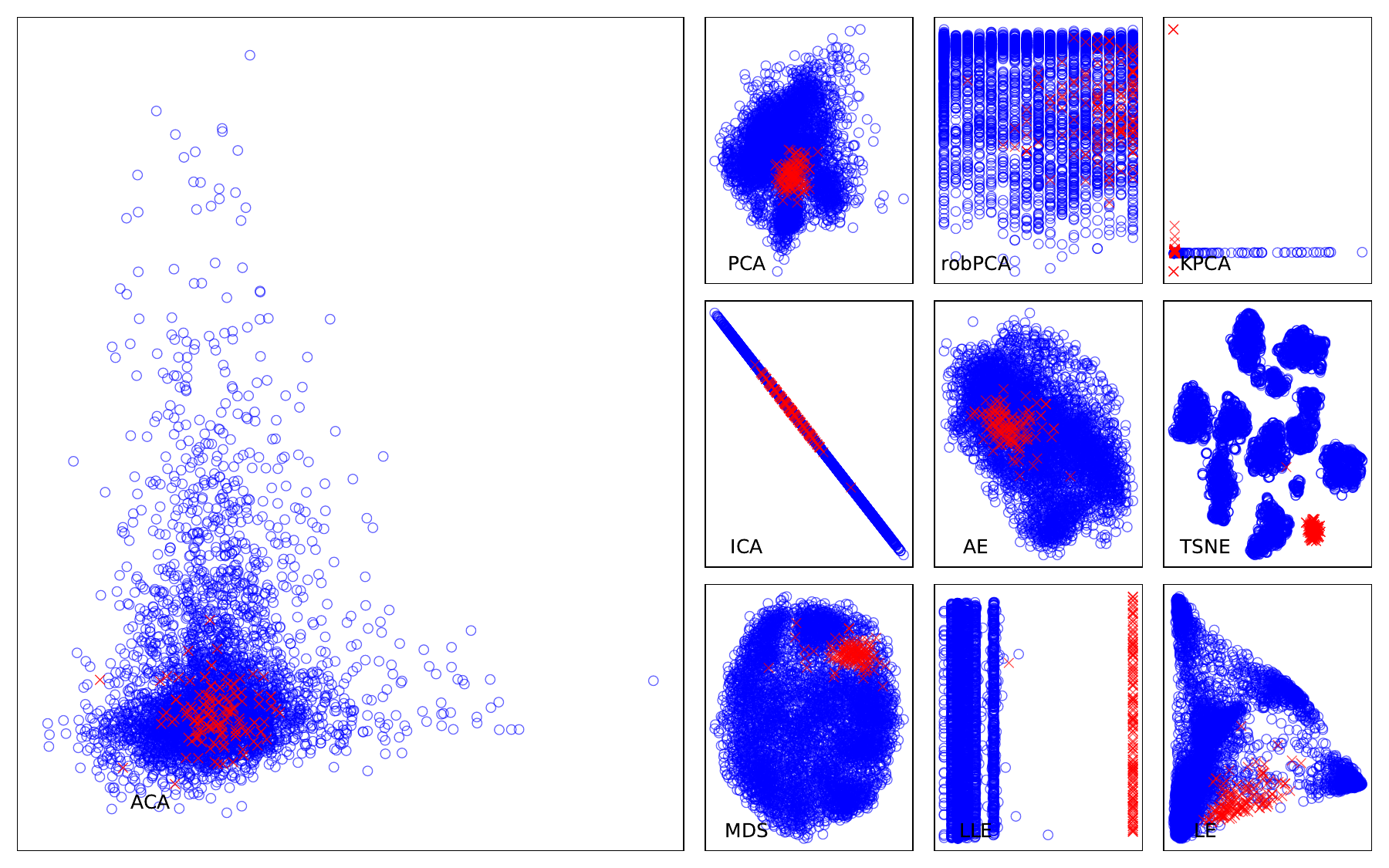}
\caption{Visualization of projection on two first components obtained by different methods for optdigits dataset.}\label{fig:realoptdigits}
\end{figure}

\begin{table}[H]
\centering
\begin{tabular}{c|cccccccc|}
\cline{2-9}
& \multicolumn{8}{c|}{Optdigits}                                                                                             \\ \cline{2-9} 
& \multicolumn{2}{c|}{PCA}                 & \multicolumn{2}{c|}{robPCA}                 & \multicolumn{2}{c|}{ICA}                 & \multicolumn{2}{c|}{ACA} \\ \cline{2-9} 
& PC1      & \multicolumn{1}{c|}{PC2}      & robPC1     & \multicolumn{1}{c|}{robPC2}    & IC1      & \multicolumn{1}{c|}{IC2}      & AC1         & AC2        \\ \hline
\multicolumn{1}{|c|}{Var1}  & 3 (4\%)  & \multicolumn{1}{c|}{14 (4\%)} & 51 (100\%) & \multicolumn{1}{c|}{47 (44\%)} & 62 (5\%) & \multicolumn{1}{c|}{62 (5\%)} & 58 (5\%)    & 8 (10\%)   \\
\multicolumn{1}{|c|}{Var2}  & 59 (4\%) & \multicolumn{1}{c|}{6 (4\%)}  & 64 (0\%)   & \multicolumn{1}{c|}{48 (32\%)} & 51 (5\%) & \multicolumn{1}{c|}{51 (5\%)} & 4 (4\%)     & 7 (10\%)   \\
\multicolumn{1}{|c|}{Var3}  & 11 (4\%) & \multicolumn{1}{c|}{15 (4\%)} & 16 (0\%)   & \multicolumn{1}{c|}{40 (3\%)}  & 30 (5\%) & \multicolumn{1}{c|}{30 (5\%)} & 61 (4\%)    & 16 (8\%)   \\
\multicolumn{1}{|c|}{Var4}  & 10 (3\%) & \multicolumn{1}{c|}{43 (4\%)} & 29 (0\%)   & \multicolumn{1}{c|}{46 (2\%)}  & 55 (5\%) & \multicolumn{1}{c|}{55 (5\%)} & 60 (4\%)    & 23 (6\%)   \\
\multicolumn{1}{|c|}{Var5}  & 4 (3\%)  & \multicolumn{1}{c|}{22 (4\%)} & 28 (0\%)   & \multicolumn{1}{c|}{39 (2\%)}  & 45 (4\%) & \multicolumn{1}{c|}{45 (4\%)} & 2 (4\%)     & 15 (4\%)   \\
\hline
\end{tabular}
\caption{Most important variables for every method applied to Optdigits dataset.}
\end{table}

\begin{figure}[!h]
\centering
\includegraphics[width=\textwidth]{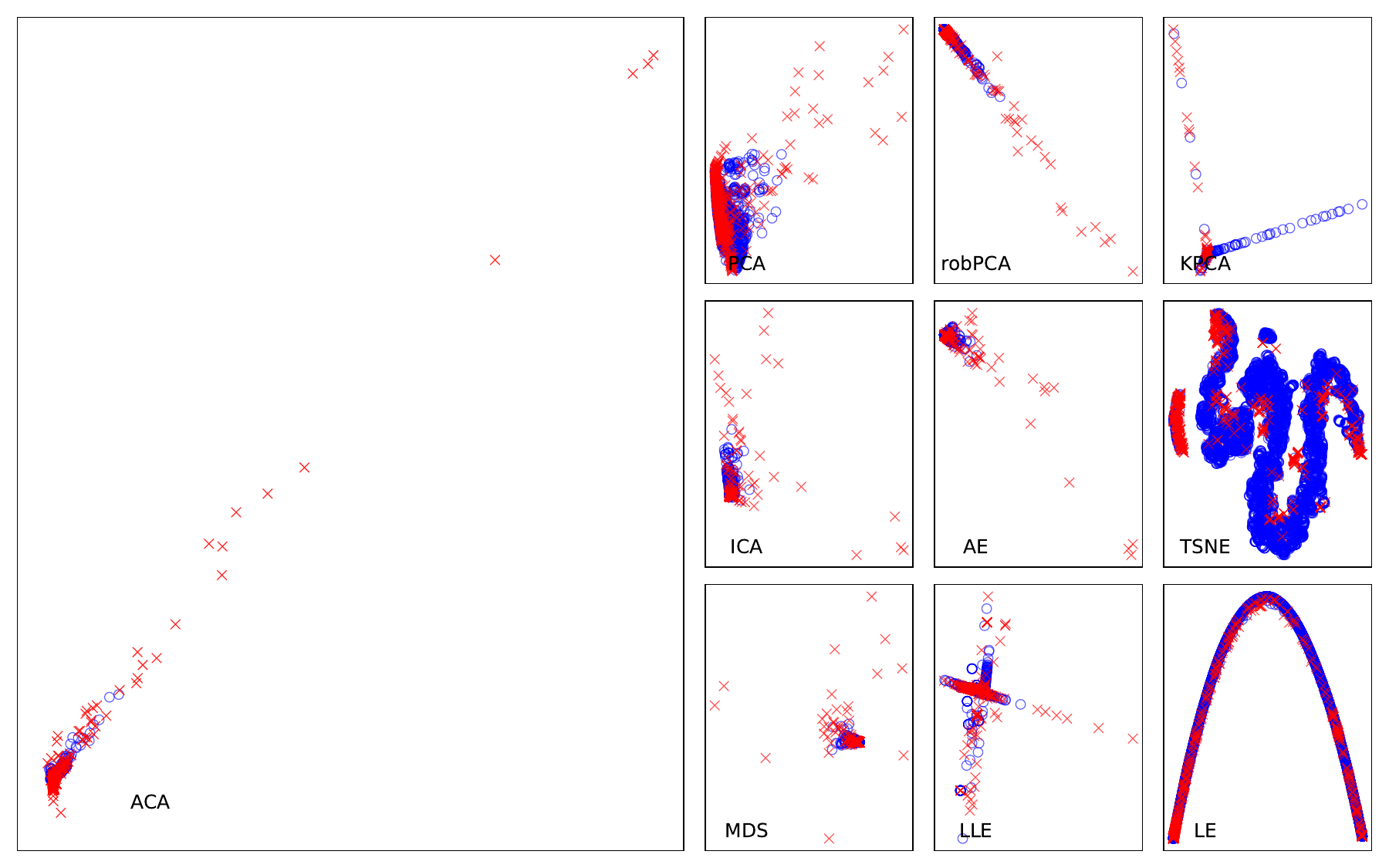}
\caption{Visualization of projection on two first components obtained by different methods for PageBlocks dataset.}\label{fig:realPageBlocks}
\end{figure}

\begin{table}[H]
\centering
\begin{tabular}{c|cccccccc|}
\cline{2-9}
& \multicolumn{8}{c|}{PageBlocks}                                                                                            \\ \cline{2-9} 
& \multicolumn{2}{c|}{PCA}                  & \multicolumn{2}{c|}{robPCA}              & \multicolumn{2}{c|}{ICA}                 & \multicolumn{2}{c|}{ACA} \\ \cline{2-9} 
& PC1       & \multicolumn{1}{c|}{PC2}      & robPC1   & \multicolumn{1}{c|}{robPC2}   & IC1      & \multicolumn{1}{c|}{IC2}      & AC1         & AC2        \\ \hline
\multicolumn{1}{|c|}{Var1}  & 9 (17\%)  & \multicolumn{1}{c|}{5 (20\%)} & 6 (68\%) & \multicolumn{1}{c|}{5 (45\%)} & 3 (39\%) & \multicolumn{1}{c|}{9 (49\%)} & 9 (26\%)    & 4 (32\%)   \\
\multicolumn{1}{|c|}{Var2}  & 10 (16\%) & \multicolumn{1}{c|}{6 (19\%)} & 5 (11\%) & \multicolumn{1}{c|}{4 (22\%)} & 9 (33\%) & \multicolumn{1}{c|}{8 (35\%)} & 3 (22\%)    & 2 (19\%)   \\
\multicolumn{1}{|c|}{Var3}  & 3 (15\%)  & \multicolumn{1}{c|}{2 (14\%)} & 7 (8\%)  & \multicolumn{1}{c|}{8 (12\%)} & 8 (25\%) & \multicolumn{1}{c|}{3 (13\%)} & 2 (16\%)    & 5 (14\%)   \\
\multicolumn{1}{|c|}{Var4}  & 8 (15\%)  & \multicolumn{1}{c|}{4 (10\%)} & 4 (5\%)  & \multicolumn{1}{c|}{6 (10\%)} & 10 (1\%) & \multicolumn{1}{c|}{10 (2\%)} & 4 (15\%)    & 7 (10\%)   \\
\multicolumn{1}{|c|}{Var5}  & 2 (12\%)  & \multicolumn{1}{c|}{8 (10\%)} & 8 (4\%)  & \multicolumn{1}{c|}{1 (4\%)}  & 2 (1\%)  & \multicolumn{1}{c|}{2 (1\%)}  & 5 (7\%)     & 3 (8\%)    \\
\hline
\end{tabular}
\caption{Most important variables for every method applied to PageBlocks dataset.}
\end{table}

\begin{figure}[!h]
\centering
\includegraphics[width=\textwidth]{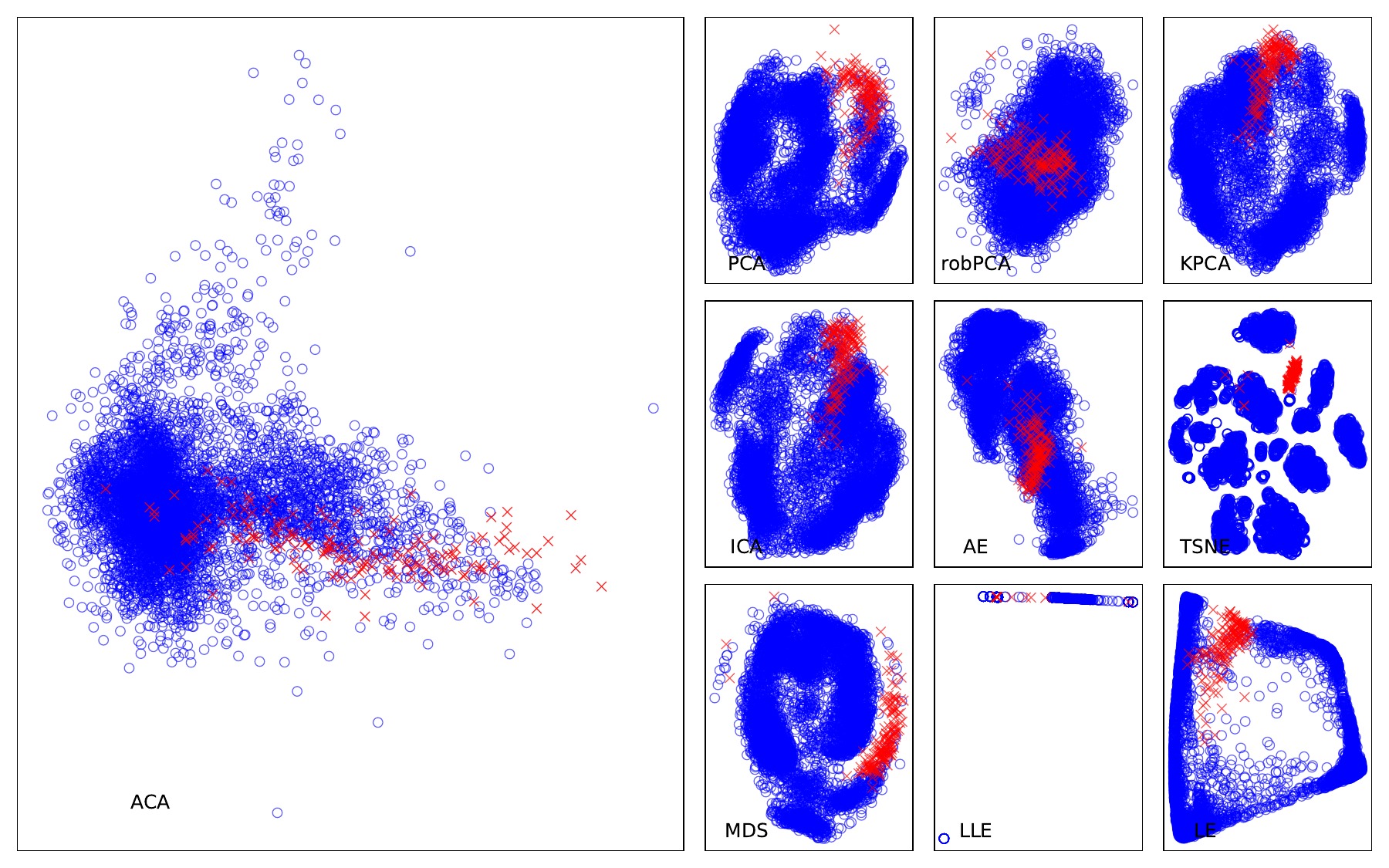}
\caption{Visualization of projection on two first components obtained by different methods for pendigits dataset.}\label{fig:realpendigits}
\end{figure}

\begin{table}[H]
\centering
\begin{tabular}{c|cccccccc|}
\cline{2-9}
& \multicolumn{8}{c|}{Pendigits}                                                                                             \\ \cline{2-9} 
& \multicolumn{2}{c|}{PCA}                   & \multicolumn{2}{c|}{robPCA}               & \multicolumn{2}{c|}{ICA}                   & \multicolumn{2}{c|}{ACA} \\ \cline{2-9} 
& PC1       & \multicolumn{1}{c|}{PC2}       & robPC1   & \multicolumn{1}{c|}{robPC2}    & IC1       & \multicolumn{1}{c|}{IC2}       & AC1         & AC2        \\ \hline
\multicolumn{1}{|c|}{Var1}  & 14 (11\%) & \multicolumn{1}{c|}{11 (13\%)} & 4 (14\%) & \multicolumn{1}{c|}{10 (22\%)} & 11 (18\%) & \multicolumn{1}{c|}{8 (11\%)}  & 6 (19\%)    & 5 (14\%)   \\
\multicolumn{1}{|c|}{Var2}  & 8 (11\%)  & \multicolumn{1}{c|}{10 (12\%)} & 8 (13\%) & \multicolumn{1}{c|}{13 (12\%)} & 15 (18\%) & \multicolumn{1}{c|}{5 (11\%)}  & 2 (15\%)    & 3 (13\%)   \\
\multicolumn{1}{|c|}{Var3}  & 6 (11\%)  & \multicolumn{1}{c|}{12 (11\%)} & 3 (12\%) & \multicolumn{1}{c|}{14 (12\%)} & 9 (10\%)  & \multicolumn{1}{c|}{14 (10\%)} & 4 (15\%)    & 7 (12\%)   \\
\multicolumn{1}{|c|}{Var4}  & 16 (10\%) & \multicolumn{1}{c|}{9 (10\%)}  & 1 (10\%) & \multicolumn{1}{c|}{11 (10\%)} & 1 (9\%)   & \multicolumn{1}{c|}{16 (9\%)}  & 15 (10\%)   & 9 (10\%)   \\
\multicolumn{1}{|c|}{Var5}  & 10 (6\%)  & \multicolumn{1}{c|}{4 (9\%)}   & 5 (10\%) & \multicolumn{1}{c|}{12 (8\%)}  & 5 (8\%)   & \multicolumn{1}{c|}{6 (9\%)}   & 11 (10\%)   & 1 (8\%)    \\
\hline
\end{tabular}
\caption{Most important variables for every method applied to Pendigits dataset.}
\end{table}

\begin{figure}[!h]
\centering
\includegraphics[width=\textwidth]{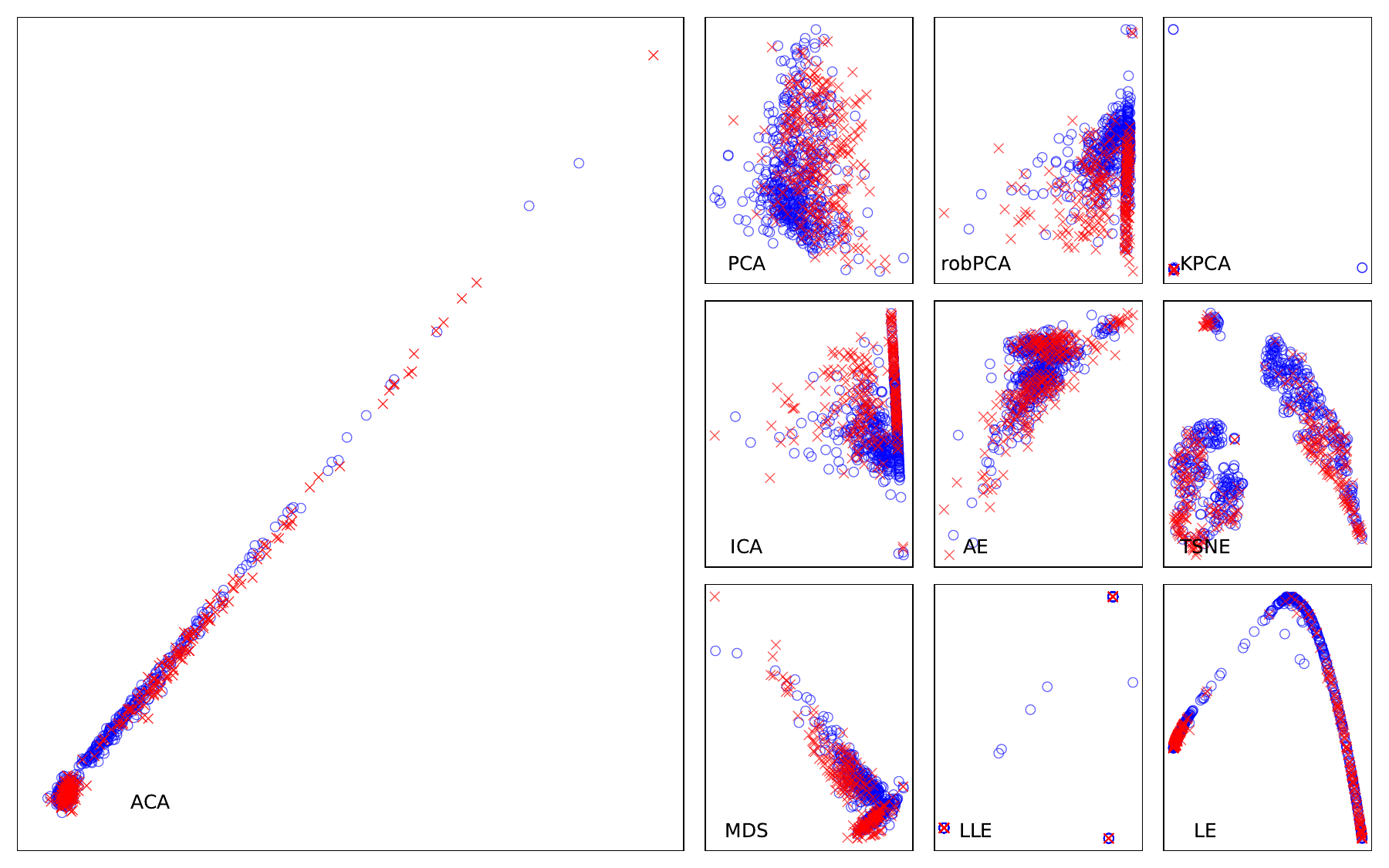}
\caption{Visualization of projection on two first components obtained by different methods for Pima dataset.}\label{fig:realPima}
\end{figure}

\begin{table}[H]
\centering
\begin{tabular}{c|cccccccc|}
\cline{2-9}
& \multicolumn{8}{c|}{Pima}                                                                                                  \\ \cline{2-9} 
& \multicolumn{2}{c|}{PCA}                 & \multicolumn{2}{c|}{robPCA}              & \multicolumn{2}{c|}{ICA}                 & \multicolumn{2}{c|}{ACA} \\ \cline{2-9} 
& PC1      & \multicolumn{1}{c|}{PC2}      & robPC1   & \multicolumn{1}{c|}{robPC2}   & IC1      & \multicolumn{1}{c|}{IC2}      & AC1         & AC2        \\ \hline
\multicolumn{1}{|c|}{Var1} & 6 (17\%) & \multicolumn{1}{c|}{8 (26\%)} & 7 (70\%) & \multicolumn{1}{c|}{2 (71\%)} & 5 (69\%) & \multicolumn{1}{c|}{2 (65\%)} & 5 (46\%)    & 7 (39\%)   \\
\multicolumn{1}{|c|}{Var2} & 4 (16\%) & \multicolumn{1}{c|}{1 (25\%)} & 5 (18\%) & \multicolumn{1}{c|}{4 (13\%)} & 2 (20\%) & \multicolumn{1}{c|}{3 (10\%)} & 7 (15\%)    & 1 (32\%)   \\
\multicolumn{1}{|c|}{Var3} & 5 (16\%) & \multicolumn{1}{c|}{4 (14\%)} & 4 (7\%)  & \multicolumn{1}{c|}{1 (9\%)}  & 4 (4\%)  & \multicolumn{1}{c|}{8 (9\%)}  & 8 (11\%)    & 5 (16\%)   \\
\multicolumn{1}{|c|}{Var4} & 2 (15\%) & \multicolumn{1}{c|}{5 (11\%)} & 2 (2\%)  & \multicolumn{1}{c|}{6 (3\%)}  & 3 (3\%)  & \multicolumn{1}{c|}{5 (7\%)}  & 2 (7\%)     & 8 (6\%)    \\
\multicolumn{1}{|c|}{Var5} & 3 (13\%) & \multicolumn{1}{c|}{3 (8\%)}  & 3 (1\%)  & \multicolumn{1}{c|}{5 (2\%)}  & 8 (2\%)  & \multicolumn{1}{c|}{4 (4\%)}  & 1 (7\%)     & 4 (4\%)    \\
\hline
\end{tabular}
\caption{Most important variables for every method applied to Pima dataset.}
\end{table}

\begin{figure}[!h]
\centering
\includegraphics[width=\textwidth]{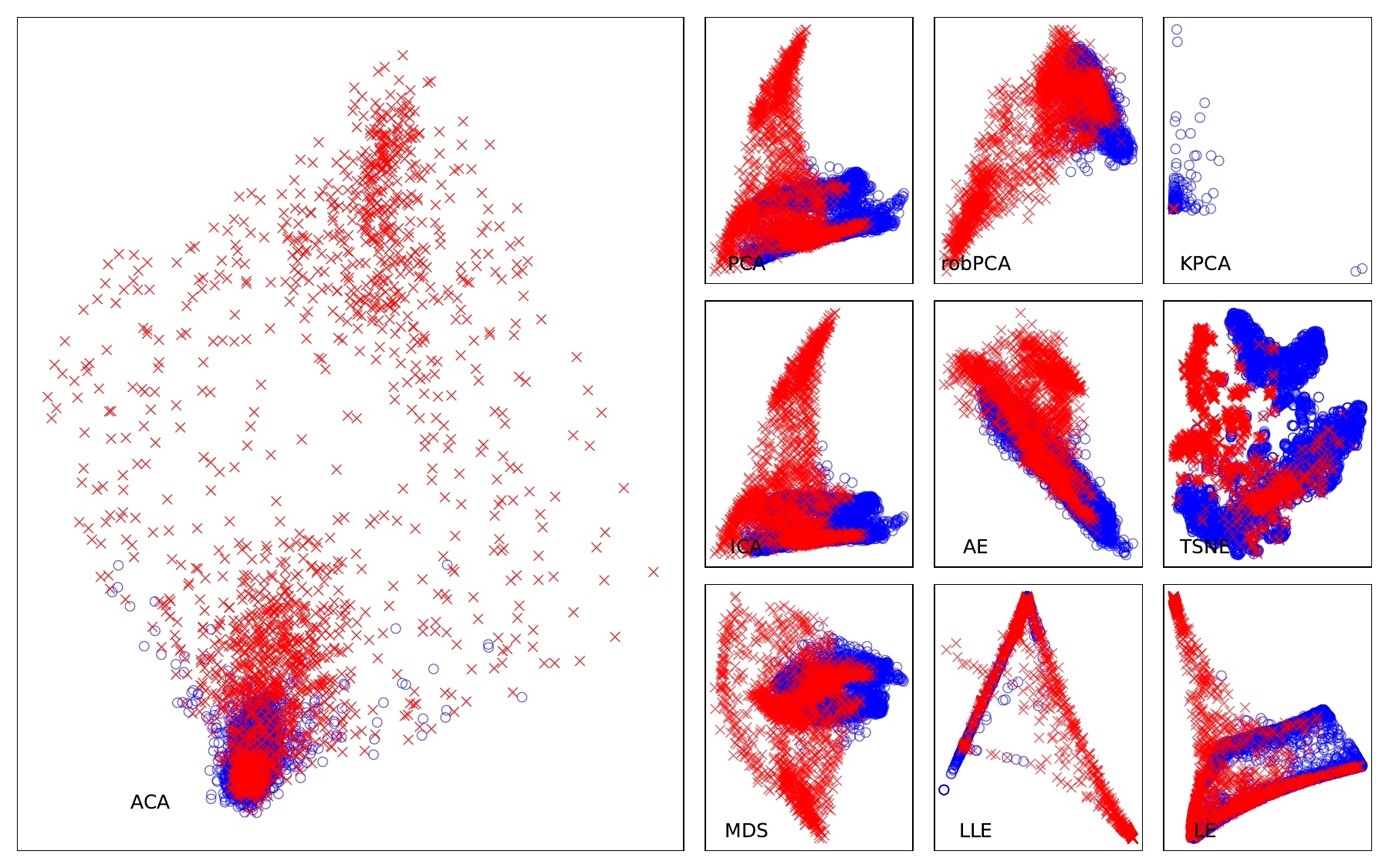}
\caption{Visualization of projection on two first components obtained by different methods for satellite dataset.}\label{fig:realsatellite}
\end{figure}

\begin{table}[H]
\centering
\begin{tabular}{c|cccccccc|}
\cline{2-9}
& \multicolumn{8}{c|}{Satellite}                                                                                             \\ \cline{2-9} 
& \multicolumn{2}{c|}{PCA}                 & \multicolumn{2}{c|}{robPCA}              & \multicolumn{2}{c|}{ICA}                 & \multicolumn{2}{c|}{ACA} \\ \cline{2-9} 
& PC1      & \multicolumn{1}{c|}{PC2}      & robPC1   & \multicolumn{1}{c|}{robPC2}   & IC1      & \multicolumn{1}{c|}{IC2}      & AC1         & AC2        \\ \hline
\multicolumn{1}{|c|}{Var1}  & 18 (3\%) & \multicolumn{1}{c|}{20 (4\%)} & 24 (7\%) & \multicolumn{1}{c|}{18 (6\%)} & 18 (4\%) & \multicolumn{1}{c|}{20 (4\%)} & 10 (11\%)   & 18 (7\%)   \\
\multicolumn{1}{|c|}{Var2}  & 14 (3\%) & \multicolumn{1}{c|}{24 (4\%)} & 10 (7\%) & \multicolumn{1}{c|}{13 (6\%)} & 22 (4\%) & \multicolumn{1}{c|}{24 (4\%)} & 12 (10\%)   & 30 (6\%)   \\
\multicolumn{1}{|c|}{Var3}  & 22 (3\%) & \multicolumn{1}{c|}{16 (4\%)} & 35 (6\%) & \multicolumn{1}{c|}{6 (6\%)}  & 14 (4\%) & \multicolumn{1}{c|}{16 (4\%)} & 6 (6\%)     & 16 (6\%)   \\
\multicolumn{1}{|c|}{Var4}  & 30 (3\%) & \multicolumn{1}{c|}{32 (4\%)} & 2 (6\%)  & \multicolumn{1}{c|}{25 (6\%)} & 6 (4\%)  & \multicolumn{1}{c|}{32 (4\%)} & 8 (6\%)     & 26 (6\%)   \\
\multicolumn{1}{|c|}{Var5}  & 6 (3\%)  & \multicolumn{1}{c|}{8 (4\%)}  & 15 (6\%) & \multicolumn{1}{c|}{17 (5\%)} & 19 (4\%) & \multicolumn{1}{c|}{36 (4\%)} & 30 (5\%)    & 34 (6\%)   \\
\hline
\end{tabular}
\caption{Most important variables for every method applied to Satellite dataset.}
\end{table}

\begin{figure}[!h]
\centering
\includegraphics[width=\textwidth]{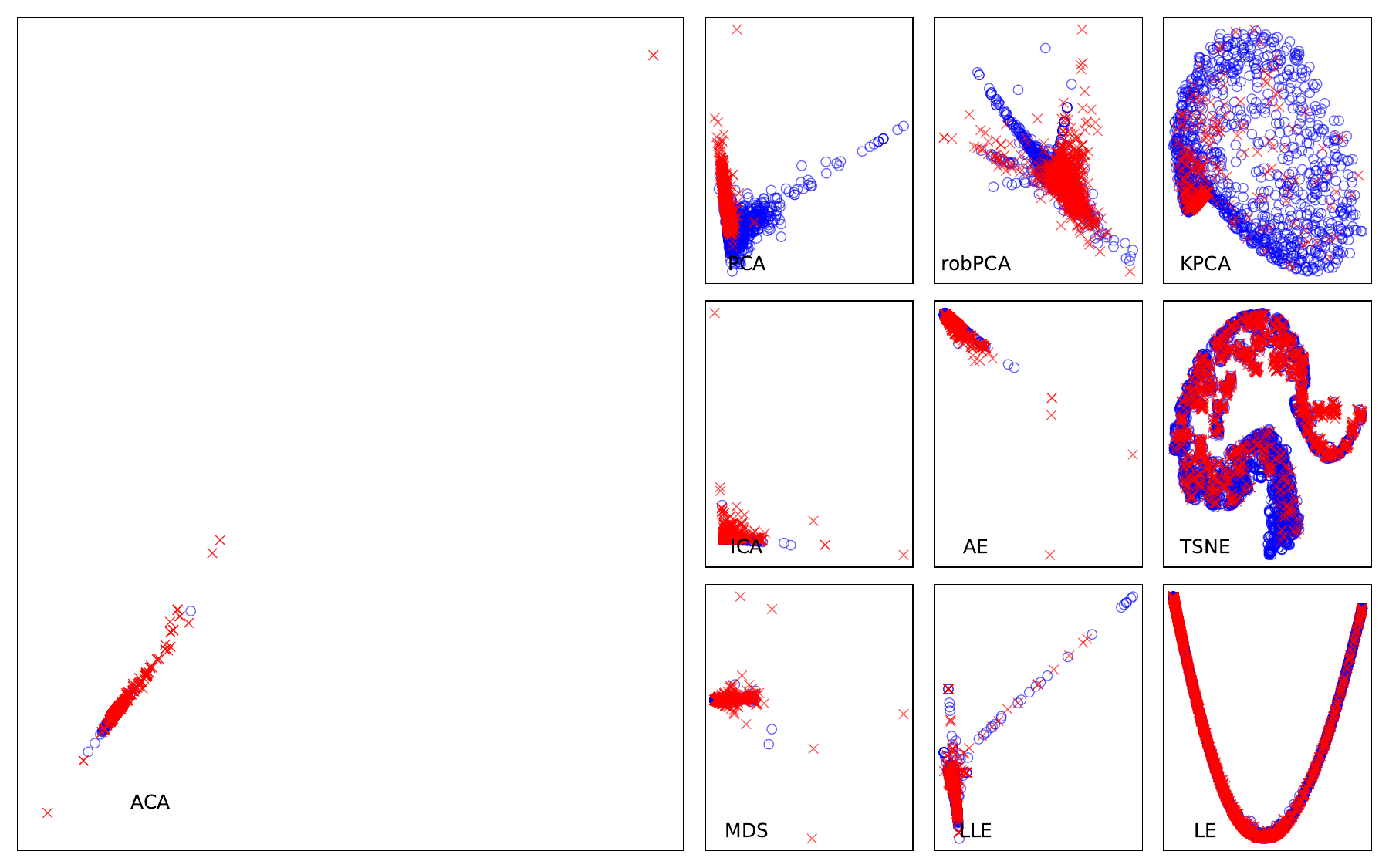}
\caption{Visualization of projection on two first components obtained by different methods for SpamBase dataset.}\label{fig:realSpamBase}
\end{figure}

\begin{table}[H]
\centering
\begin{tabular}{c|cccccccc|}
\cline{2-9}
& \multicolumn{8}{c|}{SpamBase}                                                                                              \\ \cline{2-9} 
& \multicolumn{2}{c|}{PCA}                 & \multicolumn{2}{c|}{robPCA}                & \multicolumn{2}{c|}{ICA}                   & \multicolumn{2}{c|}{ACA} \\ \cline{2-9} 
& PC1      & \multicolumn{1}{c|}{PC2}      & robPC1    & \multicolumn{1}{c|}{robPC2}    & IC1       & \multicolumn{1}{c|}{IC2}       & AC1         & AC2        \\ \hline
\multicolumn{1}{|c|}{Var1}  & 34 (7\%) & \multicolumn{1}{c|}{21 (4\%)} & 21 (15\%) & \multicolumn{1}{c|}{23 (32\%)} & 56 (49\%) & \multicolumn{1}{c|}{56 (88\%)} & 56 (22\%)   & 56 (9\%)   \\
\multicolumn{1}{|c|}{Var2}  & 32 (7\%) & \multicolumn{1}{c|}{56 (4\%)} & 22 (14\%) & \multicolumn{1}{c|}{21 (11\%)} & 57 (46\%) & \multicolumn{1}{c|}{55 (8\%)}  & 25 (7\%)    & 25 (6\%)   \\
\multicolumn{1}{|c|}{Var3}  & 40 (7\%) & \multicolumn{1}{c|}{23 (4\%)} & 23 (10\%) & \multicolumn{1}{c|}{15 (8\%)}  & 55 (5\%)  & \multicolumn{1}{c|}{57 (4\%)}  & 55 (4\%)    & 6 (4\%)    \\
\multicolumn{1}{|c|}{Var4}  & 36 (6\%) & \multicolumn{1}{c|}{53 (4\%)} & 20 (10\%) & \multicolumn{1}{c|}{25 (7\%)}  & 50 (0\%)  & \multicolumn{1}{c|}{50 (0\%)}  & 6 (3\%)     & 1 (4\%)    \\
\multicolumn{1}{|c|}{Var5}  & 31 (6\%) & \multicolumn{1}{c|}{9 (3\%)}  & 29 (8\%)  & \multicolumn{1}{c|}{17 (7\%)}  & 21 (0\%)  & \multicolumn{1}{c|}{21 (0\%)}  & 20 (3\%)    & 24 (4\%)   \\
\hline
\end{tabular}
\caption{Most important variables for every method applied to SpamBase dataset.}
\end{table}

\begin{figure}[!h]
\centering
\includegraphics[width=\textwidth]{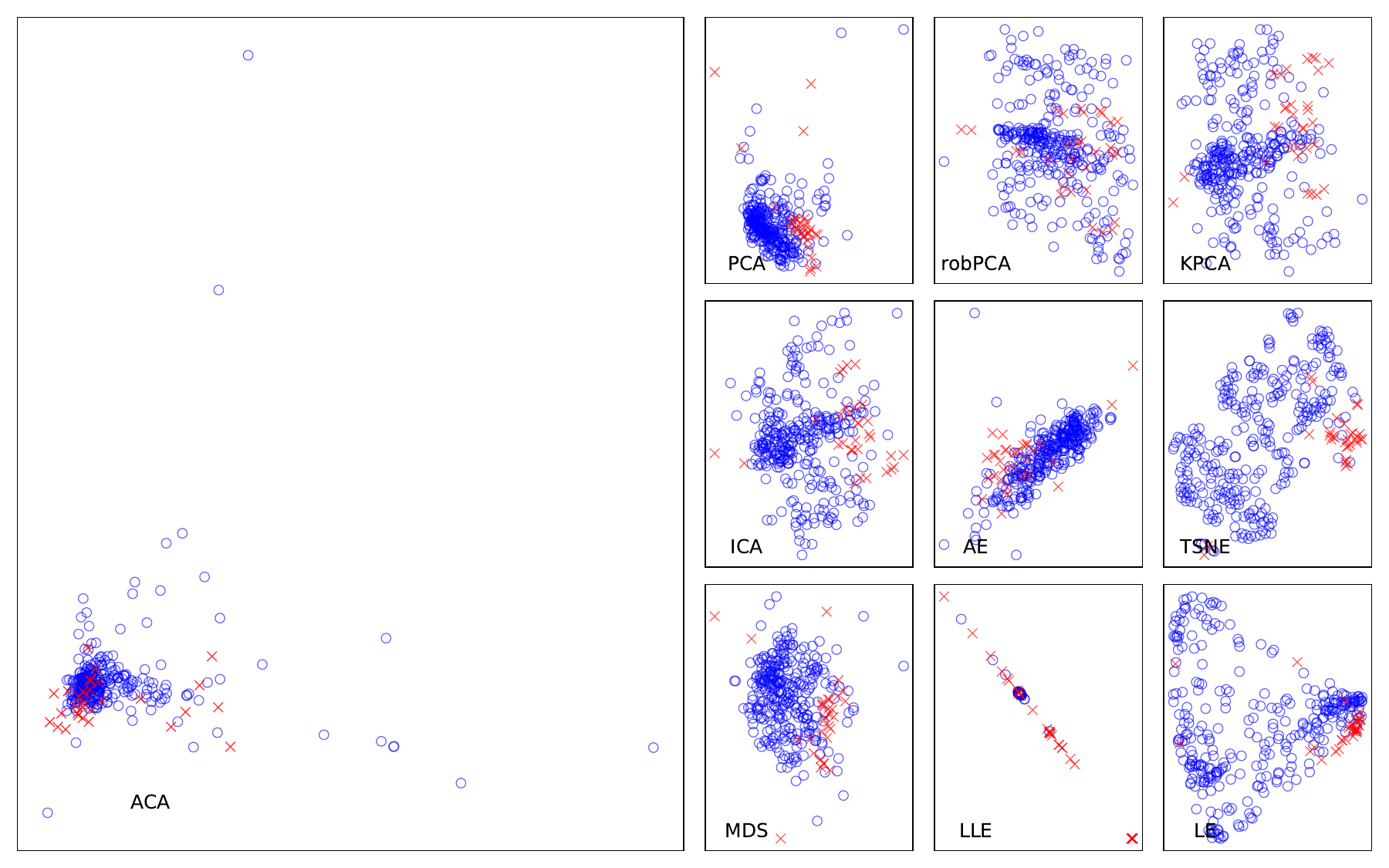}
\caption{Visualization of projection on two first components obtained by different methods for Stamps dataset.}\label{fig:realStamps}
\end{figure}

\begin{table}[H]
\centering
\begin{tabular}{c|cccccccc|}
\cline{2-9}
& \multicolumn{8}{c|}{Stamps}                                                                                                \\ \cline{2-9} 
& \multicolumn{2}{c|}{PCA}                 & \multicolumn{2}{c|}{robPCA}              & \multicolumn{2}{c|}{ICA}                 & \multicolumn{2}{c|}{ACA} \\ \cline{2-9} 
& PC1      & \multicolumn{1}{c|}{PC2}      & robPC1   & \multicolumn{1}{c|}{robPC2}   & IC1      & \multicolumn{1}{c|}{IC2}      & AC1         & AC2        \\ \hline
\multicolumn{1}{|c|}{Var1} & 4 (15\%) & \multicolumn{1}{c|}{7 (26\%)} & 5 (69\%) & \multicolumn{1}{c|}{2 (70\%)} & 9 (25\%) & \multicolumn{1}{c|}{2 (49\%)} & 3 (41\%)    & 6 (48\%)   \\
\multicolumn{1}{|c|}{Var2} & 6 (14\%) & \multicolumn{1}{c|}{8 (16\%)} & 4 (8\%)  & \multicolumn{1}{c|}{3 (14\%)} & 2 (22\%) & \multicolumn{1}{c|}{4 (13\%)} & 4 (21\%)    & 1 (18\%)   \\
\multicolumn{1}{|c|}{Var3} & 5 (13\%) & \multicolumn{1}{c|}{6 (16\%)} & 7 (7\%)  & \multicolumn{1}{c|}{4 (10\%)} & 5 (15\%) & \multicolumn{1}{c|}{5 (9\%)}  & 6 (12\%)    & 3 (10\%)   \\
\multicolumn{1}{|c|}{Var4} & 9 (13\%) & \multicolumn{1}{c|}{5 (13\%)} & 8 (4\%)  & \multicolumn{1}{c|}{1 (4\%)}  & 4 (13\%) & \multicolumn{1}{c|}{3 (8\%)}  & 8 (10\%)    & 4 (10\%)   \\
\multicolumn{1}{|c|}{Var5} & 1 (12\%) & \multicolumn{1}{c|}{2 (12\%)} & 6 (4\%)  & \multicolumn{1}{c|}{6 (0\%)}  & 3 (7\%)  & \multicolumn{1}{c|}{9 (7\%)}  & 1 (7\%)     & 8 (6\%)    \\
\hline
\end{tabular}
\caption{Most important variables for every method applied to Stamps dataset.}
\end{table}

\begin{figure}[!h]
\centering
\includegraphics[width=\textwidth]{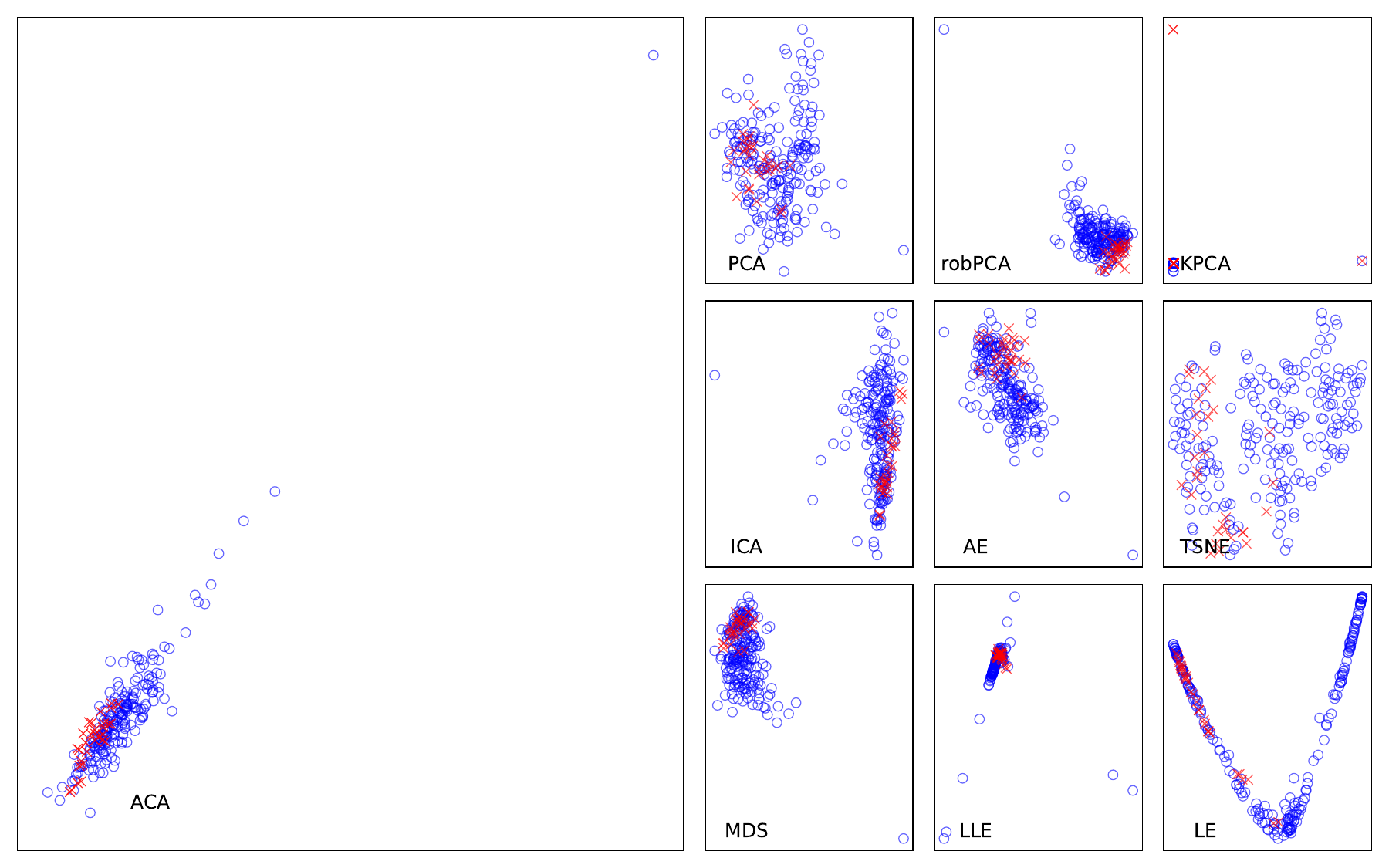}
\caption{Visualization of projection on two first components obtained by different methods for vertebral dataset.}\label{fig:realvertebral}
\end{figure}

\begin{table}[H]
\centering
\begin{tabular}{c|cccccccc|}
\cline{2-9}
& \multicolumn{8}{c|}{Vertebral}                                                                                             \\ \cline{2-9} 
& \multicolumn{2}{c|}{PCA}                 & \multicolumn{2}{c|}{robPCA}              & \multicolumn{2}{c|}{ICA}                 & \multicolumn{2}{c|}{ACA} \\ \cline{2-9} 
& PC1      & \multicolumn{1}{c|}{PC2}      & robPC1   & \multicolumn{1}{c|}{robPC2}   & IC1      & \multicolumn{1}{c|}{IC2}      & AC1         & AC2        \\ \hline
\multicolumn{1}{|c|}{Var1} & 1 (24\%) & \multicolumn{1}{c|}{5 (37\%)} & 6 (25\%) & \multicolumn{1}{c|}{4 (36\%)} & 6 (38\%) & \multicolumn{1}{c|}{3 (30\%)} & 6 (41\%)    & 1 (30\%)   \\
\multicolumn{1}{|c|}{Var2} & 3 (21\%) & \multicolumn{1}{c|}{2 (28\%)} & 2 (21\%) & \multicolumn{1}{c|}{6 (29\%)} & 3 (19\%) & \multicolumn{1}{c|}{1 (26\%)} & 3 (37\%)    & 4 (29\%)   \\
\multicolumn{1}{|c|}{Var3} & 4 (20\%) & \multicolumn{1}{c|}{4 (23\%)} & 1 (19\%) & \multicolumn{1}{c|}{2 (14\%)} & 1 (15\%) & \multicolumn{1}{c|}{4 (21\%)} & 2 (12\%)    & 2 (26\%)   \\
\multicolumn{1}{|c|}{Var4} & 6 (20\%) & \multicolumn{1}{c|}{6 (6\%)}  & 4 (16\%) & \multicolumn{1}{c|}{5 (12\%)} & 5 (14\%) & \multicolumn{1}{c|}{5 (14\%)} & 5 (5\%)     & 6 (6\%)    \\
\multicolumn{1}{|c|}{Var5} & 2 (14\%) & \multicolumn{1}{c|}{3 (4\%)}  & 5 (15\%) & \multicolumn{1}{c|}{3 (7\%)}  & 4 (13\%) & \multicolumn{1}{c|}{2 (6\%)}  & 4 (3\%)     & 3 (5\%)    \\
\multicolumn{1}{|c|}{Var6} & 5 (2\%)  & \multicolumn{1}{c|}{1 (2\%)}  & 3 (3\%)  & \multicolumn{1}{c|}{1 (3\%)}  & 2 (2\%)  & \multicolumn{1}{c|}{6 (5\%)}  & 1 (1\%)     & 5 (4\%)    \\ \hline
\end{tabular}
\caption{Most important variables for every method applied to Vertebral dataset.}
\end{table}

\begin{figure}[!h]
\centering
\includegraphics[width=\textwidth]{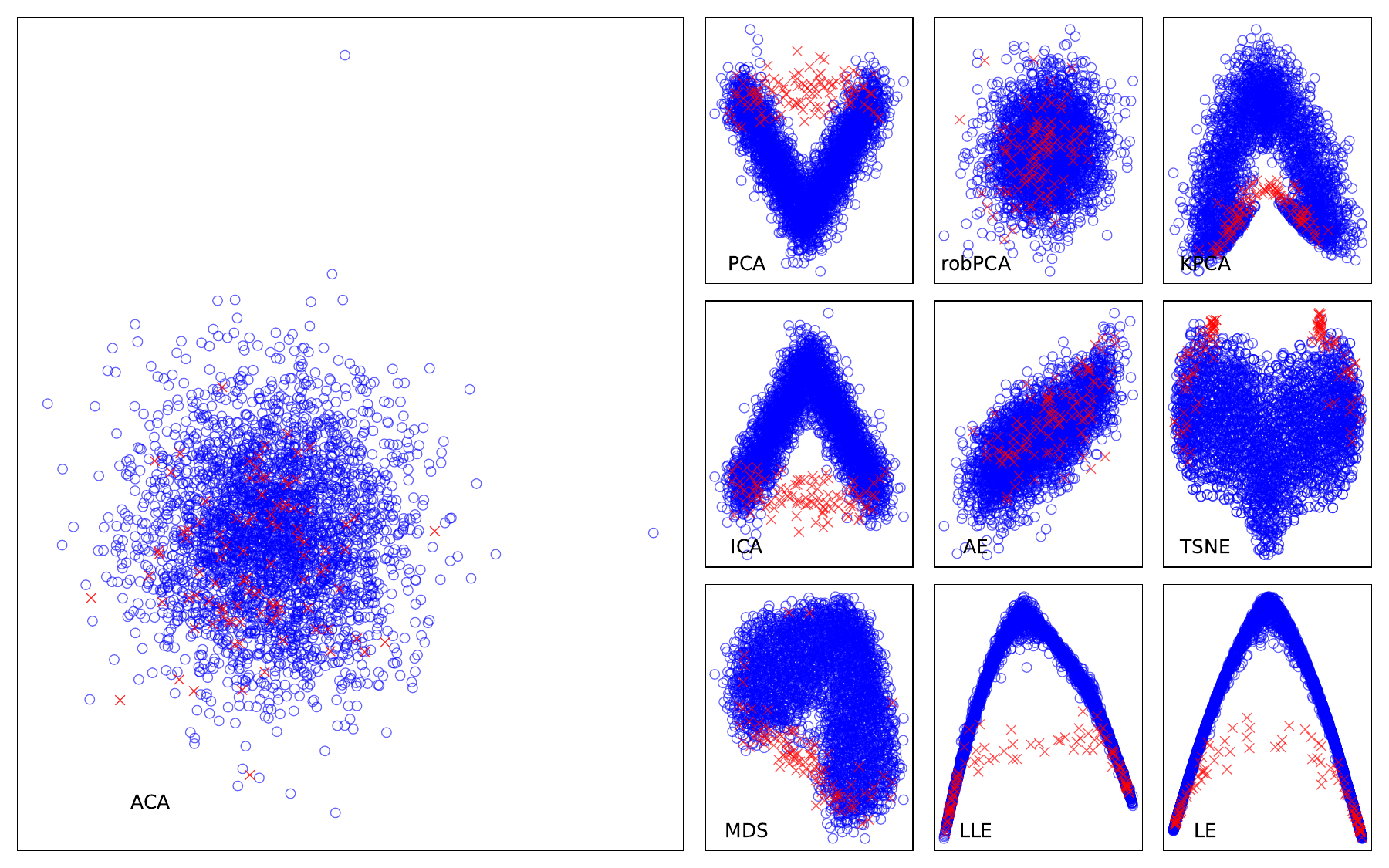}
\caption{Visualization of projection on two first components obtained by different methods for Waveform dataset.}\label{fig:realWaveform}
\end{figure}

\begin{table}[H]
\centering
\begin{tabular}{c|cccccccc|}
\cline{2-9}
& \multicolumn{8}{c|}{Waveform}                                                                                              \\ \cline{2-9} 
& \multicolumn{2}{c|}{PCA}                  & \multicolumn{2}{c|}{robPCA}                & \multicolumn{2}{c|}{ICA}                  & \multicolumn{2}{c|}{ACA} \\ \cline{2-9} 
& PC1      & \multicolumn{1}{c|}{PC2}       & robPC1    & \multicolumn{1}{c|}{robPC2}    & IC1      & \multicolumn{1}{c|}{IC2}       & AC1         & AC2        \\ \hline
\multicolumn{1}{|c|}{Var1}  & 15 (7\%) & \multicolumn{1}{c|}{11 (13\%)} & 12 (13\%) & \multicolumn{1}{c|}{15 (15\%)} & 7 (9\%)  & \multicolumn{1}{c|}{11 (13\%)} & 15 (12\%)   & 12 (12\%)  \\
\multicolumn{1}{|c|}{Var2}  & 7 (7\%)  & \multicolumn{1}{c|}{12 (10\%)} & 7 (11\%)  & \multicolumn{1}{c|}{14 (12\%)} & 15 (9\%) & \multicolumn{1}{c|}{12 (10\%)} & 10 (11\%)   & 19 (11\%)  \\
\multicolumn{1}{|c|}{Var3}  & 8 (7\%)  & \multicolumn{1}{c|}{10 (10\%)} & 4 (11\%)  & \multicolumn{1}{c|}{19 (10\%)} & 8 (8\%)  & \multicolumn{1}{c|}{10 (10\%)} & 13 (10\%)   & 14 (9\%)   \\
\multicolumn{1}{|c|}{Var4}  & 14 (7\%) & \multicolumn{1}{c|}{5 (6\%)}   & 21 (9\%)  & \multicolumn{1}{c|}{4 (9\%)}   & 6 (8\%)  & \multicolumn{1}{c|}{5 (7\%)}   & 17 (8\%)    & 16 (9\%)   \\
\multicolumn{1}{|c|}{Var5}  & 16 (7\%) & \multicolumn{1}{c|}{13 (6\%)}  & 6 (8\%)   & \multicolumn{1}{c|}{10 (8\%)}  & 14 (8\%) & \multicolumn{1}{c|}{13 (7\%)}  & 11 (8\%)    & 13 (8\%)   \\
\hline
\end{tabular}
\caption{Most important variables for every method applied to Waveform dataset.}
\end{table}

\begin{figure}[!h]
\centering
\includegraphics[width=\textwidth]{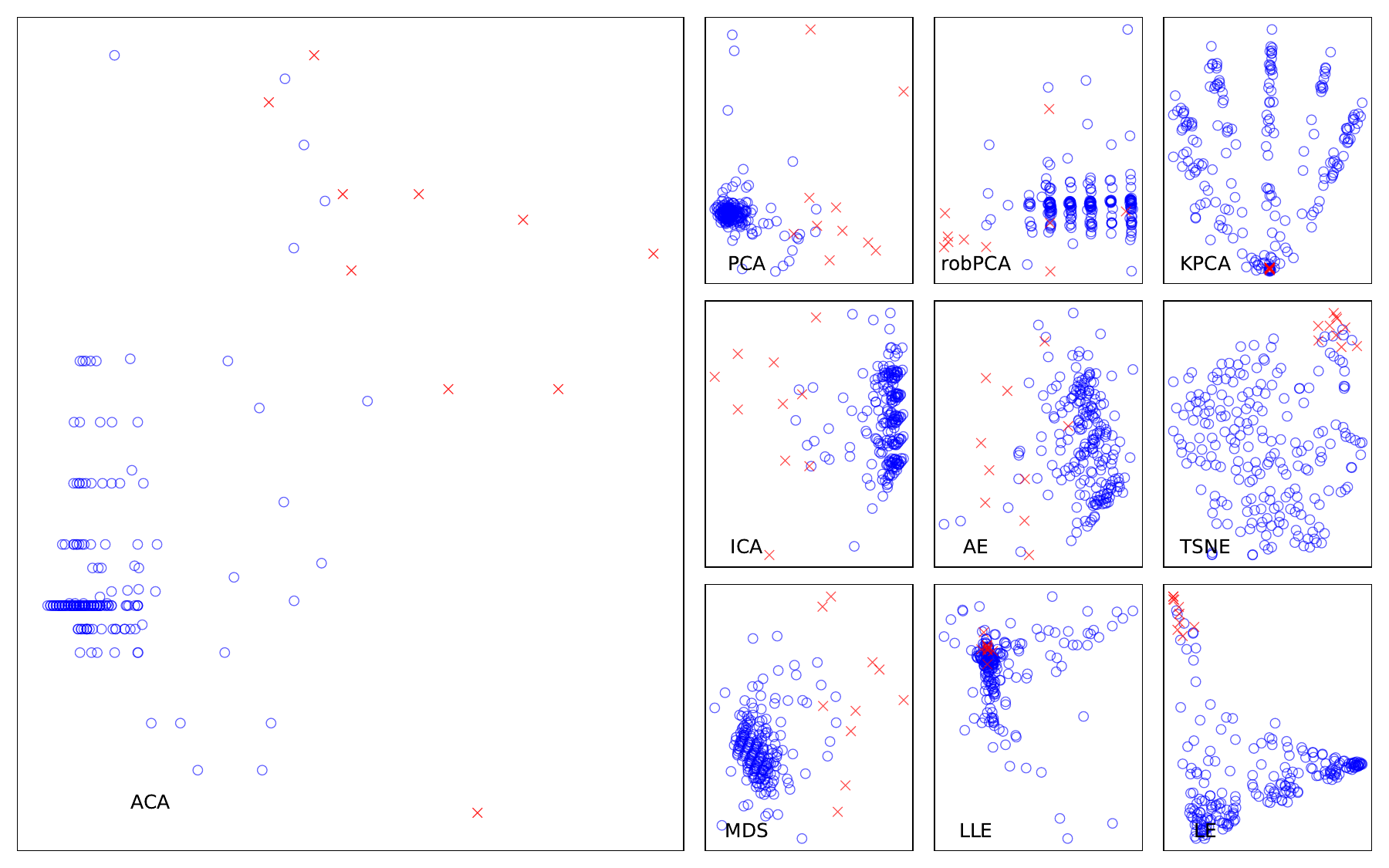}
\caption{Visualization of projection on two first components obtained by different methods for WBC dataset.}\label{fig:realWBC}
\end{figure}

\begin{table}[H]
\centering
\begin{tabular}{c|cccccccc|}
\cline{2-9}
& \multicolumn{8}{c|}{WBC}                                                                                                   \\ \cline{2-9} 
& \multicolumn{2}{c|}{PCA}                 & \multicolumn{2}{c|}{robPCA}              & \multicolumn{2}{c|}{ICA}                 & \multicolumn{2}{c|}{ACA} \\ \cline{2-9} 
& PC1      & \multicolumn{1}{c|}{PC2}      & robPC1   & \multicolumn{1}{c|}{robPC2}   & IC1      & \multicolumn{1}{c|}{IC2}      & AC1         & AC2        \\ \hline
\multicolumn{1}{|c|}{Var1} & 2 (14\%) & \multicolumn{1}{c|}{9 (48\%)} & 1 (89\%) & \multicolumn{1}{c|}{5 (58\%)} & 6 (21\%) & \multicolumn{1}{c|}{1 (55\%)} & 8 (21\%)    & 6 (70\%)   \\
\multicolumn{1}{|c|}{Var2} & 3 (14\%) & \multicolumn{1}{c|}{5 (17\%)} & 4 (4\%)  & \multicolumn{1}{c|}{6 (21\%)} & 8 (15\%) & \multicolumn{1}{c|}{6 (22\%)} & 3 (15\%)    & 8 (27\%)   \\
\multicolumn{1}{|c|}{Var3} & 8 (13\%) & \multicolumn{1}{c|}{1 (8\%)}  & 2 (4\%)  & \multicolumn{1}{c|}{2 (10\%)} & 3 (15\%) & \multicolumn{1}{c|}{7 (8\%)}  & 4 (15\%)    & 9 (3\%)    \\
\multicolumn{1}{|c|}{Var4} & 6 (12\%) & \multicolumn{1}{c|}{7 (7\%)}  & 5 (2\%)  & \multicolumn{1}{c|}{8 (3\%)}  & 2 (15\%) & \multicolumn{1}{c|}{5 (6\%)}  & 9 (13\%)    & 2 (0\%)    \\
\multicolumn{1}{|c|}{Var5} & 4 (11\%) & \multicolumn{1}{c|}{8 (7\%)}  & 3 (2\%)  & \multicolumn{1}{c|}{3 (2\%)}  & 7 (12\%) & \multicolumn{1}{c|}{8 (5\%)}  & 5 (12\%)    & 4 (0\%)    \\
\hline
\end{tabular}
\caption{Most important variables for every method applied to WBC dataset.}
\end{table}

\begin{figure}[!h]
\centering
\includegraphics[width=\textwidth]{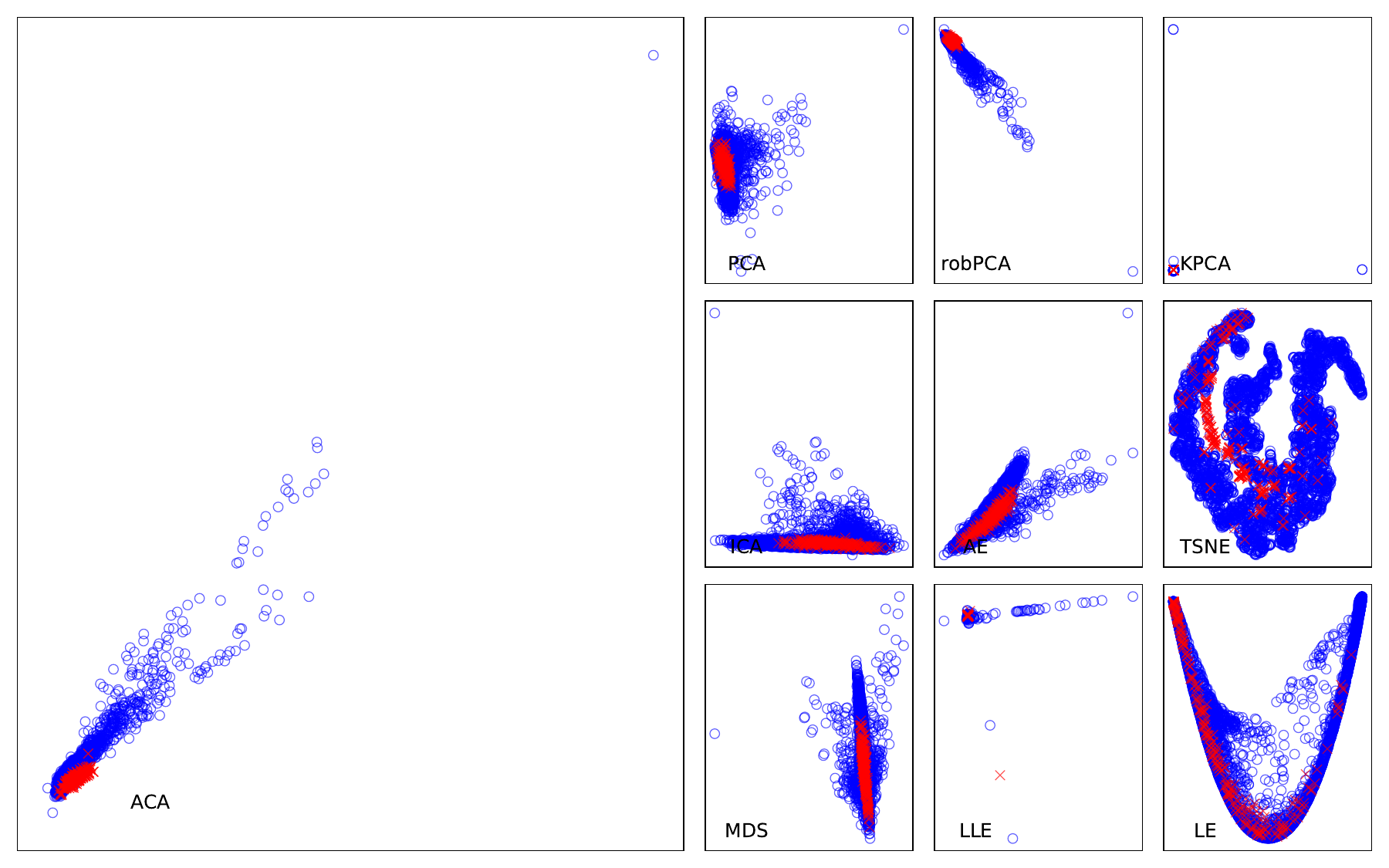}
\caption{Visualization of projection on two first components obtained by different methods for Wilt dataset.}\label{fig:realWilt}
\end{figure}

\begin{table}[H]
\centering
\begin{tabular}{c|cccccccc|}
\cline{2-9}
& \multicolumn{8}{c|}{Wilt}                                                                                                  \\ \cline{2-9} 
& \multicolumn{2}{c|}{PCA}                 & \multicolumn{2}{c|}{robPCA}              & \multicolumn{2}{c|}{ICA}             & \multicolumn{2}{c|}{ACA} \\ \cline{2-9} 
& PC1      & \multicolumn{1}{c|}{PC2}      & robPC1   & \multicolumn{1}{c|}{robPC2}   & IC1      & \multicolumn{1}{c|}{IC2}      & AC1         & AC2        \\ \hline
\multicolumn{1}{|c|}{Var1} & 2 (32\%) & \multicolumn{1}{c|}{1 (39\%)} & 3 (93\%) & \multicolumn{1}{c|}{2 (54\%)} & 4 (72\%) & \multicolumn{1}{c|}{3 (48\%)} & 3 (79\%)    & 2 (68\%)   \\
\multicolumn{1}{|c|}{Var2} & 3 (32\%) & \multicolumn{1}{c|}{4 (23\%)} & 4 (4\%)  & \multicolumn{1}{c|}{5 (36\%)} & 3 (16\%) & \multicolumn{1}{c|}{2 (47\%)} & 2 (13\%)    & 3 (10\%)   \\
\multicolumn{1}{|c|}{Var3} & 5 (15\%) & \multicolumn{1}{c|}{3 (14\%)} & 2 (2\%)  & \multicolumn{1}{c|}{1 (7\%)}  & 2 (12\%) & \multicolumn{1}{c|}{4 (3\%)}  & 5 (4\%)     & 4 (8\%)    \\
\multicolumn{1}{|c|}{Var4} & 4 (14\%) & \multicolumn{1}{c|}{5 (13\%)} & 1 (1\%)  & \multicolumn{1}{c|}{4 (1\%)}  & 1 (0\%)  & \multicolumn{1}{c|}{5 (2\%)}  & 4 (3\%)     & 5 (8\%)    \\
\multicolumn{1}{|c|}{Var5} & 1 (7\%)  & \multicolumn{1}{c|}{2 (11\%)} & 5 (0\%)  & \multicolumn{1}{c|}{3 (1\%)}  & 5 (0\%)  & \multicolumn{1}{c|}{1 (1\%)}  & 1 (1\%)     & 1 (7\%)    \\ \hline
\end{tabular}
\caption{Most important variables for every method applied to Wilt dataset.}
\end{table}

\begin{figure}[!h]
\centering
\includegraphics[width=\textwidth]{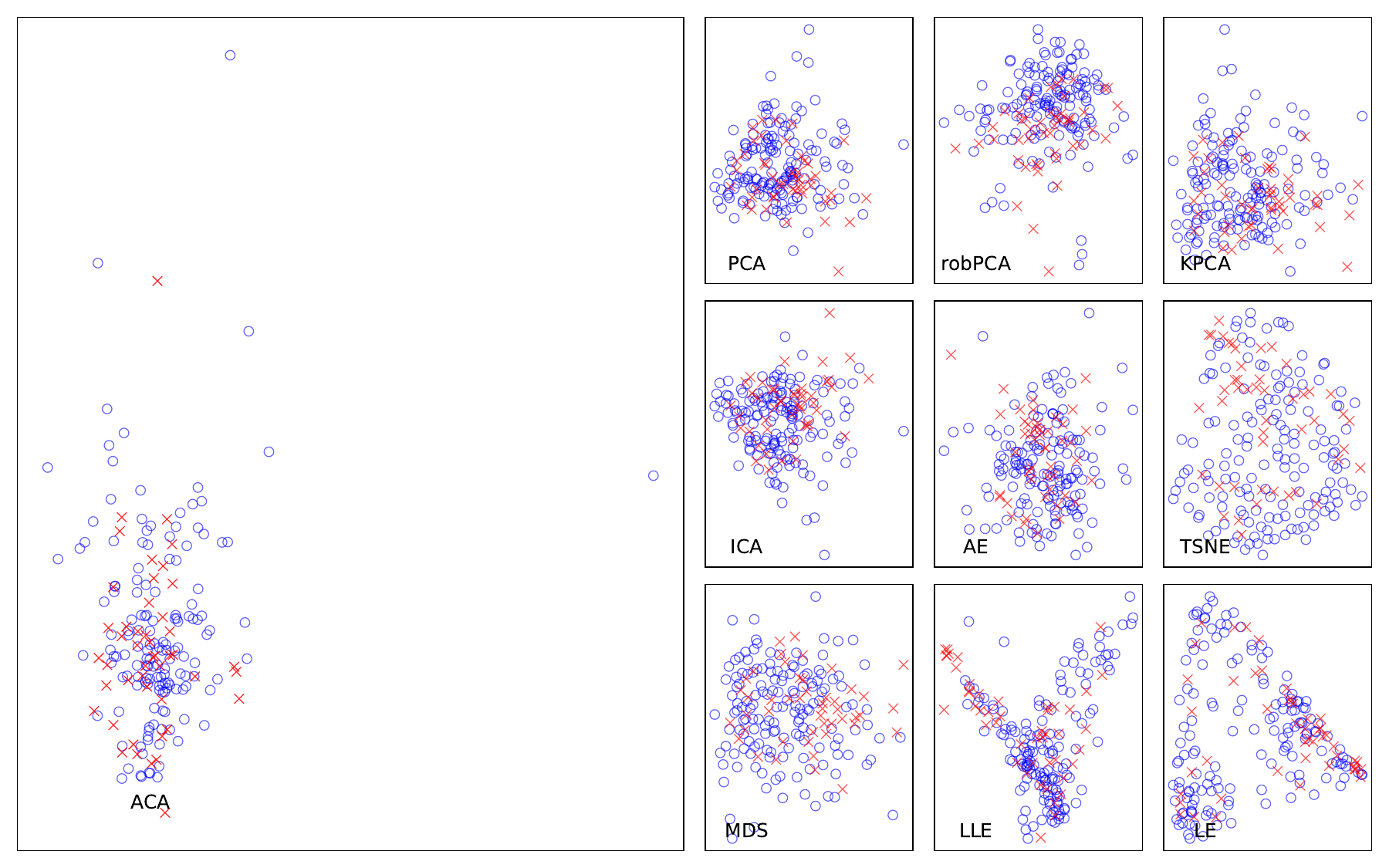}
\caption{Visualization of projection on two first components obtained by different methods for WPBC dataset.}\label{fig:realWPBC}
\end{figure}

\begin{table}[H]
\centering
\begin{tabular}{c|cccccccc|}
\cline{2-9}
& \multicolumn{8}{c|}{WPBC}                                                                                                                        \\ \cline{2-9} 
& \multicolumn{2}{c|}{PCA}                 & \multicolumn{2}{c|}{robPCA}               & \multicolumn{2}{c|}{ICA}                 & \multicolumn{2}{c|}{ACA} \\ \cline{2-9} 
& PC1      & \multicolumn{1}{c|}{PC2}      & robPC1   & \multicolumn{1}{c|}{robPC2}    & IC1      & \multicolumn{1}{c|}{IC2}      & AC1         & AC2        \\ \hline
\multicolumn{1}{|c|}{Var1}  & 9 (6\%)  & \multicolumn{1}{c|}{11 (6\%)} & 3 (26\%) & \multicolumn{1}{c|}{33 (18\%)} & 9 (6\%)  & \multicolumn{1}{c|}{11 (5\%)} & 16 (12\%)   & 13 (15\%)  \\
\multicolumn{1}{|c|}{Var2}  & 8 (6\%)  & \multicolumn{1}{c|}{31 (6\%)} & 5 (23\%) & \multicolumn{1}{c|}{1 (8\%)}   & 8 (6\%)  & \multicolumn{1}{c|}{2 (5\%)}  & 18 (10\%)   & 28 (11\%)  \\
\multicolumn{1}{|c|}{Var3}  & 24 (5\%) & \multicolumn{1}{c|}{27 (5\%)} & 2 (15\%) & \multicolumn{1}{c|}{31 (6\%)}  & 7 (6\%)  & \multicolumn{1}{c|}{6 (5\%)}  & 21 (7\%)    & 17 (7\%)   \\
\multicolumn{1}{|c|}{Var4}  & 4 (5\%)  & \multicolumn{1}{c|}{26 (5\%)} & 1 (10\%) & \multicolumn{1}{c|}{30 (6\%)}  & 29 (5\%) & \multicolumn{1}{c|}{5 (5\%)}  & 19 (7\%)    & 4 (7\%)    \\
\multicolumn{1}{|c|}{Var5}  & 14 (5\%) & \multicolumn{1}{c|}{2 (4\%)}  & 6 (7\%)  & \multicolumn{1}{c|}{13 (5\%)}  & 24 (4\%) & \multicolumn{1}{c|}{27 (5\%)} & 14 (5\%)    & 10 (6\%)   \\
\hline
\end{tabular}
\caption{Most important variables for every method applied to WPBC dataset.}
\end{table}

\begin{figure}[!h]
\centering
\includegraphics[width=\textwidth]{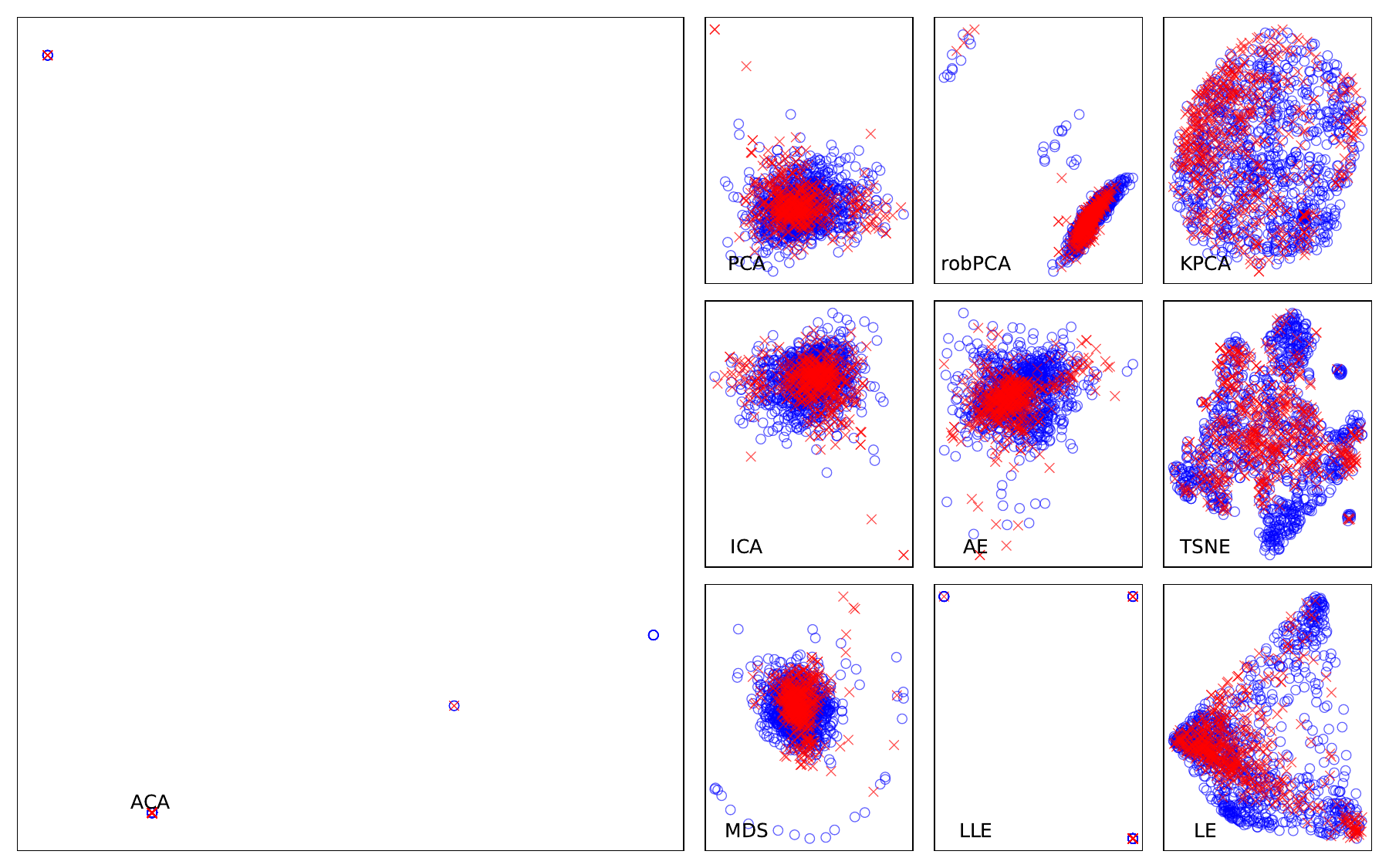}
\caption{Visualization of projection on two first components obtained by different methods for yeast dataset.}\label{fig:realyeast}
\end{figure}

\begin{table}[H]
\centering
\begin{tabular}{c|cccccccc|}
\cline{2-9}
& \multicolumn{8}{c|}{Yeast}                                                                                                 \\ \cline{2-9} 
& \multicolumn{2}{c|}{PCA}                 & \multicolumn{2}{c|}{robPCA}              & \multicolumn{2}{c|}{ICA}                 & \multicolumn{2}{c|}{ACA} \\ \cline{2-9} 
& PC1      & \multicolumn{1}{c|}{PC2}      & robPC1   & \multicolumn{1}{c|}{robPC2}   & IC1      & \multicolumn{1}{c|}{IC2}      & AC1         & AC2        \\ \hline
\multicolumn{1}{|c|}{Var1} & 2 (28\%) & \multicolumn{1}{c|}{7 (27\%)} & 5 (33\%) & \multicolumn{1}{c|}{5 (33\%)} & 2 (28\%) & \multicolumn{1}{c|}{7 (26\%)} & 6 (82\%)    & 5 (82\%)   \\
\multicolumn{1}{|c|}{Var2} & 1 (27\%) & \multicolumn{1}{c|}{4 (23\%)} & 1 (30\%) & \multicolumn{1}{c|}{1 (33\%)} & 1 (27\%) & \multicolumn{1}{c|}{4 (23\%)} & 5 (18\%)    & 6 (18\%)   \\
\multicolumn{1}{|c|}{Var3} & 3 (16\%) & \multicolumn{1}{c|}{3 (20\%)} & 3 (15\%) & \multicolumn{1}{c|}{6 (15\%)} & 3 (17\%) & \multicolumn{1}{c|}{8 (19\%)} & 8 (0\%)     & 4 (0\%)    \\
\multicolumn{1}{|c|}{Var4} & 4 (8\%)  & \multicolumn{1}{c|}{8 (19\%)} & 2 (14\%) & \multicolumn{1}{c|}{3 (9\%)}  & 7 (9\%)  & \multicolumn{1}{c|}{3 (19\%)} & 1 (0\%)     & 2 (0\%)    \\
\multicolumn{1}{|c|}{Var5} & 7 (8\%)  & \multicolumn{1}{c|}{1 (6\%)}  & 6 (8\%)  & \multicolumn{1}{c|}{2 (5\%)}  & 4 (7\%)  & \multicolumn{1}{c|}{1 (7\%)}  & 4 (0\%)     & 3 (0\%)    \\
\hline
\end{tabular}
\caption{Most important variables for every method applied to Yeast dataset.}
\end{table}

\section{Non-linear structures}\label{sup:sec:nonlinear}

\begin{figure}[H]
    \centering
    \begin{tabular}{cc}
        S-curve & ACA view\\
        \includegraphics[width=0.375\textwidth, page=1, trim=0cm 0cm 0cm 0cm, clip]{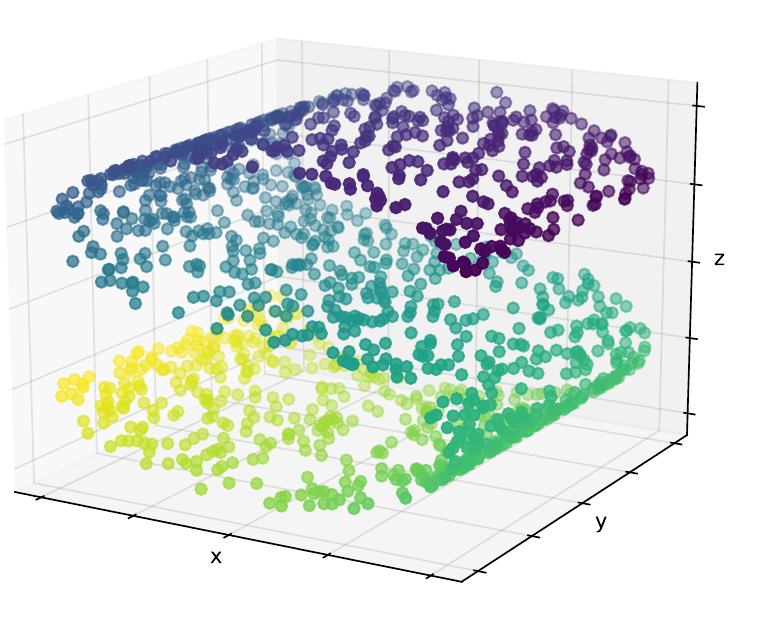} & \includegraphics[width=0.375\textwidth, page=2, trim=0cm 0cm 0cm 0cm, clip]{figures/dataset/s_curve.pdf}\\
        Swiss roll & ACA view\\
        \includegraphics[width=0.375\textwidth, page=1, trim=0cm 0cm 0cm 0cm, clip]{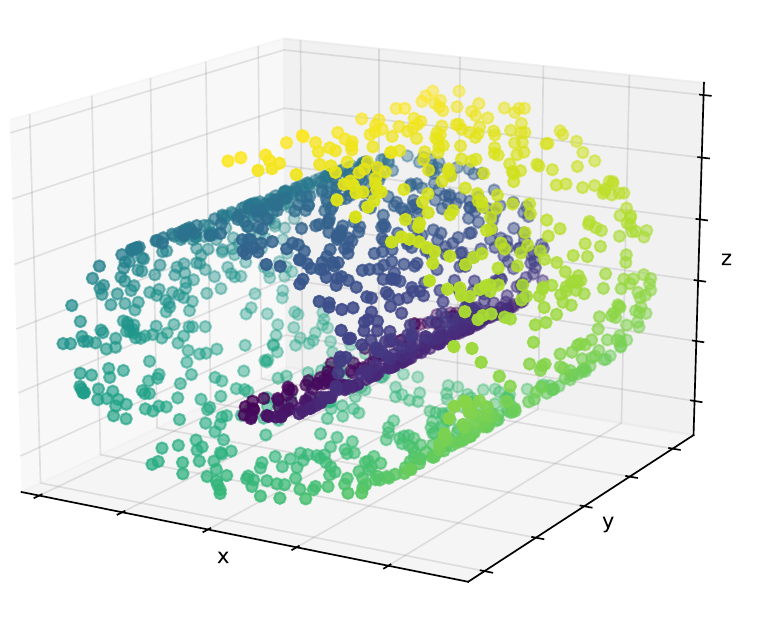} & \includegraphics[width=0.375\textwidth, page=2, trim=0cm 0cm 0cm 0cm, clip]{figures/dataset/swiss_roll.pdf}\\
        Sphere & ACA view\\
        \includegraphics[width=0.375\textwidth, page=1, trim=0cm 0cm 0cm 0cm, clip]{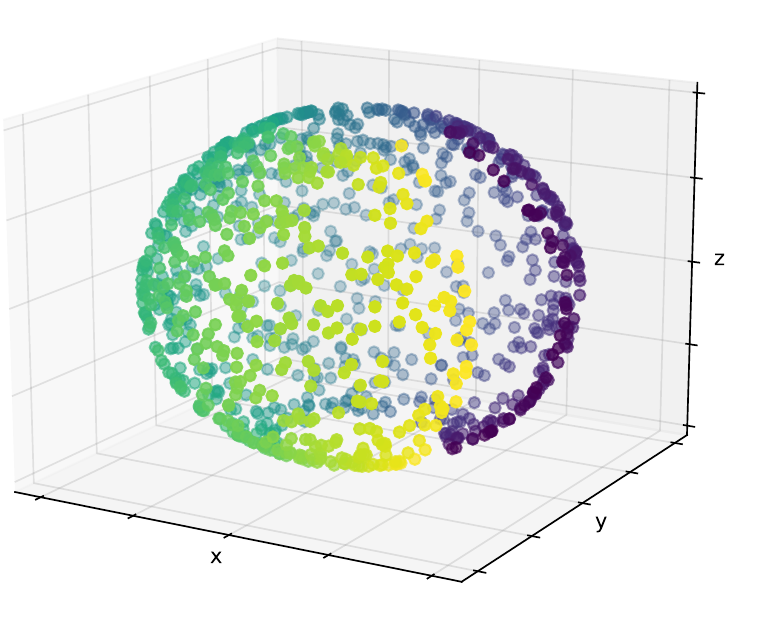} & \includegraphics[width=0.375\textwidth, page=2, trim=0cm 0cm 0cm 0cm, clip]{figures/dataset/sphere.pdf}\\
    \end{tabular}
    \caption{Three-dimensional toy datasets with non-linear structure.}
    \label{fig:nonlinear}
\end{figure}

\begin{figure}[H]
    \centering
    \begin{tabular}{cc}
        S-curve & ACA view\\
        \includegraphics[width=0.375\textwidth, page=1, trim=0cm 0cm 0cm 0cm, clip]{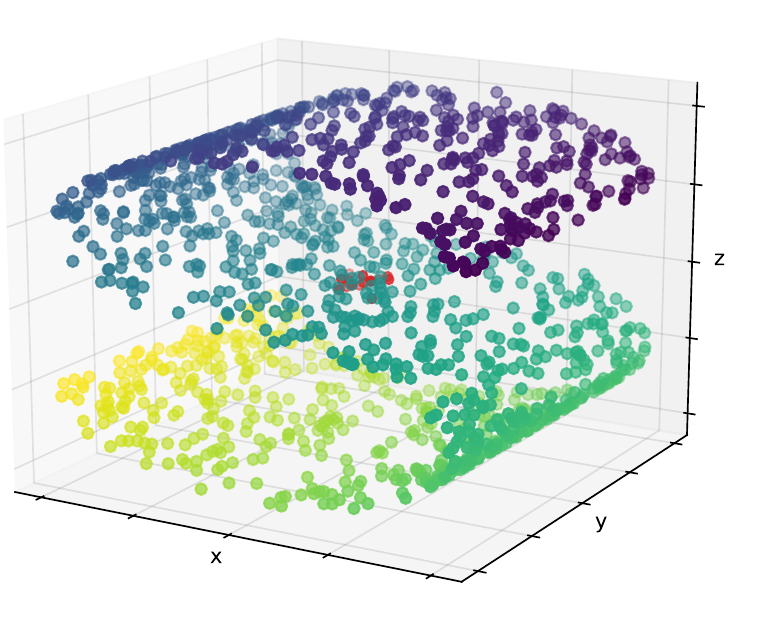} & \includegraphics[width=0.375\textwidth, page=2, trim=0cm 0cm 0cm 0cm, clip]{figures/dataset/s_curve_in.pdf}\\
        Swiss roll & ACA view\\
        \includegraphics[width=0.375\textwidth, page=1, trim=0cm 0cm 0cm 0cm, clip]{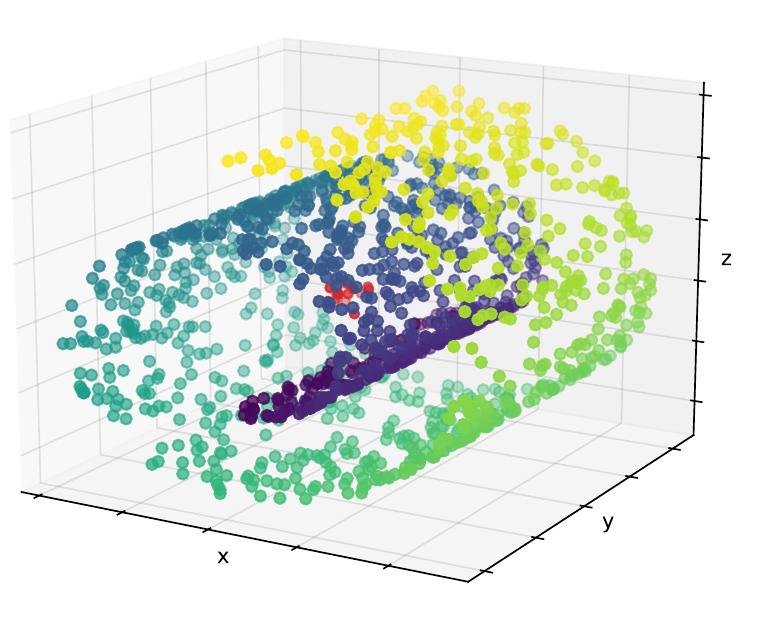} & \includegraphics[width=0.375\textwidth, page=2, trim=0cm 0cm 0cm 0cm, clip]{figures/dataset/swiss_roll_in.pdf}\\
        Sphere & ACA view\\
        \includegraphics[width=0.375\textwidth, page=1, trim=0cm 0cm 0cm 0cm, clip]{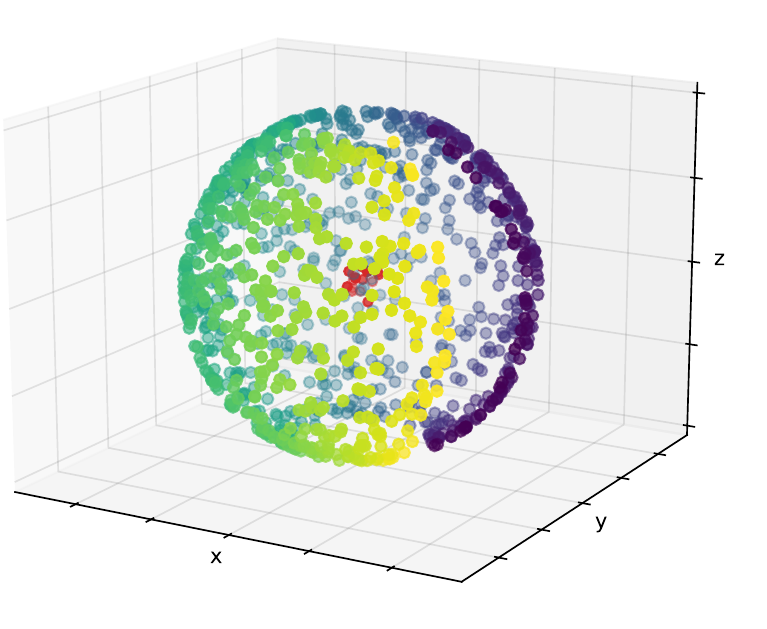} & \includegraphics[width=0.375\textwidth, page=2, trim=0cm 0cm 0cm 0cm, clip]{figures/dataset/sphere_in.pdf}\\
    \end{tabular}
    \caption{Three-dimensional toy datasets with non-linear structure contaminated with anomalies inside.}
    \label{fig:nonlinearIN}
\end{figure}

\begin{figure}[H]
    \centering
    \begin{tabular}{cc}
        S-curve & ACA view\\
        \includegraphics[width=0.375\textwidth, page=1, trim=0cm 0cm 0cm 0cm, clip]{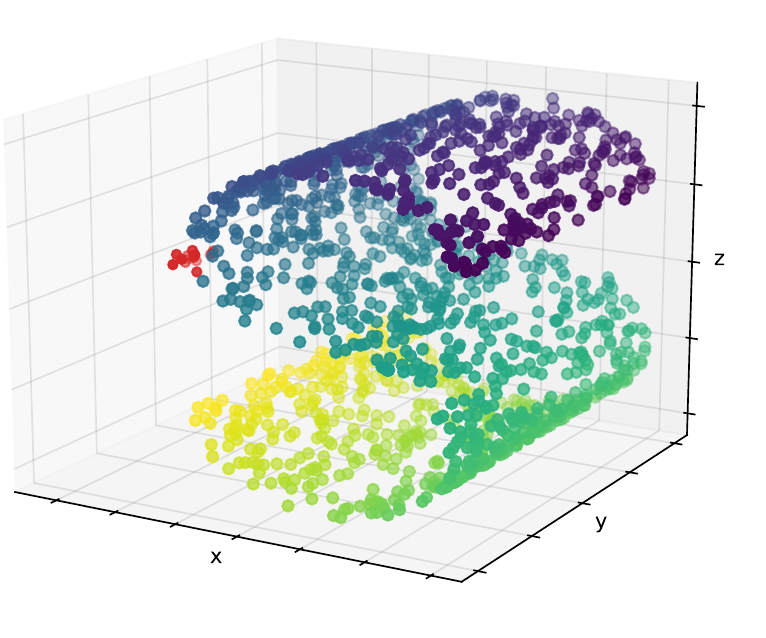} & \includegraphics[width=0.375\textwidth, page=2, trim=0cm 0cm 0cm 0cm, clip]{figures/dataset/s_curve_out.pdf}\\
        Swiss roll & ACA view\\
        \includegraphics[width=0.375\textwidth, page=1, trim=0cm 0cm 0cm 0cm, clip]{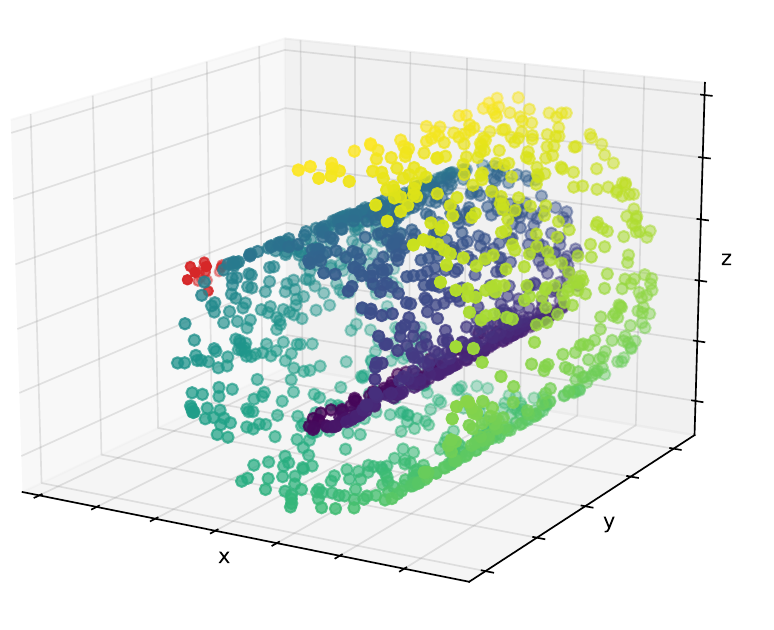} & \includegraphics[width=0.375\textwidth, page=2, trim=0cm 0cm 0cm 0cm, clip]{figures/dataset/swiss_roll_out.pdf}\\
        Sphere & ACA view\\
        \includegraphics[width=0.375\textwidth, page=1, trim=0cm 0cm 0cm 0cm, clip]{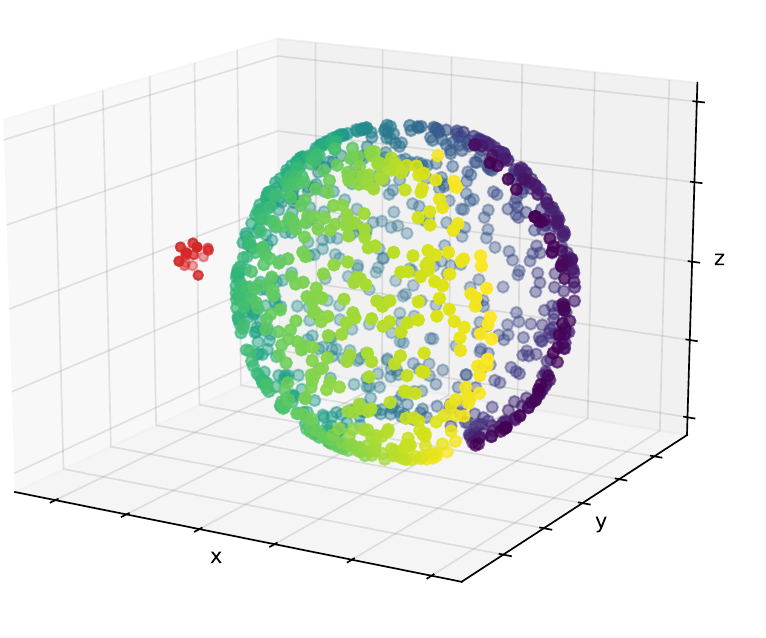} & \includegraphics[width=0.375\textwidth, page=2, trim=0cm 0cm 0cm 0cm, clip]{figures/dataset/sphere_out.pdf}\\
    \end{tabular}
    \caption{Three-dimensional toy datasets with non-linear structure contaminated with anomalies outside.}
    \label{fig:nonlinearOUT}
\end{figure}

This section proposes an application of ACA to non-linear structures with $3$-dimensional toy datasets: S-curve, Swiss roll and Sphere used in three different settings. Figure~\ref{fig:nonlinear} shows data without any contamination with $n=1500$ points while Figure~\ref{fig:nonlinearIN} and Figure~\ref{fig:nonlinearOUT} present results when $1\%$ of anomalies are added inside and outside of the structure, respectively. As expected none of the settings achieve an interesting representation of the structures but in case where anomalies are separated from the (considered) ``normal'' data, ACA proposes a good visualisation with first component directed towards the outlying cluster.

\vspace{\fill}

\section{Spherical Nelder-Mead method}\label{sup:sec:nm}

For a faster and direct reference, in this section we present the modified Nelder-Mead algorithm, adjusted to search of an abnormal component (\textit{i.e.}, a point on the unit sphere), restricted by orthogonality to already existing (assembled on the previous steps) sub-basis. Several times with random initialization each, the \texttt{projDepthNM} routine is executed in Step~4 of Algorithm~\ref{alg:ACA}. Clearly, another algorithm can be chosen instead, like coordinate descent or refined random search, with the choice left for the practitioner.

\renewcommand{\baselinestretch}{1.0}
\begin{algorithm}[H]
\caption{Spherical restricted Nelder-Mead method}\label{alg.spNelderMead}
\begin{algorithmic}[1]\footnotesize
\Function{projDepthNM}{$\bmz,\bmX,\bmB,\texttt{depth},\beta,\alpha,\gamma,\rho,\sigma$}
\If{\texttt{Start = Mn}}\ $\bmu\gets\bmz-\ol{\bmx}$\EndIf
\If{\texttt{Start = Rn}}\ $\bmu\gets\bmB\bmu_{t}\sim\mathcal{U}(\mathbb{S}^{d_e-1})$\EndIf
\State $\epsilon\gets(\pi/2)/\beta$ \Comment{size of the spherical cap}
\State $d\gets\text{rank}(\bmB)$ \Comment{effective dimension of the basis}
\For{i}{1}{d}
\Comment finding the starting simplex
\State $\bmp_i\gets\bmB\bmu_{t}\sim\mathcal{U}(\mathbb{S}^{d-1})\text{ s.t. }\arccos(\left\langle\bmu,\bmu_t\right\rangle)\le\epsilon$
\State $f_i\gets \texttt{depth}(\ip{\bmp_i}\bmz|\ip{\bmp_i}\bmX)$
\EndFor
\State $\textsc{sort}([(\bmp_1,f_1),\dots,(\bmp_d,f_d)])$
\Comment sort pairs $(\bmp_i,f_i)$ such that $f_1\le\dots\le f_d$
\Repeat
\State $\bmx_o\gets c_N(\bmp_1,\dots,\bmp_{d-1})$
\State $\bmx_r\gets\cos(-\alpha\theta) + \sin(-\alpha\theta) \frac{\bmy}{\|\bmy\|}\text{ with }\theta=\arccos(\left\langle\bmx_o,\bmp_d\right\rangle)\text{ and }\bmy=\bmp_d - \left\langle\bmx_o,\bmp_d\right\rangle\bmx_o$ \Comment{reflected point}
\State $f_r\gets \texttt{depth}(\ip{\bmx_r}\bmz|\ip{\bmx_r}\bmX)$ 
\If{$f_1\le f_r <f_{d-1}$} $(\bmp_d,f_d)\gets(\bmx_r,f_r)$
\ElsIf{$f_r<f_1$}
\State $\bmx_e\gets\cos(\gamma\theta) + \sin(\gamma\theta) \frac{\bmy}{\|\bmy\|}\text{ with }\theta=\arccos(\left\langle\bmx_o,\bmx_r\right\rangle)\text{ and }\bmy=\bmx_r - \left\langle\bmx_o,\bmx_r\right\rangle\bmx_o$ 
\Comment{expanded point}
\State $f_e\gets \texttt{depth}(\ip{\bmx_e}\bmz|\ip{\bmx_e}\bmX)$ 
\If{$f_e<f_r$} $(\bmp_d,f_d)\gets(\bmx_e,f_e)$ \textbf{else} $(\bmp_d,f_d)\gets(\bmx_r,f_r)$\EndIf
\Else \Comment{$f_{d-1}\le f_r$}
\If{$f_r<f_d$} $\bmx_h\gets\bmx_r$ \textbf{else} $\bmx_h\gets\bmp_d$\EndIf
\State $\bmx_c\gets\cos(\rho\theta) + \sin(\rho\theta) \frac{\bmy}{\|\bmy\|}\text{ with }\theta=\arccos(\left\langle\bmx_o,\bmx_h\right\rangle)\text{ and }\bmy=\bmx_h - \left\langle\bmx_o,\bmx_h\right\rangle\bmx_o$ 
\Comment{contracted point}
\State $f_c\gets \texttt{depth}(\ip{\bmx_c}\bmz|\ip{\bmx_c}\bmX)$
\If{$f_c<f_d$} $(\bmp_d,f_d)\gets(\bmx_c,f_c)$
\Else \Comment{reduction}
\For{i}{2}{d}
\State $\bmp_i\gets\cos(\sigma\theta) + \sin(\sigma\theta) \frac{\bmy}{\|\bmy\|}\text{ with }\theta=\arccos(\left\langle\bmp_1,\bmp_i\right\rangle)\text{ and }\bmy=\bmp_i - \left\langle\bmp_1,\bmp_i\right\rangle\bmp_1$
\State $f_i\gets \texttt{depth}(\ip{\bmp_i}\bmz|\ip{\bmp_i}\bmX)$
\EndFor
\State $\textsc{sort}([(\bmp_1,f_1),\dots,(\bmp_{d-1},f_{d-1})])$
\EndIf
\EndIf
\State $\textsc{inPlaceMerge}([(\bmp_1,f_1),\dots,(\bmp_{d-1},f_{d-1})],[(\bmp_d,f_d)])$
\Comment put $(\bmp_d,f_d)$ in the correct position
\Until stopping criterion is satisfied
\State $j_m\gets\argmin_{j\in\{1,...,d\}}f_j$
\State \Return $(f_{j_m},\bmp_{j_m})$
\EndFunction
\end{algorithmic}
\end{algorithm}

We refer the reader to~\cite{DYCKERHOFF21} for more information on the spherical version of Nelder-Mead as well as for further optimization algorithms.

\end{document}